\documentclass[11pt]{article}
\pdfoutput=1                    

\usepackage[utf8]{inputenc}
\usepackage{amsfonts}
\usepackage[symbol]{footmisc}
\usepackage{psfrag,epsf}
\usepackage{authblk}
\usepackage{caption}
\usepackage{fullpage}
\usepackage{graphicx, color, subcaption, float, hyperref, enumerate}
\hypersetup{
    colorlinks=true,
    linkcolor=blue,
    citecolor=blue}
\usepackage{amsmath,amssymb,mathtools,amsthm, bm, bbm}
\usepackage[ruled,vlined]{algorithm2e}

\usepackage[round]{natbib}
\allowdisplaybreaks

\usepackage{amsmath, amsthm, amssymb, amsthm}
\usepackage{subcaption}
\usepackage[flushleft]{threeparttable}
\usepackage{tabu,multirow}
\usepackage{color}
\usepackage{etoolbox}
\usepackage{float}
\usepackage{booktabs}
\usepackage[dvipsnames,svgnames,x11names]{xcolor}
\usepackage{setspace}
\usepackage{nameref}
\usepackage{listings}
\usepackage{bbm}
\usepackage{babel}
\newtheorem{theorem}{Theorem}[section]

\newtheorem{lemma}{Lemma}[section]
\newtheorem{corollary}{Corollary}[section]
\newtheorem{proposition}{Proposition}[section]
\theoremstyle{definition}
\newtheorem{assumption}{Assumption}[section]
\newtheorem{remark}{Remark}[section]

\newcommand{\ind}{\perp \!\!\! \perp}

\newcommand{\boldX}{\mathbf{X}}
\newcommand{\boldY}{\mathbf{Y}}
\newcommand{\boldZ}{\mathbf{Z}}
\newcommand{\bzero}{\mathbf{0}}
\newcommand{\bx}{\mathbf{x}}

\newcommand{\bh}{\mathbf{h}}
\newcommand{\bbeta}{\boldsymbol\beta}
\newcommand{\bgamma}{\boldsymbol\gamma}
\newcommand{\bX}{\mathbf{X}}
\newcommand{\bY}{\mathbf{Y}}

\newcommand{\Var}{\mathrm{Var}}
\newcommand{\Cov}{\mathrm{Cov}}

\newcommand{\wt}{\widetilde}

\newcommand{\calD}{\mathcal{D}}

\newcommand{\calF}{\mathcal{F}}

\newcommand{\calJ}{\mathcal{J}}
\newcommand{\calI}{\mathcal{I}}
\newcommand{\calL}{\mathcal{L}}
\newcommand{\calN}{\mathcal{N}}
\newcommand{\calP}{\mathcal{P}}

\newcommand{\calS}{\mathcal{S}}
\newcommand{\calU}{\mathcal{U}}

\newcommand{\calZ}{\mathcal{Z}}

\newcommand{\bbE}{\mathbb{E}}
\newcommand{\bbG}{\mathbb{G}}
\newcommand{\bbL}{\mathbb{L}}

\newcommand{\bbP}{\mathbb{P}}

\newcommand{\bbR}{\mathbb{R}}

\newcommand{\bbZ}{\mathbb{Z}}

\newcommand{\mbm}{\mathbf{m}}
\newcommand{\btheta}{\boldsymbol \theta}

\newcommand{\TV}{\mathrm{TV}}

\newcommand{\CF}{\mathrm{CF}}

\newcommand{\BDM}{\mathrm{BDM}}
\newcommand{\hBDM}{\mathrm{hBDM}}
\newcommand{\ORE}{\mathrm{ORE}}
\newcommand{\CI}{\mathrm{CI}}
\newcommand{\cvP}{\overset{\mathbb{P}}\to}

\newcommand{\GS}{\mathrm{GS}}

\newcommand{\Op}{O_{\mathbb{P}}}
\newcommand{\op}{o_{\mathbb{P}}}

\newcommand{\imp}{\text{imp}}

\newcommand{\wh}{\widehat}
\newcommand{\tm}{\widetilde{m}}
\newcommand{\iid}{\overset{\text{i.i.d.}}\sim}

\newcommand{\nK}{n_K}
\newcommand{\NK}{N_K}
\newcommand{\sK}{s_K}
\newcommand{\dd}{\mathrm{d}}

\newcommand{\qeds}{\strut \hfill $\blacksquare$}
\def\Lsc{\mathcal{L}}
\def\Usc{\mathcal{U}}

\usepackage{microtype}
\graphicspath{ {Pictures/} }

\begin{document}
 \date{}
\title{Bayesian Semi-supervised Inference via a Debiased Modeling Approach}

\author[1]{Gözde Sert}
\author[1]{Abhishek Chakrabortty}
\author[1,*]{Anirban Bhattacharya}
\affil[1]{Department of Statistics, Texas A\&M University}
\makeatletter
\renewcommand\AB@affilsepx{\\ \protect\Affilfont}
\renewcommand\Authsep{, }
\renewcommand\Authand{, }
\renewcommand\Authands{, }
\makeatother

\maketitle
\footnotetext[1]{Corresponding author.}
\footnotetext{{\it Email addresses:} \hyperlink{gozdesert@stat.tamu.edu}{gozdesert@stat.tamu.edu} (Gözde Sert), \hyperlink{abhishek@stat.tamu.edu}{abhishek@stat.tamu.edu} (Abhishek Chakrabortty), \hyperlink{anirbanb@stat.tamu.edu}{anirbanb@stat.tamu.edu} (Anirban Bhattacharya).}

\begin{abstract}
Inference in semi-supervised (SS) settings has received substantial attention in recent years due to increased relevance in modern big-data problems. In a typical SS setting, there is a much larger sized unlabeled data, containing observations only for a set of predictors, in addition to a moderately sized labeled data containing observations for both an outcome and the set of predictors. Such data arises naturally from settings where the outcome, unlike the predictors, is costly or difficult to obtain. One of the primary statistical objectives in SS settings is to explore whether parameter estimation can be improved by exploiting the unlabeled data. A novel Bayesian approach to SS inference for the population mean estimation problem is proposed. The proposed approach provides improved and optimal estimators both in terms of estimation efficiency as well as inference. The method itself has several interesting artifacts. The central idea behind the method is to model certain summary statistics of the data in a targeted manner, rather than the entire raw data itself, along with a novel Bayesian notion of debiasing. Specifying appropriate summary statistics crucially relies on a debiased representation of the population mean that incorporates unlabeled data through a flexible nuisance function while also learning its estimation bias. Combined with careful usage of sample splitting, this debiasing approach mitigates the effect of bias due to slow rates or misspecification of the nuisance parameter from the posterior of the final parameter of interest, ensuring its robustness and efficiency. Concrete theoretical results, via Bernstein--von Mises theorems, are established, validating all claims, and are further supported through extensive numerical studies. To our knowledge, this is possibly the first work on Bayesian inference in SS settings, and its central ideas also apply more broadly to other Bayesian semi-parametric inference problems.
\end{abstract}

\noindent{\bf Keywords:} Robustness and efficiency; Bayesian semi-parametric inference; Debiasing; Sample splitting and cross-fitting; Bernstein--von Mises theorem.

\section[Introduction and overview of contributions]{Introduction and overview of contributions}\label{introduction}

Semi-supervised (SS) learning has emerged as an exciting and active research area in statistics and machine learning in recent years. A typical SS setting involves two types of data sets: (i) a small or moderate sized \textit{labeled} (or supervised) data $\calL$ {with} observations for both an outcome (or label) $Y$ and a set of predictors $\boldX$, and (ii) a \textit{much larger} sized \textit{unlabeled} (or unsupervised) data $\calU$ containing observations only for $\boldX$. SS settings arise naturally when the outcome is difficult or costly to obtain, but observations for the predictors are plenty and easy to access. Typically, this scenario occurs in many modern big-data problems involving large (electronic) databases, such as speech recognition, text mining, and more recently, biomedical applications like electronic health records \citep{books/mit/06/CSZ2006, zhu05survey, kohane2011using, chakrabortty2018efficient}. In a standard SS setup, one of the primary statistical goals is to investigate whether and how parameter estimation and accuracy of inference can be improved by making use of the unlabeled data $\calU$, unlike supervised methods, which use only the labeled data $\calL$ and completely ignore $\calU$. SS inference in this spirit has been studied in the recent frequentist literature for various problems, including mean estimation \citep{zhang2019semi,zhang2022high} and linear regression \citep{chakrabortty2018efficient,azriel2022semi}, among others. However, Bayesian approaches for SS inference are largely lacking in the literature to the best of our knowledge.

We propose a \textit{Bayesian debiased modeling and inference} (BDMI) procedure for estimating the {\it population mean} $\theta_0:= \bbE(Y)$ of $Y$ under the SS setting, as a prototypical example. {A fundamental idea behind BDMI is to} carefully model certain {\it summary statistics} of the data in a {\it targeted} manner, rather than specifying a probability model for the raw data itself, along with developing and exploiting a novel {\it Bayesian notion of debiasing of nuisance parameters} (that are inherently involved in the procedure). Most existing SS approaches for estimating $\theta_0$ (or similar parameters/functionals of the distribution of $Y$) naturally require estimation of the possibly high dimensional {\it regression function} $m_0(\boldX):= \bbE(Y|\boldX)$ to exploit $\calU$ \citep{chakrabortty2018efficient, zhang2019semi, TonyCai2018SemisupervisedIF, zhang2022high}. $m_0(\cdot)$ therefore acts as a {\it nuisance function} here, that is needed (for exploiting $\calU$) but is not of primary interest. In general, the presence of such a nuisance parameter and its own estimation bias can drastically affect the final estimator's asymptotic behavior in the first order. In recent years, a popular frequentist debiasing procedure called double machine learning (DML) based on Neyman orthogonalization has been developed to rectify the impact of bias in learning a nuisance parameter \citep{chernozhukov2018double}. A key contribution of this work is to develop a {\it Bayesian analogue} of such debiasing procedures, that ensures {\it robust, efficient} and {\it nuisance-insensitive} Bayesian inference for $\theta_0$ (the {\it target}) while {\it allowing for slow/inefficient (or even inconsistent) learning of} $m_0$.

BDMI encapsulates a new principle of {\it disentangling} the nuisance parameter that is amenable to {\it Bayesian} modeling and inference. It crucially relies on a \textit{debiased representation} (Section~\ref{overcoming_bias})
of $\theta_0$ in terms of $m_0$ (specifically, its estimator or a posterior sample) that simultaneously exploits $\calU$ and also captures the nuisance bias incurred. Exploiting this representation, we then propose to model carefully chosen summary statistics of the data (see Section~\ref{likelihood_ind_data}). Modeling summary statistics of the data has been sporadically considered in the Bayesian literature for estimation and hypothesis testing \citep{pratt1965bayesian, savage1969nonparametric, doksum1990consistent, clarke1995posterior,johnson2005bayes,lewis2021} as well as in likelihood-free inference methods like Approximate Bayesian Computation (ABC) \citep{marjoram2003markov, fearnhead2012constructing, drovandi2015}. In the present setting, the summary statistics are exploited to: (i) carefully pinpoint the target and the bias induced from the nuisance, and (ii) learn them jointly by constructing a robust working likelihood (that can be justified under mild assumptions on the data generating mechanism) which can then be combined with default prior distributions on the model parameters to arrive at a posterior distribution. Further, a key feature of our approach is the careful usage of {\it sample-splitting} and {\it cross-fitting (CF)} \citep{chernozhukov2018double, newey2018cross} -- {\it not} just as a technical artifact (as is common in the frequentist literature) but as an {\it integral component} of the debiasing process itself.
 It helps create {\it independent} sub-folds of the entire data that crucially enable the disentangling of the nuisance estimation process from the summary statistics modeling process. Further, to ensure usage of the full data overall, we use CF by rotating the roles of the splits and using each sub-fold in turn, and thereafter {\it aggregating} the posteriors from all sub-folds using a consensus Monte Carlo type approach \citep{scott2022bayes}. It is worth mentioning that, while commonplace in the modern frequentist literature on semi-parametric inference, handling sample splitting (and CF) under a Bayesian framework is more challenging since it requires combining {\it distributions} (posteriors) and not just point estimators. Our final CF-based version of BDMI is given in Section~\ref{BCF} and summarized in Algorithm~\ref{algo}.

 We show through our theoretical results in Section~\ref{sec_theory} that the marginal posterior distribution $\Pi_{\btheta}$ for $\theta$  from BDMI inherits a {\it Bernstein--von Mises (BvM)-type limiting behavior} \citep[Chapter 10]{van2000asymptotic} with an asymptotically Gaussian shape, and contracts {\it always} around the true $\theta_0$ at a parametric $n^{-1/2}$ rate ($n$ being the size of $\Lsc$) and with a spread {\it tighter} than the supervised counterpart -- all holding {\it irrespective} of the choice/method used to obtain the nuisance posterior ($\Pi_{\mbm}$) for learning $m_0$. Further, $\Pi_{\btheta}$'s first order variability is {\it unaffected} by that of $\Pi_{\mbm}$ and is of the correct $n^{-1/2}$ rate {\it even if} the contraction rate of $\Pi_{\mbm}$ is arbitrarily {\it slow} or if it is even {\it misspecified} (i.e., does not contract around the true $m_0$). This makes BDMI \textit{first-order insensitive} \citep{chernozhukov2018double} to the nuisance estimation. Most importantly, from an SS inference perspective, $\Pi_{\btheta}$ (and its posterior mean) provably possess the desirable properties of \textit{global robustness} and \textit{efficiency improvement}: we show (i) the symmetric Bayesian credible intervals (CIs) from $\Pi_{\btheta}$ possess asymptotically correct frequentist coverage and sizes (of order $n^{-1/2}$) guaranteed to be {tighter} than their supervised counterpart; and (ii) the posterior mean is \textit{always} $\sqrt{n}$-consistent, asymptotically Normal and \textit{more efficient or at least as efficient} as the supervised estimator. Furthermore, when $\Pi_{\mbm}$ is correctly specified (with arbitrary contraction rate), $\Pi_{\btheta}$ and its posterior mean attain {\it optimal} efficiency, with variance matching the {\it semi-parametric efficiency bound}. All our claims above are validated through extensive simulations as well as a real data application in Section~\ref{simulations}. It is also worth noting that BDMI is {\it computationally scalable,} with all ingredient posteriors (from each fold) in $\Pi_{\btheta}$ being convolutions of $t$-distributions (hence {easy to sample} from). To our knowledge, BDMI is the {\it first work} on Bayesian inference (with provable guarantees) in SS settings.

Aside from SS inference itself, this work also contributes more generally to the growing literature on {\it Bayesian semi-parametric inference} in modern big-data settings. The SS setting has a distinct semi-parametric flavor, with $m_0(\cdot)$ being the (potentially high dimensional) nuisance parameter and functionals like $\theta_0$ being the target. There is a growing literature on frequentist properties of Bayesian semi-parametric inference procedures; see,
e.g., \citet{bickel2012semiparametric, rivoirard2012bernstein,castillo2015bernstein, norets2015bayesian, ray2019debiased}; where the quantity of interest is the marginal posterior of the parameter of interest obtained upon marginalizing out the nuisance parameter. Under delicate conditions on the prior distribution of the nuisance parameter, BvM results have been established for the parameter of interest in some of these works. Moreover, there have been some recent developments in the Bayesian semi-parametric literature (primarily for missing data or causal inference problems) aimed at alleviating bias arising from the nuisance estimation with slow rates \citep{ray2020semiparametric,luo2023semiparametric,breunig2022double,yiu2023semiparametric}. Most of these are based on careful prior selection/modification, or tailored posterior updating, to mimic the flavors of their frequentist counterparts.
BDMI adds to this literature by considering a different perspective and a principled approach to mitigate the bias of nuisance parameters. Another key feature of the approach is that it leaves the nuisance estimation method {\it entirely} to the user, and the nuisance posterior (or prior) does {\it not} require any form of adjustment or updating. While proposed albeit under the auspices of the SS inference problem, we believe the fundamental ideas of BDMI -- Bayesian debiasing and targeted modeling via summary statistics -- will also apply more generally to other Bayesian semi-parametric inference problems.

The rest of the article is organized as follows. We discuss the problem setup and some key preliminaries in Section~\ref{problem_setup}. Our proposed methodology is presented in Section~\ref{bayesian_SS}, with its various facets distributed across Sections~\ref{overcoming_bias}--\ref{BCF}. The theoretical properties of our method, including our main results (Theorems~\ref{bvm_on_first_half_data}--\ref{main_thm}), are presented in Section~\ref{sec_theory}, along with an alternative hierarchical version of our method and its theoretical properties discussed in Section~\ref{standard_bayesian_approach}. Finally, extensive simulation studies and real data analysis are presented in Section~\ref{simulations} to illustrate its empirical performance, followed by a concluding discussion in Section~\ref{conclusion_discussion}. All technical materials, including the proofs of all the main theoretical results, along with supporting lemmas and their proofs, as well as additional numerical results and methodological discussions that could not be accommodated in the main paper, are collected in the \hyperref[sec:supplementary]{Supplementary Material} (Sections~\ref{supp_sec_simulation}--\ref{supp:prelim_proof}).

\section[The problem setup and key preliminary ideas]{The problem setup and key preliminary ideas}\label{problem_setup}
Let $Y \in \bbR$ be the outcome variable, $\boldX \in \bbR^p$ be the covariate (or predictor) vector, and $\bbP_{\boldZ} \equiv \bbP_{Y \mid \boldX} \otimes \bbP_{\boldX}$ be the unknown joint distribution of $\boldZ := (Y, \boldX')'$, where $\bbP_{Y \mid \boldX}$ and $\bbP_{\boldX}$ denote the conditional distribution of $Y\mid \boldX$ and the marginal distribution of $\boldX$, respectively. The {\it available data} under the SS setting is denoted as: $\calD := \calL \hspace{0.7mm} \cup \hspace{0.7mm} \calU$, with $\calL := \{ \boldZ_i \equiv (Y_i, \boldX_i')' : i = 1, \dots , n \}$ being the labeled data containing $n$ independent and identically distributed (i.i.d.) samples of $\boldZ \sim \bbP_{\boldZ}$, and $\calU := \{\boldX_i : i= n+1, \dots, n+N \}$ being the unlabeled data containing $N$ i.i.d. samples of $\boldX \sim \bbP_{\boldX}$, and $\calL$ and $\calU$ are independent, denoted as $\calL \ind \calU$.

\begin{assumption}[{Standard features of SS settings}] \label{SS-assumptions}
We assume throughout that: (i) the unlabeled data size $N$ grows at least as fast as (and typically faster than) the labeled data size $n$, such that $n/N \to c$ as $n, N \to \infty$, where $0 \leq c < 1$ ($c=0$ being a key focus); and (ii) the observations for $\boldZ$ in $\calL$ and those for $\boldZ$ underlying the unlabeled $\boldX$ in $\calU$ arise from the same distribution $\bbP_{\boldZ}$ above,
and $\boldZ$ has finite second moments.
\end{assumption}

\remark
\label{remark_discussion_assumption}
Assumption~\ref{SS-assumptions} is fairly standard in the SS inference literature \citep{books/mit/06/CSZ2006, kawakita2013semisupervised}. The condition (i)
encodes a key (and {\it unique}) feature of SS settings, allowing for disproportionate sizes of $\calL$ and $\calU$. For example, while the size of $\calL$ may be of the order of hundreds, the size of $\calU$ could be of the order of tens of thousands. Further, since the outcome $Y$ is missing in $\calU$, one can view SS inference as a missing data problem by assuming $Y$ is `missing completely at random' \citep{tsiatis2006semiparametric}. However, since $\lim_{n, N \to \infty} n/N \to c = 0$ is allowed, it naturally violates the positivity assumption (on the proportion of $Y$ observed) standard in the missing data literature \citep{tsiatis2006semiparametric}, and makes the SS setting fundamentally {\it different} and more challenging (due to {\it non-standard asymptotics}) from the missing data setup. The condition (ii)
asserts that the underlying distributions of $\calL$ and $\calU$ are the same, which is standard and often implicit in the SS inference literature \citep{kawakita2013semisupervised, chakrabortty2018efficient, zhang2019semi, zhang2022high}, along with a mild moment assumption on $\boldZ$ to ensure $\bbE(Y\mid\boldX)$ and $\bbE(Y)$ exist. Finally, we clarify that we allow {\it high dimensional} settings throughout ($p$ can diverge with $n$).

\subsection[Preliminaries: Notational conventions and the supervised approach]{Preliminaries: Notational conventions and the supervised approach}\label{sec_sup}
We use the following {\it notational conventions} throughout the paper. Let $\bbE(\cdot) \equiv \bbE_{\boldZ}(\cdot)$, $\bbE_{Y\mid\boldX}(\cdot)$ and $\bbE_{\boldX}(\cdot)$ denote expectations under the distributions $\bbP \equiv \bbP_{\boldZ}$, $\bbP_{Y\mid\boldX}$ and $\bbP_{\boldX}$, respectively. For any dataset/collection (or its subset/functions) $\cal{C}$ on $\boldZ$, let $\bbE_{\cal{C}}(\cdot)$ and $\bbP_{\cal{C}}(\cdot)$ denote expectations and probability under the joint distribution of $\cal{C}$. Let $W$ be a generic random variable (or vector) with an underlying probability distribution $\bbP_{W}$, and let $f$ be any measurable $\bbR$-valued deterministic function of $W$. Then, the expectation of $f(W)$ is defined as: $\bbE_{W}\{f(W)\} \equiv \bbE_{W \sim \bbP_W}\{f(W)\} := \int f(w) \dd \bbP_{W}(w)$, whenever the Lebesgue integral exists. Further, for any \( d \geq 1 \), let \(\bbL_d(\bbP_{\boldZ})\) and \(\bbL_d(\bbP_{\boldX})\) denote the spaces of all \(\mathbb{R}\)-valued measurable functions \(g\) of $\boldZ$, and \(h\) of $\boldX$, such that \( \|g(\boldZ)\|^d_{\bbL_d( \bbP_{\boldZ})} := \bbE_{\boldZ}\{| g(\boldZ) |^d \} < \infty\) and \( \|h(\boldX)\|^d_{\bbL_d( \bbP_{\boldX})} := \bbE_{\boldX}\{| h(\boldX) |^d \} < \infty\), respectively. Let $\calN(\mu, \sigma^2)$ denote the Normal (Gaussian) distribution with mean $\mu$ and variance $\sigma^2$, and $t_\nu(\mu, c^2)$ denote the $t$-distribution with degrees of freedom $\nu > 0$, center $\mu$ and scale $c$. We also use $\calN(x; \mu, \sigma^2)$ and $t_\nu(x; \mu, c^2)$ to denote their respective probability density functions (pdfs) evaluated at $x \in \mathbb{R}$. For given probability measures $P$ and $Q$ on a measurable space $(\Omega, \calF)$, the total variation (TV) distance between $P$ and $Q$ is $\| P - Q \|_{\TV}:= \sup_{B \in \calF} |P(B) - Q(B)|$. For a sequence $b_n > 0$ and a sequence of random variables $X_n$, we say $X_n = \op(b_n)$ if and only if (iff) $|X_n|/b_n \cvP 0$ as $n \to \infty$. If $X_n \cvP 0$, we write $X_n = \op(1)$. Similarly, a sequence of random variables $W_n = O_{\bbP}(b_n)$ iff for any $\varepsilon > 0$, there exist $B_\varepsilon > 0$ and $n_\varepsilon$ such that $\bbP(|W_n| \leq B_{\varepsilon} \hspace{0.5mm} b_n) > 1 - \varepsilon$ for all $n \geq n_{\varepsilon}$. Furthermore, $W_n = \op(1)$ iff for some sequence $b_n \to 0$, $W_n = \Op(b_n)$. Lastly, for $\psi_0 \equiv \psi_0(\bbP)$ denoting any functional of interest for any distribution $\bbP$, we let $\psi$ represent the corresponding {\it random variable} (or vector, function, etc., as applicable) in a Bayesian framework, and denote its posterior distribution by $\Pi_{\boldsymbol{\psi}}$. {\it This convention is used consistently, without mention, throughout the paper.}

Before discussing any SS approaches, we first introduce the standard {\it supervised}
Bayesian approach for estimating $\theta_0$ using $\Lsc$ only, to set a benchmark. In the supervised setting, one can adopt a Bayesian framework by modeling $\calL$ (i.e., the $Y_i$'s $\in \calL$) with a working Gaussian likelihood with mean $\theta$ and variance $\sigma^2$, combined with a joint prior on $(\theta, \sigma^2)$. This yields a marginal {\it posterior} $\Pi_{\sup}$ for $\theta$ which, under mild regularity conditions on the prior, satisfies a BvM result \citep[Chapter 10.2]{van2000asymptotic}: $\Pi_{\sup} \approx  \calN(\wh \theta_{\sup}, \sigma^2_{Y}/n)$ as $n \to \infty$, where $\wh \theta_{\sup}:= \overline Y \equiv n^{-1}\sum_{i = 1}^n Y_i$ and $\sigma^2_Y:= \Var(Y)$. Thus, $\Pi_{\sup}$ yields $\wh \theta_{\sup} \equiv \overline Y$ as a natural (supervised) {\it point estimator} of $\theta_0$, as well as CIs of sizes $\propto \sigma_Y/\sqrt{n}$. Further, $\sigma^2_Y$ is the {\it best} achievable variance in the {\it supervised} setting and attains the semi-parametric efficiency bound under a fully non-parametric model \citep[Chapter 25.3]{van2000asymptotic} for estimating $\theta_0$. We will therefore use the limiting supervised posterior $\calN(\wh \theta_{\sup}, \sigma^2_{Y}/n)$ as a {\it benchmark} for asymptotic estimation/inference efficiency comparisons with BDMI later.

\subsection[A motivating imputation-type Bayesian SS approach]{A motivating imputation-type Bayesian SS approach}\label{motivation}

The construction of the supervised posterior $\Pi_{\sup}$ (and $\wh\theta_{\sup}$) naturally does not utilize the large unlabeled data $\Usc$ on $\boldX$ available in the SS setting. By virtue of its large size, $\calU$ essentially informs us on the distribution, $\bbP_{\boldX}$, of $\boldX$. Thus, whenever $\bbP_{\boldX}$ is informative about the parameter of interest \citep{zhang2000value, seeger2000learning}, one may hope to utilize $\calU$ and come up with an improved SS Bayesian estimation procedure with a more efficient, i.e., tighter posterior contracting around $\theta_0$ (albeit at a $\sqrt{n}$-rate, since information on $Y$ is still limited to $n$ observations), and accordingly a $\sqrt{n}$-consistent point estimator of $\theta_0$ that is more efficient than $\wh\theta_{\sup}$. We now discuss such an intuitive {\it imputation-based} approach with a natural Bayesian flavor, along with its potential drawbacks, which form a crucial basis for our final formulation of the BDMI method in Section~\ref{bayesian_SS}.

Recalling $m_0(\boldX) \equiv \bbE(Y \mid \boldX)$, the functional $\theta_0 \equiv \theta_0(\bbP_{\boldZ}) = \bbE(Y)$ can be written via iterated expectations as: $\theta_0 \equiv \theta_0(\bbP_{\boldX}; m_0) = \bbE_{\boldX}\{\bbE_{Y \mid \boldX}(Y \mid \boldX)\} =\bbE_{\boldX}\{m_0(\boldX)\}$. This representation clearly explains the {\it connection} between $\bbP_{\boldX}$ and $\theta_0$, and the potential for $\Usc$ to be exploited through bringing in the {\it nuisance} function $m_0$ (unknown but {\it estimable} via $\Lsc$). One can then construct an imputation-based Bayesian SS approach as follows.

Suppose one learns $m_0(\cdot)$ from $\calL$ via {\it any} reasonable Bayesian regression method (see Remark~\ref{remark_choice_of_nuisance_method} for some examples) that provides a {\it nuisance posterior} $\Pi_{\mbm} \equiv \Pi_{\mbm}(\cdot\,;\calL)$ for $m$. Then, using the identity $\theta_0 = \bbE_{\boldX}\{m_0(\boldX)\}$, and replacing $\bbE_{\boldX}$ therein with an empirical average over $\calU$, one may obtain an {\it induced posterior} $\Pi_{\imp}$ for $\theta$ via a natural {\it imputation} approach, i.e., for samples $\wt m \sim \Pi_{\mbm}$, we let $\theta_{\imp} \equiv \theta_{\imp}(\wt m) := {N^{-1}}\sum_{\boldX_i \in \calU} \wt m (\boldX_i) \sim \Pi_{\imp}$. Further, by linearity of expectation, it is easy to show that, $\wh\theta_{\imp}:=  N^{-1} \sum_{\boldX_i \in \calU} \wh m(\boldX_i)$ is the posterior mean of $\Pi_{\imp}$ (and hence, a point estimate of $\theta_0$), where $\wh m(\cdot) :=\bbE_{\wt m \sim \Pi_{\mbm}}\{\wt m(\cdot) \mid \calL\}$ is the posterior mean of $\Pi_{\mbm}$.

There are two major issues with this approach: (i) potential {\it misspecification} of $\Pi_{\mbm}$ in learning the true $m_0$;
and (ii) more importantly,
effect of the {\it nuisance} $\Pi_{\mbm}$'s {\it first-order properties (its rate/bias and variability) directly impacting} the {\it target} $\Pi_{\imp}$'s {\it first-order behavior}.
To illustrate, consider the {\it ideal} case: $N = \infty$. Then, the posterior sample $\theta_\imp$ equals $\bbE_{\boldX}\{\wt m(\boldX) \hspace{0.01in} | \hspace{0.01in} \wt m\} \equiv \theta_0 + \bbE_{\boldX}\{\wt m(\boldX) - m_0(\boldX) \hspace{0.01in} | \hspace{0.01in} \wt m\}$. Thus, when misspecification is allowed, i.e., $\bbE_{\wt m \sim \Pi_{\mbm}} \{ \|\wt m(\boldX) - m^*(\boldX) \|_{\bbL_2(\bbP_{\boldX})} \mid \calL\} \cvP 0$ (under $\bbP_{\calL}$) for {\it some} function $m^*(\cdot) \in \bbL_2(\bbP_{\boldX})$ possibly $\neq m_0(\cdot)$, then $\Pi_{\imp}$ may become {\it inconsistent} (i.e., not contracting around the true $\theta_0$).
More fundamentally, {\it even if} $m^*(\cdot) = m_0(\cdot)$, the {\it entire} first-order behavior (rate, shape, and variability) of $\Pi_{\imp}$ depends {\it directly} on the corresponding behavior of (posterior of): $\wt m (\cdot) - m_0 (\cdot)$, the `bias term', making $\Pi_{\imp}$ {\it sensitive}, in the {\it first order}, to $\Pi_{\mbm}$'s first order properties, and accordingly, the choice of the {\it method} used therein. In particular, if $\Pi_{\mbm}$ has a contraction rate, $a_n$, {\it slower} than $n^{-1/2}$, then so will $\Pi_{\imp}$. More importantly, the {\it variability} of $\Pi_{\imp}$ itself (after scaling by its rate) will be directly impacted by that of $\Pi_{\mbm}$. Overall, this indicates that to obtain a BvM-type result on $\Pi_{\imp}$ -- necessary to ensure provably valid estimation and inference on $\theta_0$ -- one {\it requires} the
availability of a corresponding semi-parametric BvM-type result under the nuisance $\Pi_{\mbm}$, which may necessitate delicate conditions/control on specifics of $\Pi_{\mbm}$'s construction. This becomes especially challenging when using non-smooth or complex methods, e.g., sparse regression in high dimensions or non-parametric machine learning methods, as nuisance estimators. These methods, while highly relevant and popular, have rates slower than $n^{-1/2}$, as well as unclear first-order properties with often intractable posteriors and limited availability (or feasibility) of corresponding BvM results. In general, this first-order sensitivity of $\Pi_{\imp}$ and its reliance on such intricate aspects of $\Pi_{\mbm}$, therefore, jeopardizes rate-optimal and provably valid inference on $\theta_0$ with the correct variance. In Section~\ref{supp:imputation_approach} of the \hyperref[sec:supplementary]{Supplementary Material}, we present a detailed case study on $\Pi_{\imp}$ (and also
compare it to BDMI)
showcasing its sensitivity and failure to provide a valid inference on $\theta_0$.

\section[Bayesian debiased modeling and inference: BDMI]{ Bayesian debiased modeling and inference: BDMI}\label{bayesian_SS}

This section introduces the BDMI approach, which addresses the limitations of the imputation approach discussed in Section~\ref{motivation}, by appropriately accounting for nuisance estimation bias within a Bayesian likelihood framework. BDMI is based on the principle of disentangling the nuisance parameter, and jointly {\it learning} its bias with the parameter of interest via targeted summary statistics {\it amenable} to Bayesian modeling.
Incorporating this debiasing idea and the targeted modeling approach are our key methodological contributions towards {\it Bayesian semi-parametric inference}, in general, for robust and efficient inference in the presence of high dimensional nuisances, drawing parallels to the recent frequentist DML literature \citep{chernozhukov2018double}.

\subsection[Bayesian debiasing: Overcoming the bias from nuisance estimation within the Bayesian framework]{Bayesian debiasing: Overcoming the bias from nuisance estimation within the Bayesian framework}\label{overcoming_bias}

For exposition of the BDMI approach and its salient features, we assume for the time being that there exists a dataset $\calS$ which is an {\it independent copy} of the labeled data $\calL$. The sample size $s_n$ of $\calS$ is assumed to be of the same order as $n$; see Section~\ref{BCF} for more details. Suppose the nuisance estimation is performed on this $\calS$, using {\it any} reasonable Bayesian (or frequentist) method by constructing a likelihood for the nuisance parameter $m$ on $\calS$, combining with a suitable prior on $m$, to obtain a posterior $\Pi_{\mbm}$ for $m$. For our primary goal of inference on $\theta_0$, the specific construction of $\Pi_{\mbm}$ is not crucial, provided it satisfies some basic regularity conditions (see Section~\ref{sec_theory} for details). Henceforth, we assume access to a {\it generic} posterior $\Pi_{\mbm}$ for $m$, noting that $\Pi_{\mbm}(\cdot) \equiv \Pi_{\mbm}(\cdot; \calS)$ is itself a {\it random} distribution dependent on $\calS$. For simplicity, this dependence is suppressed in the notations whenever clear from context. The dataset $\calS$ can be viewed as {\it training data}, used {\it solely} to obtain the nuisance posterior $\Pi_{\mbm}$ for $m$. In contrast, $\calD = \calL \cup \calU$ serves as {\it test data}, used to obtain the posterior for the parameter of interest $\theta$ via the BDMI procedure. In practice, we construct such pairs of independent training and test datasets from the original data $\calD$ itself via {\it sample splitting}; see Section~\ref{BCF}.

Let $\tm : \bbR^p \to \bbR$ be any {\it random function} \citep[Ch.~19.4]{van2000asymptotic}
output from $\calS$ (e.g., a posterior sample from a Bayesian regression model fitted to $\calS).$ More formally, $\tm: (\Omega_{\calS}, \bbP_{\calS}) \times \bbR^p \to \bbR$ is a measurable map, i.e., $\{\tm(x)\}_{x \in \bbR^p} \equiv \{\tm(\omega;x)\}_{x \in \bbR^p}$ is a {\it stochastic process}, with sample paths $\tm (\omega; \cdot)$ for $\omega \in \Omega_{\calS}$, where $(\Omega_{\calS}, \bbP_{\calS})$ denotes the probability space underlying the randomness of $\calS$ and any derived measures (e.g., posteriors) from it. Suppose now the argument $\bx$ (or domain) of $\wt m(\bx) \equiv \wt m(\omega; \bx)$ is {\it measurized} (randomized) {\it independently} as: $\boldX \sim \bbP_{\bX} \ind \bbP_{\calS}$, e.g., $\bX \in \calD$ meets this requirement, since $\calD \ind \calS$ by construction. Consider the {\it doubly} random variable $\tm(\boldX)$ -- having two sources of randomness that are independent -- (i) the process $\tm(\cdot)$ {\it itself} from $\calS$, and (ii) its random argument $\boldX \sim \bbP_{\boldX}$ from $\calD$ ($\ind \calS$). We can then write $\theta_0 \equiv \bbE(Y)$ as:
\begin{align}
   \theta_0 & ~=~ \bbE_{\boldX}\{\bbE(Y \mid \boldX)\} ~\equiv~
   \bbE_{\boldX}\{ m_0(\boldX)\}
   ~= \underbrace{\bbE_{\boldX  \in \calD}[\{ m_0(\boldX) - \wt m(\boldX)\} \mid \wt m]}_{:= ~ b(\tm) \, \leadsto \, \mbox{\small Bias induced from $\wt m(\cdot)$}} +
   \underbrace{\bbE_{\boldX \in \calD}\{ \tm(\boldX) \mid \wt m \}}_{\mbox{\small Imputation via $\wt m(\cdot)$}};
   \label{robust_mtilde} \\
   & ~\equiv~ b(\wt m) +  \bbE_{\boldX \in \calD}\{ \tm(\boldX) \hspace{0.2mm} | \hspace{0.2mm} \wt m \} ~=~  \bbE_{\boldZ \in \calL}\{  Y - \wt m (\boldX) \hspace{0.2mm} | \hspace{0.2mm} \wt m\}  +   \bbE_{\boldX \in \calU}\{ \tm(\boldX) \hspace{0.2mm} | \hspace{0.2mm} \wt m \} ~~ \mbox{[$\calD \equiv \calL \cup \calU \ind \wt m(\cdot)$].}
   \label{robust_mtilde-PART2}
\end{align}
The steps in both \eqref{robust_mtilde}--\eqref{robust_mtilde-PART2} use $\tm(\cdot)$ from $\calS$ is $\ind$ of $\boldX$ (and $\boldZ)$ $\in \calD$. This independence is \textit{crucial} and necessary to derive \eqref{robust_mtilde}, which we refer to as the {\it debiased} {\it representation} of $\theta_0$. For notational clarity, we emphasize that for a given $\tm$, $b(\tm)$ should be interpreted as a parameter dependent on $\tm$, i.e., a {\it function} of $\tm$. Finally, we reiterate that the above representations \eqref{robust_mtilde}--\eqref{robust_mtilde-PART2} remain valid if $\tm(\cdot)$ is a random draw from the posterior $\Pi_{\mbm}$, and
$\boldX = \boldX_i \in \calD$ ($i = 1, \ldots, n + N$), and $\boldZ = \boldZ_i \in \calL$ ($i=1,\ldots,n$), since $\Pi_{\mbm}$ is constructed from $\calS$ which is independent of $\calD$. Subsequent references to \eqref{robust_mtilde}--\eqref{robust_mtilde-PART2} are with respect to (w.r.t.) these particular choices.

Note that the first term $b(\tm)$ in \eqref{robust_mtilde} is essentially the expected {\it bias}, which is the price of replacing $m_0(\cdot)$ with a random sample $\tm(\cdot)$. As noted in Section~\ref{motivation}, this is precisely the primary cause of the issues with the imputation approach. Modeling this $b(\tm)$ itself, along with $\theta_0$, is the central idea of BDMI. Note further that:
\begin{align*}
b(\tm) \ \equiv \  \bbE_{\boldX \in \calD}\big[\{m_0(\boldX) - \tm(\boldX)\} \hspace{0.5mm} | \hspace{0.5mm} \tm \big] \ = \ \bbE_{\boldX}\{m_0(\boldX) - m^*(\boldX)\} + \bbE_{\boldX \in \calD}\big[\{m^*(\boldX) - \tm(\boldX)\} \hspace{0.5mm} | \hspace{0.5mm} \tm \big].
\end{align*}
This shows $b(\tm)$ captures two pivotal aspects: (i) when $m^*(\cdot) \neq m_0(\cdot)$, the first term measures its average deviation from $m_0(\cdot)$, and (ii) the second term importantly reflects the {\it variability} of $\tm(\cdot)$ itself as a sample from $\Pi_{\mbm}$ (which is further random through $\calS$).
From the perspective of statistical learning theory \citep{Vapnik1998}, one could think of the first term as {\it approximation error} and the second term as {\it estimation error}.

Most importantly, observe that \eqref{robust_mtilde-PART2} implies we also have {\it i.i.d. replicates} $\{Y_i - \tm(\boldX_i)\}_{i \in \Lsc}$ and $\{\tm(\boldX_i)\}_{i \in \Usc}$ from {\it conditionally (given $\wt m$) independent sources} that {\it target} {$b(\tm)$} and {$\theta_0 - b(\wt m)$,} respectively, through their expectations. Thus, $b(\tm)$ and $\theta_0 - b(\wt m)$ can be seen as {\it functionals} of the underlying distribution of $\calL$ and $\calU$, specifically depending on the {\it summary statistics} (means) of $Y - \tm(\boldX)$ in $\calL$ and $\tm(\boldX)$ in $\calU$ (given $\tm$ from an \textit{independent} source), respectively.
The {\it basic premise} of BDMI is: to model the data for these {\it target-specific parameters} --
 $b(\tm)$ and $\theta_0 - b(\tilde m)$ -- via summary statistics, since they {\it directly inform us on $\theta_0$, while also learning the bias} induced by $\tm$. This {\it targeted modeling} of summary statistics (instead of the entire data as in traditional Bayesian approaches) is a salient feature of BDMI. Further, its modeling of the bias $b(\wt m)$ {\it encodes a Bayesian form of debiasing} which plays a crucial role in ensuring nuisance-insensitive inference for $\theta_0$.

\subsection[Targeted modeling of summary statistics: Likelihood construction and final posterior]{Targeted modeling of summary statistics: Likelihood construction and final posterior}\label{likelihood_ind_data}
We are now ready to introduce the target-specific model construction discussed in the previous section. Given $\tm \sim \Pi_{\mbm}$ (from $\calS$), the i.i.d. replicates $\{ Y_i - \tm(\boldX_i) \}_{i = 1}^n$ and $\{\tm(\boldX_i)\}_{i = n+1}^{n+N}$ from $\calD$ ($\ind \calS$) target $b(\tm)$ and $\theta_0 - b(\tm)$, respectively, in terms of their means. These variables are now treated as our `observables' on the data $\calD \mid \tm$, and we now present a working likelihood construction for these observables on this data. To proceed, let us first define $\sigma^2_{1}(\tm) := \Var_{\boldZ}\{Y - \tm(\boldX)\}$ and $\sigma^2_{2}(\tm) := \Var_{\boldX}\{\tm(\boldX)\}$. Then, {\it given} $\tm$, $Y_i - \tm(\boldX_i)$ are i.i.d. with mean $b(\tm)$ and variance $\sigma^2_{1}(\tm)$ for $i \in \{1, \dots , n\}$, and $\tm(\boldX_i)$ are i.i.d. with mean $\theta_0 - b(\tm)$ and variance $\sigma^2_{2}(\tm)$ for $i \in \{n + 1, \dots , n + N\}$. Since these observables are i.i.d., a natural choice of a working model for such data could be based on Normal distributions with unknown variances, as follows:
\begin{align}\label{model_construction_indp_data}
Y_i - \tm(\boldX_i) \mid  \tm, b(\tm), \sigma^2_{1}(\tm) & \ \iid \ \calN(b(\tm), \sigma^2_{1}(\tm)), \quad i \in \{1, \dots, n \}; ~~\mbox{and} \nonumber \\
\tm(\boldX_i)   \mid  \tm, b(\tm), \theta_0, \sigma^2_{2}(\tm)  & \ \iid \ \calN(\theta_0 - b(\tm), \sigma^2_{2} (\tm)), \quad i \in \{n + 1, \dots, n + N \}.
\end{align}
Then, the likelihood as a function of the parameters $\{\theta, b(\tm), \sigma^2_{1}(\tm), \sigma^2_{2}(\tm)\}$ is given by:
\begin{align}\label{likelihood_function_indp_data}
L\{\theta, b(\tm), \sigma^2_{1}(\tm), \sigma^2_{2}(\tm)\} \ \propto \ \prod_{i = 1}^n \calN(Y_i-\tm(\boldX_i); b(\tm), \sigma^2_{1}(\tm)) \prod_{i = n+1}^{n+N}  \calN(\tm(\boldX_i);\theta -b(\tm), \sigma^2_{2}(\tm)).
\end{align}
The (pseudo-) likelihood constructed above can be combined with a prior distribution on the model parameters $\{\theta, b(\tm), \sigma^2_{1}(\tm), \sigma^2_{2}(\tm)\}$ using Bayes' formula to yield a posterior, and thereafter a {\it marginal posterior} $\Pi_{\btheta}$ of $\theta$.

We note that the Normal distributions in \eqref{model_construction_indp_data} above are only chosen as {\it working}, i.e., not necessarily correctly specified, distributions. Since a posterior depends on the data only through sufficient statistics, one could directly model the sample averages of $Y - \tm(\boldX)$ and $\tm(\boldX)$ as Normally distributed with appropriate parameters under modeling assumptions similar in spirit to \eqref{model_construction_indp_data}, operationally leading to the same posterior. In that case, one could simply treat the sample means as the `derived' observations, and since, given a sufficiently large number of observations, the sample averages are approximately Normal following the Central Limit Theorem (CLT), the Normality assumption on the sample averages would therefore be quite reasonable.

As a concrete {\it prior choice}, for the sake of theoretical and computational simplicity, we recommend using an improper prior on the model parameters $\{\theta, b(\tm)$, $\sigma^2_{1}(\tm), \sigma^2_{2}(\tm)\}$ in \eqref{model_construction_indp_data}, given by:
\begin{equation}
\label{eq:improper_prior_construction}
    \pi\big\{\theta, b(\tm) \mid \sigma^2_{1}(\tm), \sigma^2_{2}(\tm) \big\} \propto 1, ~~ \pi\big\{\sigma^2_{1}(\tm)\big\} \propto \{\sigma^2_{1}(\tm)\}^{-1} ~~ \text{and} ~~\pi\big\{\sigma^2_{2}(\tm)\big\} \propto \big\{\sigma^2_{2}(\tm)\big\}^{-1},
\end{equation}
with $\sigma^2_{1}(\tm)$ and $\sigma^2_{2}(\tm)$ being independent. We note that more general prior choices could also be employed here (see Remark~\ref{remark_algo_discussion} for a discussion) without altering the asymptotic conclusions, such as the limiting posterior and related properties of the procedure, established in Section~\ref{sec_theory}. For instance, by defining $\delta := \theta - b(\tm)$, one could place independent conjugate Normal-Inverse Gamma priors on $\{b(\tm), \sigma^2_{1}(\tm)\}$ and $\{\delta, \sigma^2_{2}(\tm)\}$. The proposed improper prior in \eqref{eq:improper_prior_construction} can then be viewed as a limiting (diffused) version of such a proper prior.

We now explicitly compute the marginal posterior $\Pi_{\btheta}$ of $\theta$ under \eqref{model_construction_indp_data} and the prior choice \eqref{eq:improper_prior_construction}, as follows.

\begin{proposition}\label{prop_ind_data} Given the likelihood function
$L\{\theta, b(\tm), \sigma^2_{1}(\tm), \sigma^2_{2}(\tm)\}$ in \eqref{likelihood_function_indp_data} and the improper prior in \eqref{eq:improper_prior_construction}, the marginal posterior distribution
$\Pi_{\btheta}$ of $\theta$ is the convolution of two $t$-distributions with the pdf $
\pi_{\btheta}(\theta) = (f * g)(\theta) := \int f(\theta - w)g(w) \dd w$, where $\pi_{\btheta}(\cdot), f(\cdot)$ and $g(\cdot)$ are the pdfs of $\Pi_{\btheta}$,
$t_{\nu_{n}}(\mu_{n}(\tm), \wh \sigma^2_{1, n}(\tm)/n)$ and $t_{\nu_{N}}(\mu_{N}(\tm), \wh \sigma^2_{2, N}(\tm)/N)$, respectively, where the parameters are given by: $\nu_{n} :=  n - 1$, $\nu_{N} := \ N - 1$,
\begin{align}\label{notations_ind_data}
& \mu_{n}(\tm) \ := \ \frac{1}{n}\sum_{i = 1}^n \big\{Y_i - \tm(\boldX_i)\big\}  ~~~\mbox{and}~~~ \frac{\wh \sigma^2_{1, n}(\tm)}{n} \ := \ \frac{\sum_{i = 1}^n \big[\{Y_i - \tm(\boldX_i)\} - \mu_{n}(\tm)\big]^2}{n(n - 1)}; \nonumber \\
& \mu_{N}(\tm) \ := \ \frac{1}{N}\sum_{i = n +1}^{n + N} \tm(\boldX_i) ~~~\mbox{and}~~~ \frac{\wh \sigma^2_{2, N}(\tm)}{N} \ := \ \frac{\sum_{i = n+1}^{n +N}\big\{ \tm(\boldX_i) - \mu_{N}(\tm)\big\}^2}{N(N - 1)}.
\end{align}
\end{proposition}
Note that $\Pi_{\btheta}$, being a convolution of two $t$-distributions, is {\it easy to sample} from (e.g., for constructing CIs). Further, the {\it posterior mean:} $\wh \theta_{\BDM}(\wt m)$ of $\Pi_{\btheta}$ can be considered as a natural point estimator of $\theta_0$. Note that the $\wt m$ in $\wh \theta_{\BDM}(\wt m)$ reflects that the estimator (and the posterior $\Pi_{\btheta} \equiv \Pi_{\btheta}(\wt m)$ itself) fundamentally depends on the nuisance posterior sample $\wt m \sim \Pi_{\mbm}$ used. From Proposition~\ref{prop_ind_data}, it follows that $
\wh \theta_{\BDM}(\wt m) = \mu_{n}(\tm) + \mu_{N}(\tm)$. Note that $\wh \theta_{\BDM}(\wt m)$ (and $\Pi_{\btheta}$, in general) utilize {\it both} $\calL$ and $\calU$, thereby justifying its billing as an SS approach. Also, as $n, N \to \infty$, it converges to $\theta_0$ {\it even if} $\Pi_{\mbm}$ is misspecified. This is because the first term in $\wh \theta_{\BDM}(\wt m)$ targets $\bbE_{\boldZ}
[\{Y - \tm(\boldX) \}\mid \tm]$, while the second term
targets $\bbE_{\boldX}\{\tm(\boldX) \mid \tm\}$, hence canceling out $\wt m$'s effect. Thus, BDMI gives a posterior mean that is {\it always} a consistent point estimator. Moreover, one would expect the spread of the posterior $\Pi_{\btheta}$ to be of the correct rate $n^{-1/2}$, and also tighter than the supervised counterpart. These claims, along with other desirable properties of BDMI, are formally established later in Section~\ref{sec_theory}.

\begin{remark}\label{rem_hbdmi}
A notable feature of BDMI is that it needs \textit{only one} sample $\tm$ from the nuisance posterior $\Pi_{\mbm}$. However, one could also consider a more conventional version of BDMI based on a {\it hierarchical} construction, requiring use of {\it multiple} samples of $\tm$. Section~\ref{standard_bayesian_approach} rigorously discusses this alternative version, which we call {\it hierarchical-BDMI} (h-BDMI), and shows that it inherits the same BvM result as BDMI, but under a stronger assumption; see Theorem~\ref{bvm_standard}. Even empirically, based on extensive simulation studies, we observed that the two versions have mostly similar performances, both in estimation and inference; see Section~\ref{simulations} for details. Therefore, given that it is computationally simpler, we recommend the original BDMI as the final approach.
\end{remark}

\subsection[Sample splitting based version: BDMI with cross-fitting (BDMI-CF)]{Sample splitting based version: BDMI with cross-fitting (BDMI-CF)}\label{BCF}

To practically implement the ideas introduced in Sections~\ref{overcoming_bias} and~\ref{likelihood_ind_data}, we need to construct independent training and test dataset pairs $(\calS, \calD)$ such that $\calS \ind \calD$. To achieve this from the original data $\calD = \calL \cup \calU$, we employ a $K$-fold sample splitting (with cross-fitting) procedure, where $K \geq 2$ is {\it fixed} (relative to $n,N$) and we assume without loss of generality (w.l.o.g.), that $|\calL| = n$ and $|\calU| = N$ are divisible by $K$. To construct independent training and test datasets required for the debiasing representation in \eqref{robust_mtilde}, we perform $K$-fold sample splitting by randomly partitioning the indices $\{1, \dots , n\}$ (for $\calL$) and $\{ n+1, \dots n+N\}$ (for $\calU$) into $K$ disjoint folds $\{\calI_k\}_{i = 1}^K$ and $\{\calJ_k\}_{i = 1}^K$, respectively, with each fold $\calI_k$ of size $\nK := n/K$ and $\calJ_k$ of size $\NK := N/K$, for each $k \in \{1, \dots , K\}$, define $\calI_k^- := \{1, \dots , n\} \backslash I_k$. Then, using these partitions, we construct pairs of training and test data folds $\{(\calS_k, \calD_k)\}_{k = 1}^K$, where $\calS_k := \{\boldZ_i: i \in \calI_k^- \}$ $\ind$ $\calD_k := \calL_k \cup \calU_k$, with $\calL_k := \{\boldZ_i: i \in \calI_k \}$ and $\calU_k := \{\boldX_i: i \in \calJ_k \}$. This provides $K$ such (training, test) data pairs for constructing the BDMI approach on each pair. Importantly, the test datasets $\calD_1, \dots, \calD_K$ are all disjoint and {\it independent}.

Adopting the BDMI construction from Section~\ref{likelihood_ind_data}, we now detail the BDMI procedure for one pair $(\calS_k, \calD_k)$. Since $\calS_k \ind \calD_k$, we use the training subfold $\calS_k$ to obtain the nuisance posterior $\Pi_{\mbm}^{(k)}$ for $m$, as detailed in Section~\ref{overcoming_bias}. Let $\tm_k$ be {\it one} random sample from $\Pi_{\mbm}^{(k)}$. Following the same model construction in Section~\ref{likelihood_ind_data}, we use the same likelihood formulation for the test subfold $\calD_k$ as given in equations \eqref{model_construction_indp_data}--\eqref{likelihood_function_indp_data}:
\begin{equation} \label{supp:likelihood_func_half_data}
\begin{aligned}
 Y_i -\tm_k(\boldX_i) \mid \tm_k & ~\iid ~ \calN(b(\tm_k), \sigma^2_{1}(\tm_k)), ~
 i \in \calI_k;
~\ \mbox{and}~ \\
\tm_k(\boldX_i) \mid \tm_k & ~\iid ~ \calN(\theta - b(\tm_k), \sigma^2_{2}(\tm_k)),
~i \in \calJ_k.
\end{aligned}
\end{equation}
Using the same improper prior on the model parameters $\{\theta, b(\tm_k), \sigma^2_{1}(\tm_k), \sigma^2_{2}(\tm_k)\}$ from \eqref{eq:improper_prior_construction}, and applying Proposition~\ref{prop_ind_data} with $(\calS,\calD)$ therein set as $ (\calS_k, \calD_k)$, we derive the marginal posterior $\Pi_{\btheta}^{(k)}$ for $\theta$ as follows:
\begin{proposition}\label{prop_half_data}
 Given the model construction in \eqref{supp:likelihood_func_half_data} and the improper prior in \eqref{eq:improper_prior_construction}, the marginal posterior distribution $\Pi_{\btheta}^{(k)}$ of $\theta$ given $\{\calD_k, \tm_k \}$ is a convolution of the $t$-distributions: \\ $t_{\nu_{\nK}}(\mu_{\nK}(\tm_k),\wh \sigma^2_{1, \nK}(\tm_k)/\nK)$ and $t_{\nu_{\NK}}(\mu_{\NK}(\tm_k), \wh \sigma^2_{2, \NK}(\tm_k)/\NK)$, where the parameters are given by: $\nu_{\nK} := \nK - 1$, $\nu_{\NK} := \NK - 1$,
 \begin{equation}\label{supp:notations_half_data}
\begin{aligned}
& \mu_{\nK}(\tm_k) := \frac1\nK \sum_{i \in \calI_k} \{Y_i - \tm_k(\boldX_i)\} ~~\mbox{and}~~ \frac{\wh \sigma^2_{1, \nK}(\tm
_k)}{\nK} :=  \frac{\sum_{i \in \calI_k} [\{Y_i - \tm_k(\boldX_i)\} - \mu_{\nK}(\tm_k)]^2}{\nK(\nK - 1)}; \\
&
\mu_{\NK}(\tm_k)  :=  \frac1\NK \sum_{i \in \calJ_k} \tm_k(\boldX_i) ~~\mbox{and}~~
\frac{\wh \sigma^2_{2, \NK}(\tm_k)}{\NK}  := \frac{\sum_{i \in \calJ_k} \big\{ \tm_k(\boldX_i) - \mu_{\NK}(\tm_k) \big\}^2}{\NK(\NK - 1)}.
\end{aligned}
 \end{equation}
\end{proposition}
\noindent Consistent with our earlier notation, let $\wh \theta_{\BDM}^{(k)}(\tm_k)$ denote the posterior mean of $\Pi_{\btheta}^{(k)}$. From Proposition~\ref{prop_half_data}, we have $
\wh \theta_{\BDM}^{(k)}(\tm_k) = \mu_{\nK}(\tm_k) + \mu_{\NK}(\tm_k)
$ and it retains the same properties as $\wh \theta_{\BDM}(\tm)$ from Section~\ref{likelihood_ind_data}.

While sample splitting enables us to obtain the debiased representation in \eqref{robust_mtilde}, which is crucial for the BDMI approach, it uses only a subset $\calD_k$ of the full dataset $\calD$ to obtain a posterior for $\theta$. This causes a notable lack of efficiency. Since sample splitting produces $K$ splits, each data fold pair $(\calS_k, \calD_k)$ can be utilized to obtain a posterior $\Pi_{\btheta}^{(k)}$ of $\theta$ for $k = 1, \dots, K$. We now introduce a method for combining these posteriors of $\theta$, referred to as {\it BDMI with cross-fitting} (BDMI-CF), to construct an {\it aggregated} full-data posterior for $\theta$. This approach addresses the efficiency loss discussed earlier by fully utilizing the available data and ensuring that the variance and contraction rates of the final procedure depend directly on $n$, as shown in Theorem~\ref{main_thm}.

BDMI-CF is inspired by the frequentist cross-fitting (CF) idea \citep{chernozhukov2018double}, addressing challenges in high dimensional nuisance parameter estimation. The conventional CF approach has been used to (i) relax strong assumptions, e.g., Donsker class conditions \citep[Chapter 19]{van2000asymptotic}, and (ii) make the sample splitting process efficient utilizing the full data in a `cross-fitted' manner \citep{chernozhukov2018double}. CF techniques are widely used in the modern semi-parametric inference literature, where a combined estimator is obtained by averaging the estimators obtained from each split to regain full efficiency \citep{chernozhukov2018double, newey2018cross}. In a {\it Bayesian} framework, however, additional care is required during the combination step, since entire {\it distributions (posteriors)} must be aggregated rather than point estimates. BDMI-CF addresses this issue by employing a consensus Monte Carlo-type approach \citep{Scott02042016} to suitably aggregate the posteriors from the sub-folds. This type of usage of cross-fitting (CF) for combining posteriors in Bayesian semi-parametric inference problems is not common. In the existing Bayesian literature, sample splitting has primarily been used to improve computational efficiency when handling large datasets \citep{scott2022bayes}. However, BDMI leverages sample splitting in a novel way: to ensure independence between the estimation of the nuisance parameter and the parameter of interest, and further via CF based aggregation, ensures efficient usage of the {\it entire} data. We now discuss the CF procedure.

Let $\theta_1, \dots , \theta_K $ be independent random variables drawn from the corresponding posteriors $\Pi_{\btheta}^{(1)}, \dots, \Pi_{\btheta}^{(K)}$ which are obtained from $(\calS_1,\calD_1), \dots , (\calS_K,\calD_K)$, respectively. We then define a new random variable:
\begin{align}
& \theta_{\BDM} \ := \ \frac{1}{K} \sum_{k = 1}^K \theta_k, ~~\mbox{and let}~ \Pi_{\btheta} ~\mbox{be the corresponding distribution of}~\theta_{\BDM}. \label{eq:FINALposteriorBDMI}
\end{align}

The distribution $\Pi_{\btheta}$ in \eqref{eq:FINALposteriorBDMI} is referred to as the {\it final (aggregated)} posterior of $\theta$ from BDMI, specifically BDMI-CF. This final posterior $\Pi_{\btheta}$ is a (scaled) {\it convolution} of the posteriors $\Pi_{\btheta}^{(1)}, \dots, \Pi_{\btheta}^{(K)}$ obtained from each data fold pair $(\calS_1,\calD_1), \dots , (\calS_K,\calD_K)$. Hence, samples from $\Pi_{\btheta}$ can be easily generated by construction.

Further, by linearity of expectation, the posterior mean $\wh \theta_{\BDM}(\tm_{\CF})$ of $\Pi_{\btheta}$ is the average of the posterior means $\wh \theta_{\BDM}^{(1)}(\tm_1), \dots, \wh \theta_{\BDM}^{(K)}(\tm_K)$ from the corresponding posteriors $\Pi_{\btheta}^{(1)}, \dots, \Pi_{\btheta}^{(K)}$. More explicitly,
\begin{equation}\label{eqn_postMean_n}
    \wh \theta_{\BDM}(\tm_{\CF}) \ = \ \mu_n(\tm_{\CF}) + \mu_N(\tm_{\CF}) \ := \ \frac1n\sum_{i = 1}^n \big\{Y_i - \tm_{\CF}(\boldX_i) \big\} ~+~ \frac1N\sum_{i = n+1}^{n+N}\tm_{\CF}(\boldX_i),
\end{equation}
where $\tm_{\CF}(\boldX_i) := \tm_k(\boldX_i)$ for $i \in \calI_k$ or $i \in \calJ_k$ where $\tm_k$ is a random sample from the respective posterior $\Pi_{\mbm}^{(k)}$ of $m$ for $k = 1, \dots, K$. Naturally, we consider the posterior mean $\wh \theta_{\BDM}(\tm_{\CF})$ as a point estimator of $\theta_0$. Furthermore, Theorem~\ref{main_thm} guarantees the $\sqrt{n}$-consistency of $\wh \theta_{\BDM}(\tm_{\CF})$ as an estimator of $\theta_0$. Detailed properties of $\wh \theta_{\BDM}(\tm_{\CF})$, and more generally the posterior $\Pi_{\btheta}$ in \eqref{eq:FINALposteriorBDMI}, are further examined in Section~\ref{sec_theory}. We now present the final algorithm for our BDMI (specifically, BDMI-CF) approach in Algorithm~\ref{algo}.

\begin{algorithm}
\caption{
The BDMI (with cross-fitting) procedure for SS mean estimation}\label{algo}
\vspace{0.05in}
 \KwIn{Data $\calD = \calL \cup \calU$, $K =$ the number of folds to use for CF, $M =$ number of samples to draw from $\Pi_{\btheta}$ (the final posterior \eqref{eq:FINALposteriorBDMI} from BDMI-CF), and the improper prior as in \eqref{eq:improper_prior_construction}.}
 \KwOut{Posterior samples $\theta_1 \dots, \theta_M$ from $\Pi_{\btheta}$, the posterior mean $\wh\theta_{\BDM}(\tm_{\CF})$ as a point estimate of $\theta_0$, and a $100\times(1-\alpha)\%$ credible interval (CI) for $\theta_0$, for a given $\alpha \in (0, 1)$.}
\vspace{0.05in}
 Split $\calD$ randomly
 into $K$ disjoint sets: $(\calD_k)_{k=1}^K \equiv (\calL_k \cup \calU_k)_{k =1}^K$, as in Section~\ref{BCF}, and let $\calS_k = \calL \backslash \calL_k$.

 \vspace{0.05in}
 \For{$k = 1$ \KwTo $K$:}
{
Pick any Bayesian (or frequentist) regression method to obtain a posterior $\Pi_{\mbm}^{(k)}$ for $m$ based on $\calS_k$. \\
Draw {\it one} sample $\tm^{(k)} \sim \Pi_{\mbm}^{(k)}$. Given $\tm^{(k)}$, compute $\Pi_{\btheta}^{(k)}$ for $\theta$ based on $\calD_k$ as in Proposition~\ref{prop_half_data}.\\
Draw $M$ many samples of $\theta$ from $\Pi_{\btheta}^{(k)}$: $\{ \theta_1^{(k)}, \dots, \theta_M^{(k)}\}$, for {\it each} $k =1, \ldots, K$.
}

\par\smallskip
Obtain the {\it samples} $\theta_1, \dots \theta_M \sim \Pi_{\btheta}$ as: $\theta_j := K^{-1}\sum_{k = 1}^K \theta_j^{(k)}$ for $j =1, \ldots, M$, using $\theta_j^{(k)}$ from Step 5.\\
Obtain $\wh\theta_{\BDM}(\tm_{\CF}) = \mu_n(\tm_{\CF}) + \mu_N(\tm_{\CF})$ as in \eqref{eqn_postMean_n} $\leadsto$ posterior mean of BDMI-CF {\it (point estimate)}.
\par\smallskip
Use the $(\alpha/2)^{th}$ and $(1-\alpha/2)^{th}$ sample quantiles of $\theta_1, \dots \theta_M$ as a $(1-\alpha)$-level {\it CI} of $\theta_0$ via BDMI-CF.
(We use Monte Carlo (MC) approximations to calculate the posterior quantiles of $\theta$ using a sufficiently large number $M$ of samples of $\theta$ so that the statistical error margin dominates the MC error.)
\end{algorithm}

\begin{remark}[Discussion on Algorithm~\ref{algo}]\label{remark_algo_discussion} We first clarify that MC approximations are employed in Algorithm~\ref{algo}, particularly in the last step, to calculate posterior quantiles of $\theta$. This involves using a sufficiently large number $M$ of $\theta$-samples to ensure that the statistical error margin dominates the MC error. Also, as detailed in Proposition~\ref{prop_half_data}, we calculated the posteriors $\{\Pi_{\btheta}^{(k)}\}_{k = 1}^K$ for $\theta$ under the improper prior given in \eqref{eq:improper_prior_construction}. Alternatively, users may pick a {\it different prior} (possibly non-conjugate) for $(\theta, b(\tm_k), \sigma^2_{1}(\tm_k), \sigma^2_{2}(\tm_k))$. Using the same likelihood construction in \eqref{supp:likelihood_func_half_data}, one can compute the posteriors $\{\Pi_{\btheta}^{(k)}\}_{k = 1}^K$ of $\theta$ under the chosen prior. It is important to note that these posteriors $\{\Pi_{\btheta}^{(k)}\}_{k = 1}^K$ would differ (possibly, not having a closed form) from those in Proposition~\ref{prop_half_data}. Despite such differences, one can {\it still} define a corresponding posterior mean $\wh \theta_{\BDM}(\tm_{\CF})$ (the average of the posterior means of the corresponding posteriors $\{\Pi_{\btheta}^{(k)}\}_{k = 1}^K$) and use it as a valid point estimator for $\theta_0$. When an exact expression for $\wh \theta_{\BDM}(\tm_{\CF})$ is unavailable (so \eqref{eqn_postMean_n} no longer holds), an MC average $M^{-1} \sum_{j = 1}^M \theta_j$ of the $M$ $\theta$-samples (as obtained in Step 7 of Algorithm~\ref{algo}) can approximate $\wh \theta_{\BDM}(\tm_{\CF})$. To construct a $100 \times (1 - \alpha) \%$ CI for $\theta_0$, we still use MC approximations to calculate posterior quantiles of $\theta$. Lastly, we highlight that BDMI provides a computationally efficient procedure for obtaining samples for $\theta$. The primary computational cost lies in sampling from the nuisance posterior for $m$, as the remaining step of sampling $\theta$ from a convolution of two $t$-distributions is negligible. Moreover, by leveraging parallel computing, Steps 3--5 in Algorithm~\ref{algo} can be executed in parallel to accelerate computation further.
\end{remark}

\begin{remark}[Recommendation for the choice of $K$] \label{remark_choice_of_K}
As established in Section~\ref{sec_theory}, the choice of $K$ does {\it not} impact {\it asymptotic} properties or performance of BDMI-CF, provided that $K$ is {\it fixed} (relative to $n$). However, in finite samples, $K$ may influence performance and should be chosen carefully. The parameter $K$ can be interpreted as a `tuning parameter' that embodies the {\it variance-bias trade-off}. Specifically, as $K$ increases, the training data size grows, leading to more stable nuisance estimation (reducing bias). However, this comes at the cost of smaller test data sizes, which may increase finite-sample variance. Thus, selecting $K$ involves balancing these competing factors to achieve optimal performance.
Based on extensive simulations under various settings (see Section~\ref{simulations}), we observed that $K = 5$ or $10$ generally provides (near-)optimal (and fairly robust) performance in terms of {\it both} estimation and inference. We therefore recommend such a $K$ in practice.
\end{remark}

\begin{remark}[Choice of methods for the nuisance posterior $\Pi_{\mbm}$]\label{remark_choice_of_nuisance_method}
We conclude by discussing the choice of methods to obtain the nuisance posterior $\Pi_{\mbm}$. Firstly, BDMI is fully flexible in that it allows $\Pi_{\mbm}$ to be {\it any user-chosen off-the-shelf approach} that can be used {\it without} any modifications/adjustments to the posterior (or its prior). Therefore, it allows most standard Bayesian (or frequentist) regression approaches, {\it parametric} and {\it non-parametric}, provided they only satisfy some reasonable (and {\it high-level}) contraction
conditions (formalized in Assumption~\ref{assumption_for_half_fold}).
Parametric methods include traditional linear regression approaches such as Bayesian ordinary or ridge regression (corresponding to improper and Gaussian priors on the regression parameters), or their frequentist counterparts. Further, {\it sparsity} (or shrinkage) based parametric methods, commonly adopted in {\it high dimensional} settings can also be used, including sparse Bayesian linear regression based on spike-and-slab type priors \citep{mitchell1988bayesian,george1993variable,johnson2012bayesian,rovckova2018spike} or continuous shrinkage priors \citep{carvalho2010horseshoe,bhattacharya2015dirichlet}, along with their frequentist counterparts such as LASSO \citep{hastie2015, wainwright_2019} or its variants. On the other hand, non-parametric methods may include Gaussian process regression \citep{williams1998prediction}, kernel smoothing-based methods \citep{tsybakov2009, simonoff2012smoothing}, reproducing kernel Hilbert space based methods
\citep{berlinet2011reproducing}, like smoothing splines \citep{green1994nonparametric},
as well as modern black-box machine learning (ML) methods such as random forest \citep{breiman2001random, wager2018estimation}, Bayesian additive regression trees (BART) \citep{bart2010}, and neural networks \citep{specht1991general, farrell2021deep}. These non-parametric methods are better suited for low dimensional (or fixed $p$) settings. Overall, BDMI affords notable flexibility to adapt to various modeling scenarios for $\Pi_{\mbm}$.
\end{remark}

\section[Theoretical properties of the BDMI procedure]{Theoretical properties of the BDMI procedure} \label{sec_theory}
In this section, we analyze in detail the theoretical underpinnings of our proposed BDMI procedure. Under mild regularity conditions, we show (in Theorems~\ref{bvm_on_first_half_data}--\ref{main_thm}) that the BDMI posteriors $\{\Pi_{\btheta}^{(k)}\}_{k = 1}^K$ (the `one fold' versions) and $\Pi_{\btheta}$ (the final aggregated version via CF) all inherit BvM-type limiting behaviors with asymptotically Gaussian posteriors contracting around the true $\theta_0$ at a $\sqrt{n}$-rate, along with various desirable properties on robustness, efficiency and nuisance insensitivity, which are all discussed in detail subsequently.

\begin{assumption}\label{assumption_for_half_fold}
We assume throughout that the number of folds $K$ (for CF) is {\it fixed}. Further, we make the following {\it high-level} assumptions on the nuisance posterior $\Pi_{\mbm}$ (or its versions $\Pi_{\mbm}^{(k)}$ for any $k=1,\ldots,K$):
\begin{itemize}
\item[(i)] For any sample $\tm_k \sim \Pi_{\mbm}^{(k)}(\cdot) \equiv \Pi_{\mbm}^{(k)}(\cdot ;\calS_k)$, we assume that $\| \tm_k(\boldX) \|_{\bbL_4(\bbP_{\boldX})} = O_{\bbP}(1)$ and $\| Y - \tm_k(\boldX) \|_{\bbL_4(\bbP_{\boldZ}}) = O_{\bbP}(1)$, where $\bbP$ denotes the joint probability distribution $\Pi_{\mbm}^{(k)}(\calS_k)$ for any $k = 1,\ldots, K$.
\item[(ii)] The posterior $\Pi_{\mbm}^{(k)}$ of $m$ satisfies the {\it nuisance posterior contraction condition} (NPCC): $\Pi_{\mbm}^{(k)}$ contracts (at {\it some} rate $a_n$) around 
{\it some} non-random limiting function $m^*(\cdot)  \in \bbL_2(\bbP_{\boldX})$ (with $m^*(\cdot)$ {\it not} necessarily equal to the true $m_0(\cdot)$). That is, for {\it some} (non-negative) sequence $a_n \to 0$, and for any $k = 1,\ldots, K$,
\begin{equation}\label{eqn_HLBC}
\Pi_{\mbm}^{(k)}\big[\{m:\| m(\boldX) - m^*(\boldX) \|_{\bbL_2(\bbP_{\boldX})} > a_n \} \mid \calS_k\big] \ \cvP \ 0 ~ \text{ under } \bbP_{\calS_k}, \text{ as }  n \to \infty.
\end{equation}
\end{itemize}
\end{assumption}

\begin{remark}[Discussion on Assumption~\ref{assumption_for_half_fold}] \label{remark_discussion_assump_bvm} The assumption on $K$ and the condition (i) above are both fairly mild and reasonable. The condition (ii) is the {\it only} required assumption on the nuisance posterior $\Pi_{\mbm}^{(k)}$ for our Theorems~\ref{bvm_on_first_half_data}--\ref{main_thm}. It embodies one of the key features of BDMI: it does {\it not} impose any restrictions on the distributional form or properties of $\Pi_{\mbm}^{(k)}$, nor the regression method (left entirely to the user's choice) used to obtain $\Pi_{\mbm}^{(k)}$. Typically, most of the existing Bayesian semi-parametric methods \citep{ray2020semiparametric,  luo2023semiparametric, breunig2022double, yiu2023semiparametric} crucially rely on {prior selection/modification} or {tailored posterior updates} to mitigate nuisance estimation bias and achieve the $n^{-1/2}$ contraction rate for the target
parameter. However, as Theorems~\ref{bvm_on_first_half_data}--\ref{main_thm} will demonstrate, the posterior convergence {\it rate} of $\theta$ and its {\it variability} are entirely unaffected by the posterior
contraction rate and variability of $\Pi_{\mbm}^{(k)}$, or even the method used to obtain $\Pi_{\mbm}$, provided Assumption~\ref{assumption_for_half_fold} holds (for a given $m^*$). This flexibility is largely due to our Bayesian debiasing approach presented in Section~\ref{overcoming_bias}, and its exploitation under the Bayesian framework via targeted modeling of summary statistics, as in Section~\ref{likelihood_ind_data}. It is worth noting that the condition (ii) is similar in spirit to $\bbL_2$-consistency conditions on nuisance estimators that (along with usage of CF) have become quite prevalent in the recent frequentist literature on debiased semi-parametric inference; see, e.g., \citet{chernozhukov2018double}. The NPCC can be viewed as an appropriate (and suitable) {\it analogue} in the {\it Bayesian} framework.
\end{remark}

\begin{remark}[Examples of contraction rate $a_n$ of the nuisance posterior $\Pi_{\mbm}$ and misspecification of $m_0(\cdot)$]\label{remark_nuisance_contraction_rate}

As detailed in Remark~\ref{remark_choice_of_nuisance_method}, Assumption~\ref{remark_discussion_assump_bvm} (ii) allows BDMI significant flexibility in accommodating a wide range of methods for estimating $m$. Specifically, $\Pi_{\mbm}^{(k)}$ can contract around a non-random function $m^*(\cdot)$, not necessarily equal to $m_0(\cdot)$, allowing misspecification. Further, regardless of $m^*(\cdot) = m_0(\cdot)$ or not (i.e., correctly specified or misspecified), the \textit{posterior contraction rate} $a_n$ of $\Pi_{\mbm}^{(k)}$ is \textit{not} restricted, and it
can be {\it any} rate that goes $0$, potentially slower than the parametric rate (see Remark~\ref{remark_bvm_half_data}). For parametric methods in low-dimensional settings ($p$ fixed or $p = o(n)$), contraction rates are typically $a_n = \sqrt{p/n}$. In high-dimensional settings ($p \gg n$), sparsity-based methods achieve rates of $a_n = \sqrt{s\log(p)/n}$, where $s$ is the sparsity level of the regression parameter $\bbeta$ \citep{wainwright_2019}. Non-parametric methods generally exhibit slower rates; for instance, kernel smoothing or smoothing splines achieve $a_n = n^{-q/(2q + p)}$, where $q$ represents the smoothness level of $m_0(\cdot)$ \citep{tsybakov2009}. Modern machine learning methods often achieve rates of $a_n = n^{-\alpha}$ for some $\alpha < 1/2$ \citep{chernozhukov2018double}. Finally, as noted above, BDMI remains robust even in misspecified cases, allowing for $\Pi_{\mbm}$ to contract around some function $m^*(\cdot) \neq m_0(\cdot)$. For instance, when $m_0(\cdot)$ is non-linear but a linear model is fitted, $\Pi_{\mbm}$ contracts around $m^*(\boldX) := \wt\boldX'\bbeta^*$, where $\wt\boldX = (1, \boldX')'$ and $\bbeta^* := \arg\min_{\bbeta} \bbE
\|Y - \wt\boldX'\bbeta\|^2$ or equivalently, $\bbeta^* = \{\bbE(\wt\boldX\wt\boldX')\}^{-1}\bbE(\wt\boldX Y)$ and $m^*(\boldX)$ is the {\it best linear predictor} of $Y$ given $\boldX$, i.e., the $\bbL_2(\bbP_{\boldX})$-projection of $m_0(\cdot)$ onto the linear span of $\boldX$. This functional misspecification does not affect BDMI’s ability to maintain $\sqrt{n}$-consistency/contraction for $\theta_0$, as shown in Theorems~\ref{bvm_on_first_half_data}--\ref{main_thm}.
\end{remark}

\begin{theorem}\label{bvm_on_first_half_data}
Under Assumptions~\ref{SS-assumptions} and~\ref{assumption_for_half_fold}, the marginal posterior $\Pi_{\btheta}^{(k)}$ of $\theta$ (as in Proposition \ref{prop_half_data}) obtained from one pair $(\calS_k,\calD_k)$ inherits a BvM-type limiting behavior as follows: for each $k = 1,\ldots, K$,
$$
\left \|\hspace{0.7mm}  \Pi_{\btheta}^{(k)} - \calN \left(\wh \theta_{\BDM}^{(k)}(m^*), \tau^2_{\nK,\NK}(m^*)\right) \right\|_{\TV} \ \cvP \ 0 \text{ in probability under }\ \bbP_{\wt\calD_k}, \text{ as } n, N \to \infty,
$$
where, with $\sigma^2_{1}(m^*) := \Var_{\boldZ}\{Y - m^*(\boldX)\}$ and $  \sigma^2_{2}(m^*) := \Var_{\boldX}\{m^*(\boldX)\}$, $\wh \theta_{\BDM}^{(k)}(m^*)$ and $\tau^2_{\nK,\NK}(m^*)$ are:
$$
\wh \theta_{\BDM}^{(k)}(m^*) \ := \  \frac{1}{\nK} \sum_{i \in \calI_k}\big\{Y_i - m^*(\boldX_i)\big\} +  \frac{1}{\NK} \sum_{i \in \calJ_k} m^*(\boldX_i) \ \text{ and } \ \tau^2_{\nK,\NK}(m^*) \ := \ \frac{\sigma^2_{1}(m^*)}{\nK} + \frac{\sigma^2_{2}(m^*)}{\NK}.
$$
Further, let $h := \sqrt{\nK}(\theta - \theta_0)$ and $\Pi_{\mathbf{h}}^{(k)}$ be the posterior of $h$. Then, under Assumptions~\ref{SS-assumptions} and~\ref{assumption_for_half_fold},
$$
 \big\| \hspace{0.7mm} \Pi_{\mathbf{h}}^{(k)} - \calN\left(\sqrt{\nK} \big\{ \wh \theta_{\BDM}^{(k)}(m^*) - \theta_0 \big\}, \nK\tau^2_{\nK,\NK}(m^*)\right) \big\|_{\TV} \ \cvP \ 0 \ \text{ in probability under } \ \bbP_{\wt\calD_k}.
$$
\end{theorem}

\begin{theorem}[Main
result] \label{main_thm}
Under Assumptions~\ref{SS-assumptions} and~\ref{assumption_for_half_fold}, the final (aggregated) posterior $\Pi_{\btheta}$ of $\theta$, as defined in \eqref{eq:FINALposteriorBDMI}, from the BDMI-CF procedure
inherits a BvM-type limiting behavior as follows:
\begin{align*}
\left\| \Pi_{\btheta} - \calN(\wh \theta_{\BDM}(m^*), \tau^2_{n,N}(m^*)) \right\|_{\TV} \ \cvP \ 0 \ \text{ in probability w.r.t. } \bbP_{\calD}, \text{ as } n, N \to \infty,
\end{align*}
where $\wh \theta_{\BDM}(m^*) := \mu_{n}(m^*) + \mu_{N}(m^*)$ as defined in \eqref{eqn_postMean_n}
with $\tm_{\CF}$ therein substituted by
$m^*$,
and $\tau^2_{n,N}(m^*) := \{\sigma^2_{1}(m^*)/n\} + \{\sigma^2_{2}(m^*)/N\}$ with $\sigma^2_{1}(m^*)$ and $\sigma^2_{2}(m^*)$ as defined in Theorem~\ref{bvm_on_first_half_data}.
\end{theorem}

The BDMI-CF procedure provides the posterior $\Pi_{\btheta}$ with the posterior mean $\wh\theta_{\BDM}(\tm_{\CF})$ as defined in \eqref{eqn_postMean_n}. Naturally, $\wh\theta_{\BDM}(\tm_{\CF})$ can be considered as a valid {\it SS point estimator} for $\theta_0$. Beyond direct implications of Theorem~\ref{main_thm}, the asymptotic behavior of the SS estimator $\wh\theta_{\BDM}(\tm_{\CF})$ inherently is of separate interest. Towards that, in Corollary~\ref{corollory_asymp_equiv_postmean}, we rigorously establish an asymptotically linear representation of $\wh\theta_{\BDM}(\tm_{\CF})$.

\begin{corollary}
[Asymptotically linear representation of the posterior mean $\wh \theta_{\BDM}(\tm_{\CF})$ of BDMI-CF] \label{corollory_asymp_equiv_postmean}
Under Assumptions~\ref{SS-assumptions} and~\ref{assumption_for_half_fold}, the posterior mean $\wh \theta_{\BDM}(\tm_{\CF})$ of $\Pi_{\btheta}$ as in \eqref{eqn_postMean_n} is asymptotically equivalent to the mean $\wh \theta_{\BDM}(m^*)$ of the limiting distribution in Theorem~\ref{main_thm} at a $1/\sqrt{n}$ rate. In particular,
\begin{align}\label{eq:asymp_equiv_pmean_proof}
\sqrt{n}\hspace{0.5mm}\{\hspace{0.5mm}\wh\theta_{\BDM}(\tm_{\CF}) - \theta_0 \hspace{0.5mm} \} & \ = \ \sqrt{n}\hspace{0.5mm}\{\hspace{0.5mm} \wh \theta_{\BDM}(m^*) - \theta_0  \hspace{0.5mm} \} + o_{\bbP_{\calD} \hspace{0.5mm}}(1) \\
& \ \equiv \ \sqrt{n}\left[\frac1n\sum_{i = 1}^n \big\{Y_i - m^*(\boldX_i) \big\} + \frac1N\sum_{i = n+1}^{n+N}m^*(\boldX_i)  - \theta_0\right] + o_{\bbP_{\calD}}(1). \nonumber
\end{align}
\end{corollary}

\begin{remark}[Asymptotic properties of the posteriors $\Pi_{\btheta}^{(k)}$ and $\Pi_{\btheta}$] \label{remark_bvm_half_data}
Theorem~\ref{main_thm} establishes a BvM-type result for the final BDMI-CF procedure presented in Section~\ref{BCF}. Firstly, it shows that the posterior $\Pi_{\btheta}$ of $\theta$ behaves as Gaussian and concentrates around the true $\theta_0$ at a rate $1/\sqrt{n}$ with $|\calL| = n$. Importantly, while this rate is parametric in the labeled data size $n$, it is {\it non-standard} in the full data size $(n+N)$, particularly when $n/N \to 0$, making SS settings unique and their technical analyses substantially more challenging. Secondly, Theorem~\ref{main_thm} demonstrates that for large $n, N$, the posterior $\Pi_{\btheta}$ is approximately Normal with mean $\wh \theta_{\BDM}(m^*)$ and variance $\tau_{n, N}^2(m^*)$, which {\it matches} the asymptotic theory for corresponding existing frequentist approaches applied to the full data in recent SS inference literature \citep{zhang2019semi, zhang2022high}. Furthermore, it is important to note that all properties of the posterior $\Pi_{\btheta}$ discussed here, and all subsequent discussions in Section~\ref{remark_eff_and_contraction} below
in the context of Theorem~\ref{main_thm}, also apply to Theorem~\ref{bvm_on_first_half_data}  and $\Pi_{\btheta}^{(k)}$, with appropriate modifications for the one-fold data pair $(\calS_k, \calD_k)$ where $\calD_k = \calL_k \cup \calU_k$ and $|\calL_k| = \nK$. Since these extensions are straightforward and analogous, we refrain from restating them anywhere for brevity.
\end{remark}

\begin{remark}[Proof techniques and subtleties]\label{remark_proof_techniques}
It is worth mentioning that while Theorems~\ref{bvm_on_first_half_data}--\ref{main_thm} have clear and strong implications,
their proofs (deferred to the \hyperref[sec:supplementary]{Supplement}
in the interest of space) are {non-trivial}, and involve a {\it synergy} of ideas and techniques from disparate literatures. Handling the theoretical underpinnings of BDMI and its key features: debiasing and the use of CF -- both under a {\it Bayesian} framework -- require bridging classical Bayesian tools/techniques for BvM-type results with those from the modern frequentist literature on debiased semi-parametric inference \citep{chernozhukov2018double}. Central to the proofs is the {\it interplay} between empirical process theory (along with CF), to handle the nuisance debiasing, and the {\it probabilistic structure} of Bayesian posteriors, to guarantee strong and nuisance-insensitive properties of BDMI while allowing $\Pi_{\mbm}$ to be generic throughout. In addition, the use of sample splitting and {\it posterior aggregation} via CF, though both crucial, introduce further technical subtleties that require novel adaptations under the Bayesian paradigm.
\end{remark}

\subsection[Robustness, efficiency and nuisance insensitivity of BDMI]{Robustness, efficiency and nuisance insensitivity of BDMI}\label{remark_eff_and_contraction}

Theorem~\ref{main_thm} establishes that, under the SS setting, the posterior $\Pi_{\btheta}$ concentrates around the true parameter $\theta_0$ at the parametric rate $1/ \sqrt{n}$ (ensuring usage of the {\it full} data) and possesses {\it universal robustness}
to the choice of the nuisance estimation method. This robustness manifests in two ways: (i) {\it global robustness} w.r.t. the limiting function $m^*(\cdot)$, ensuring that $\Pi_{\btheta}$ contracts around $\theta_0$ at a rate $1/\sqrt{n}$ {\it regardless} of the contraction rate $a_n$ of $\Pi_{\mbm}$ and {\it even if} $m^*(\cdot) \neq m_0(\cdot)$; and (ii) {\it insensitivity} to the nuisance estimation bias, as $\Pi_{\btheta}$ is {\it not} affected by slower convergence rates $a_n$ of $\Pi_{\mbm}$, nor by $\Pi_{\mbm}$'s {\it own} first order properties like its shape, variability etc. (even after scaling by $a_n$). $\Pi_{\btheta}$ depends on $\Pi_{\mbm}$ {\it only} through its limit $m^*$, and validity/properties of $\Pi_{\btheta}$ as in Theorem~\ref{main_thm} requires only $a_n \to 0$. Hence, BDMI effectively addresses the primary issue of the imputation approach (see Section~\ref{motivation}), where nuisance estimation bias directly characterizes the first-order behavior/properties of the posterior for $\theta$, and offers substantial {\it flexibility} in choosing regression methods to obtain $\Pi_{\mbm}$. In particular, it paves the way for using non-smooth or complex methods, like sparse regression (in high dimensions) or non-parametric ML methods, both of which may unavoidably have slow or unclear first order behaviors (refer to Remarks~\ref{remark_choice_of_nuisance_method} and~\ref{remark_nuisance_contraction_rate} for examples of these methods and their contraction rates).
Moreover, BDMI-CF achieves {\it efficiency improvement} over the supervised approach based on $\calL$, irrespective of whether $m^*(\cdot) = m_0(\cdot)$. While both $\Pi_{\btheta}$ and $\Pi_{\sup}$ converge to $\theta_0$ at the parametric rate $1/ \sqrt{n}$, the variance $\tau^2_{n, N}(m^*)$ of the limiting distribution is {\it always smaller} than the variance of the supervised approach as we will show in Remark~\ref{remark_variance_comparison}, and further achieves the {\it semi-parametric efficiency bound} when $m^*(\cdot) = m_0(\cdot)$ (correctly specified case). These results align with frequentist asymptotic theory in recent SS inference literature \citep{zhang2019semi, zhang2022high}. Moreover, these desirable properties of $\Pi_{\btheta}$ also naturally extend to posterior summaries. In particular, the {\it posterior mean} $\wh\theta_{\BDM}(\tm_\CF)$, as a valid SS point estimator of $\theta_0$, inherits these properties. As Corollary~\ref{remark_eff_and_contraction} shows, it remains $\sqrt{n}$-consistent, asymptotically Normal, and asymptotically linear regardless of the nuisance estimation method, and its expansion is unaffected by the estimation bias/error of the nuisance, showing its first-order insensitivity. Finally, its asymptotic variance also equals the posterior variance $\tau^2_{n, N}(m^*)$ (see Remark~\ref{remark_variance_comparison} below), ensuring {\it valid and accurate inference} for $\theta_0$.

\begin{remark}[Variance comparison]
\label{remark_variance_comparison}
Theorem~\ref{main_thm} establishes that the posterior $\Pi_{\btheta}$ is asymptotically Normal with mean $\wh \theta_{\BDM}(m^*)$ and variance $\tau^2_{n,N}(m^*)$, which is also the variance of $\wh \theta_{\BDM}(m^*)$. Specifically, using the definition of $\wh \theta_{\BDM}(m^*)$ in Theorem~\ref{main_thm}, and due to the independence between $\calL$ and $\calU$, we have:
\begin{align}
& \Var \{ \wh \theta_{\BDM}(m^*) \}
~=~ \frac{\Var\{Y -m^*(\boldX)\}}{n} + \frac{\Var\{m^*(\boldX)\}}{N}  ~\equiv~  \frac{\sigma^2_{1}(m^*)}{n} + \frac{\sigma^2_{2}(m^*)}{N} ~=~ \tau^2_{n,N}(m^*). \label{eq:varcomp:pvarequality}
\end{align}

\noindent This equality is crucial for ensuring valid inference for $\theta_0$. Using the asymptotic equivalence in Corollary~\ref{corollory_asymp_equiv_postmean}, we can consider the asymptotic variance of $\wh \theta_{\BDM}(m^*)$ to {\it compare} the asymptotic variance of $\wh \theta_{\BDM}(\tm_{\CF})$ with the asymptotic variance of $\wh \theta_{\sup} \equiv \overline{Y}$ (based on $\calL$). Further, for any non-random $g(\cdot) \in \bbL_2(\bbP_{\boldX})$, we have:
\begin{align}
\sigma^2_{\sup} & ~\equiv ~ \lim_{n \to \infty}\Var[\sqrt{n}\{\wh \theta_{\sup} - \theta_0\}] ~=~ \Var(Y) \nonumber \\
& ~=~ \Var\{Y - g(\boldX)\} ~+~ \Var\{g(\boldX)\} ~+~ 2\hspace{0.03cm}\Cov\{Y - g(\boldX), g(\boldX)\}. \label{eq:varcomp:avarsup}
\end{align}
Under Assumption~\ref{SS-assumptions} (i), where $\lim_{n, N \to \infty}n/N \to c \in [0,1)$ and setting $g(\cdot) = m^*(\cdot)$ in \eqref{eq:varcomp:avarsup}, we obtain
\begin{align}
\sigma^2_{\BDM} & \equiv  \lim_{n, N \to \infty}\! \Var\big[\sqrt{n}\{\wh \theta_{\BDM}(\tm_{\CF}) - \theta_0 \}\big] = \! \lim_{n, N \to \infty}\! \Var\big[\sqrt{n} \{ \wh \theta_{\BDM}(m^*) - \theta_0\} \big] =  \lim_{n, N \to \infty}\! \tau^2_{n,N}(m^*)  \nonumber \\
& = \Var\{Y - m^*(\boldX)\} + c \hspace{0.3mm} \Var\{ m^*(\boldX)\} \ \equiv \ \sigma^2_{1}(m^*) + c \hspace{0.3mm} \sigma^2_{2}(m^*) \leq  \sigma^2_{1}(m^*) + \sigma^2_{2}(m^*) = \sigma^2_{\sup}.\label{eq:varcomp:avarBDMI}
\end{align}
This inequality holds if either: (i) $m^*(\boldX) = m_0(\boldX)$ (i.e., correctly specified model), or (ii) $m^*(\boldX) \neq m_0(\boldX)$ (misspecified model) but $\Cov \{Y-m^*(\boldX), m^*(\boldX)\} = 0$. Moreover, the inequality in \eqref{eq:varcomp:avarBDMI} is {\it strict} unless $m^*(\cdot)$ is a constant function. Hence, in either case, the SS estimator $\wh\theta_{\BDM}(\tm_{\CF})$ {\it outperforms} the supervised estimator $\wh\theta_{\sup}$ in terms of (asymptotic) variance and efficiency (see Table~\ref{table_efficiency}). Finally, note that the condition $\Cov \{Y-m^*(\boldX), m^*(\boldX)\} = 0$, represents a natural requirement on {\it orthogonality (in the population)} between the model-based predictions/target function $m^*(\boldX)$ and the residuals $\{Y-m^*(\boldX)\}$. This condition is satisfied by most reasonable regression procedures, including least squares-type methods, where the target functions (even if they are misspecified) $m^*(\cdot)$ {\it can be viewed as the $\bbL_2(\bbP_{\boldX})$-projection of $m_0(\cdot)$ onto the working model space}. For correctly specified models, i.e., $m^*(\cdot) = m_0 (\cdot)$, this condition, of course, holds trivially.
\end{remark}

\begin{table}[!ht]
\centering
\caption{Full characterization of efficiency improvement with BDMI and its robustness in terms of rate and the pair ($\Pi_{\mbm}, m^*$).}
\label{table_efficiency}
\begin{tabular}{p{3.2cm}p{2.0cm}p{2.1cm}p{7.3cm}}
\toprule
  \multicolumn{4}{c}{Comparison of the supervised versus SS estimators (BDMI) regarding efficiency and robustness} \\
\midrule
  Estimators & Rate of \newline convergence & Limiting \newline distributions & Asymptotic variance comparison \\
\midrule
Supervised \newline estimator: $\wh\theta_{\sup}$ & \hspace{0.4cm} $\frac{1}{\sqrt{n}}$ & $\calN(\theta_0, \frac{\sigma^2_{\sup}}{n})$
& $\sigma^2_{\sup} = \sigma^2_{1}(m^*) + \sigma^2_{2}(m^*) + 2\Cov[Y -m^*(\boldX), m^*(\boldX)]$ \\
  \hline
SS estimator with \newline BDMI: $\wh \theta_{\BDM}(\tm_{\CF})$ & \hspace{0.4cm} $\frac{1}{\sqrt{n}}$ & $\calN(\theta_0, \frac{\sigma^2_{\BDM}}{n})$ &
\parbox{5cm}{\vspace{-0.3cm}\begin{align*}
    \sigma^2_{\BDM} &  \equiv \sigma^2_{1}(m^*) + c \hspace{0.05cm} \sigma^2_{2}(m^*)
    ~\leq~ \sigma^2_{\sup} ~\mbox{[see \eqref{eq:varcomp:avarBDMI}]}
  \end{align*}}
  if either: {\bf (i)} $m^*(\boldX) = m_0(\boldX)$, or {\bf (ii)}
 $m^*(\boldX) \neq m_0(\boldX)$ and $\Cov\{Y - m^*(\boldX), m^*(\boldX)\} = 0$ hold.\newline
 ({\bf Note:} {\it Strict} inequality unless
 $m^*(\cdot)$ is constant.) \\
\bottomrule
 \end{tabular}
\end{table}

\begin{remark}[Adapting BDMI when $N < n$]\label{rem_N_less_than_n}
Our main focus is on scenarios where $N$ is substantially larger than $n$, as reflected in Assumption~\ref{SS-assumptions} (i): $\lim_{n, N \to \infty} n/N = c \in [0, 1)$. BDMI -- in its current form -- requires $c < 1$ (i.e., $N > n$) to guarantee efficiency improvement, as Remark~\ref{remark_variance_comparison} shows. While it still applies when $N < n$, the improvement is not guaranteed. However, it is theoretically {possible} to {\it adapt} BDMI to guarantee it even if $c > 1$ (i.e., $N < n$) as well, by slightly modifying our modeling and likelihood construction \eqref{model_construction_indp_data}--\eqref{likelihood_function_indp_data} in Section~\ref{likelihood_ind_data}. The primary reason behind this `discontinuity' (in behavior w.r.t. $c$) is due to the second model in \eqref{model_construction_indp_data} for $\wt m(\boldX_i)$ being considered over $\boldX_i \in \calU$ ($i=n+1,\ldots,n+N)$ only. One may alternatively consider this for $\boldX_i$'s over the {\it entire} $\calD \equiv \calL \cup \calU$. Our current approach conveniently ensures that the two components (from the two models) forming the product in the likelihood \eqref{likelihood_function_indp_data} are actually based on {\it independent} sources of data, $\calL$ and $\calU$, ensuring the likelihood's probabilistic validity as a {\it joint} likelihood, and that $\theta$ and $b(\wt m)$ can be learnt {\it simultaneously}. On the other hand, if the second component now includes all $\boldX_i \in \calD$ $(i=1,\ldots, n+N$), then this product formulation is lost and one needs to consider an alternative {\it hierarchical} approach to learn the two parameters, as follows. For ease of exposition here, we keep the hyperparameters $\sigma_1^2$ and $\sigma_2^2$ implicit in the notations below. Let $L_1\{b(\wt m); \calL\}$ denote the first component in the likelihood \eqref{likelihood_function_indp_data}. Then, we {\it first} learn a posterior for $b(\wt m)$ based on $L_1(\cdot)$, and then {\it given} a sample of $b(\wt m)$, we learn $\theta \mid b(\wt m)$ hierarchically using the `conditional' likelihood $L_2\{ \theta; \calD \mid b(\wt m)\}$, where $L_2(\cdot)$ is the {\it modified version} of the second component in \eqref{likelihood_function_indp_data} with {\it all} the $\boldX_i's \in \calD$ being now included (i.e., $i=1,\ldots, n+N$). Collecting samples of $b(\wt m)$ and $\theta \mid b(\wt m)$ across this hierarchical approach eventually leads to the final posterior. Though technically more nuanced and also computation-intensive, this approach can be shown to have all the desirable properties of BDMI, while also allowing for $c > 1$. Nevertheless, given that our general focus is mostly on cases where $N \gg n$, we prefer to stick to our original BDMI formulation due to its simplicity, both technically and computationally.
\end{remark}

\subsection{A hierarchical variant of BDMI: h-BDMI}
\label{standard_bayesian_approach}

Recall that the original BDMI procedure, as described in Section~\ref{bayesian_SS}, is constructed using a {\it single} random sample $\tm \sim \Pi_{\mbm}$. Alternatively, a more traditional Bayesian approach can be adapted by considering multiple samples of $\tm$ through a hierarchical construction, as briefly mentioned in Remark~\ref{rem_hbdmi}. This section presents this alternative version of BDMI, referred to as the {\it hierarchical-BDMI} (henceforth h-BDMI), which constructs a joint posterior of $(\theta, m)$ and then marginalizes over $m$ to obtain the marginal posterior of $\theta$. This differs from the original BDMI procedure, and h-BDMI aligns more closely with traditional hierarchical Bayesian modeling principles. For its exposition, we focus on only one data fold, say $\wt\calD_k := \calD_k \cup S_k$, where $\calD_k$ and $\calS_k$ are as defined in Section~\ref{BCF} for some $k = 1, \dots, K$. Following the conventional Bayesian idea of integrating out the nuisance parameter $m$, we proceed as follows. Using $\calS_k$ as a training data, we obtain a posterior $\Pi_\mbm^{(k)} \equiv \Pi_\mbm^{(k)}(\cdot; \calS_k) $ for $m$. By the conditional independence between $m \sim \Pi_\mbm^{(k)}$ and $\calD_k$, the joint posterior of $(\theta, m)$ has the pdf $
\pi(\theta, m  \mid \wt\calD_k) \ = \ \pi(\theta \mid m, \calD_k) \hspace{0.7mm}\pi_{\mbm}^{(k)}(m),
$
where $\pi_{\mbm}^{(k)}(\cdot)$ is the pdf of the nuisance posterior $\Pi_\mbm^{(k)}$ of $m$. The pdfs $\pi(\theta \mid m, \calD_k)$ and $\pi_{\mbm}^{(k)}(m)$ remain as defined in Section~\ref{BCF}. By integrating out $m$, we obtain the marginal posterior of $\theta$, denoted $\wt \Pi_{\btheta}^{(k)}$, with corresponding pdf $\pi_{\btheta}^{(k)}(\cdot)$, based on h-BDMI as follows:
$$
\pi_{\btheta}^{(k)}(\theta) \ = \ \int \pi\big(\theta, m \mid  \wt\calD_k\big) \hspace{0.4mm}\dd m \ = \ \int \pi\big(\theta  \mid  m, \calD_k\big)\hspace{0.4mm} \pi_{\mbm}^{(k)}(m) \hspace{0.4mm} \dd m.
$$
Estimation and inference on the true parameter $\theta_0 \equiv \bbE(Y)$ using h-BDMI can be performed based on this posterior $\wt \Pi_{\btheta}^{(k)}$, for any $k = 1,\ldots, K$. Using iterated expectations, the posterior mean $\wh \theta_{\hBDM}^{(k)} \equiv \bbE_{\theta \sim \wt \Pi_{\btheta}^{(k)}}(\theta)$ of $\wt \Pi_{\btheta}^{(k)}$ can be expressed as
$
\wh \theta_{\hBDM}^{(k)} ~\equiv ~ \int \left\{ \int \theta \hspace{0.5mm} \pi(\theta \mid m, \calD_k) \hspace{0.5mm} \dd \theta \right\} \pi_{\mbm}^{(k)} \hspace{0.5mm} \dd m,
$
where the inner integral is the conditional mean of $\theta$ given $m$ and $\calD_k$, i.e., $\bbE(\theta \mid m, \calD_k)$. Under the prior choice in \eqref{eq:improper_prior_construction}, $\wh \theta_{\hBDM}^{(k)}$ is explicitly given by:
$$
\wh \theta_{\hBDM}^{(k)} ~\equiv ~ \frac{1}{\nK} \sum_{i \in \calI_k} \big\{Y_i - \wh m^{(k)}(\boldX_i)\big\} + \frac{1}{\NK} \sum_{i \in \calJ_k} \wh m^{(k)}(\boldX_i), ~\text{ where $\wh m^{(k)}$ is the posterior mean of $\Pi_{\mbm}^{(k)}$}.
$$
Also, it is easy to draw samples from the posterior $\wt \Pi_{\btheta}^{(k)}$ of $\theta$ to construct credible intervals. Specifically, for sufficiently large $M$, we first draw samples $\tm_1, \dots , \tm_M$ from the posterior $\Pi_{\mbm}^{(k)}$, and for each sample, we draw a sample $\theta \mid \tm_j, \calD_k$ from the posterior $\Pi^{(k)}_{\btheta}$ as described in Proposition~\ref{prop_half_data} for $j = 1, \dots , M$. This process yields $M$ samples of $\theta$ from the posterior $\wt \Pi_{\btheta}^{(k)}$. Finally, applying the h-BDMI procedure to each $\wt \calD_1 , \dots , \wt \calD_K$, we obtain the corresponding posteriors $\wt \Pi_{\btheta}^{(1)}, \dots \wt \Pi_{\btheta}^{(K)}$. Following the aggregation approach detailed in Section~\ref{BCF}, we can construct a CF-based aggregated posterior $\wt \Pi_{\btheta}$. These modifications can be incorporated into Algorithm~\ref{algo}, which we omit for brevity. We next present the result on the theoretical properties of h-BDMI on one data fold $\wt\calD_k$, followed by a discussion on the differences between BDMI and h-BDMI.

\begin{theorem}\label{bvm_standard}
Suppose Assumptions~\ref{SS-assumptions} and~\ref{assumption_for_half_fold} hold, except that the NPCC \eqref{eqn_HLBC} in Assumption~\ref{assumption_for_half_fold} (ii) is replaced with a modified NPCC as follows: assume that the posterior $\Pi_{\mbm}^{(k)}$ of $m$ satisfies the nuisance Bayes risk condition: $
\bbE_{m \sim \Pi_{\mbm}^{(k)}} \{ \| m(\boldX) - m^*(\boldX)\|^2_{\bbL_2(\bbP_{\boldX})} \hspace{0.7mm} | \hspace{0.7mm} \calS_k \} \cvP 0$ under $\bbP_{\calS_k}$. Then, the posterior $\wt \Pi_{\btheta}^{(k)}$ of $\theta$ from the h-BDMI procedure on any pair $(\calD_k, \calS_k)$ as above inherits a BvM-type limiting behavior as follows:
\begin{equation}
\left \|\wt \Pi_{\btheta}^{(k)} - \calN(\wh\theta_{ \BDM}^{(k)}(m^*), \tau^2_{\nK, \NK}(m^*)) \right \|_{\TV} \ \cvP \ 0 \ \text{ in probability w.r.t.} \ \bbP_{\wt\calD_k} \ \text{ as } \ n, N \to \infty,
\end{equation}
for any $k = 1,\ldots, K$, where $\wh\theta_{ \BDM}^{(k)}(m^*)$ and $\tau^2_{\nK,\NK}(m^*)$ are the same as defined in Theorem~\ref{bvm_on_first_half_data}.
\end{theorem}

We conclude this section with a brief comparison between the BDMI and h-BDMI approaches. Firstly, Theorem~\ref{bvm_standard} establishes a corresponding BvM-type result for h-BDMI, similar to Theorem~\ref{bvm_on_first_half_data} for the `one data fold' version of the original BDMI procedure described in Section~\ref{BCF}.
While both theorems demonstrate that the marginal posteriors of $\theta$ inherit a BvM-type limiting behavior with the same limiting posterior, they {\it do} have some important differences. Notably, Theorem~\ref{bvm_standard} requires a {\it stronger} $L_1$-type (Bayes risk) convergence condition on the contraction of the posterior $\wt \Pi_{\mbm}^{(k)}$ around $m^*(\cdot)$, while Theorem~\ref{bvm_on_first_half_data} relies on the much weaker in-probability type condition \eqref{eqn_HLBC}. In practice, our simulation results in Section~\ref{simulations} reveal that the difference between BDMI and h-BDMI is less pronounced. In most cases, the two methods perform similarly in estimating $\theta_0$, as illustrated in Table~\ref{table_n500_mse_try}. Occasionally, h-BDMI tends to give slightly conservative coverages compared to BDMI (see Table~\ref{table_n500_p50_cov}), which is not unexpected since h-BDMI involves multiple samples (hence more noise) as it integrates out the nuisance parameter $m$ rather than conditioning on a single draw.

A key advantage of BDMI lies in its simplicity and computational efficiency. Unlike h-BDMI, which requires multiple samples from the nuisance posterior $\Pi_{\mbm}$ of $m$, BDMI relies on
only a single sample, reducing computation burden. Thus, we recommend the original BDMI approach for achieving {\it both} efficient estimation and reliable inference for the true parameter $\theta_0$. For further details and discussions, we refer to Section~\ref{simulations}.

Finally, while we have used a `one fold' version of h-BDMI here for clarity, it also admits a CF-based full data version (`h-BDMI-CF', if we may) analogous to the BDMI-CF procedure in Section~\ref{BCF}. This version inherits similar theoretical properties as Theorem~\ref{main_thm} (with the same distinctions as above). In our simulations in Section~\ref{simulations}, we implemented h-BDMI via this CF-based full data version to ensure a fair comparison with the BDMI-CF and supervised approaches. The notation `h-BDMI' therein refers to this CF-based version.

\section[Numerical studies]{Numerical studies}\label{simulations}

We conducted extensive simulation studies to investigate the finite sample performance, both in estimation and inference, for our proposed SS approach(es) and the supervised approach under various settings. In particular, as point estimators, we compare the supervised estimator $\wh\theta_{\sup} \equiv \overline{Y}$ based on $\calL$, the posterior mean $\wh\theta_{\BDM} \equiv \wh\theta_{\BDM}(\tm_{\CF})$ of $\Pi_{\btheta}$ from the final BDMI-CF procedure (as in Algorithm \ref{algo}) and the posterior mean $\wh\theta_{\hBDM}$ of $\wt \Pi_{\btheta}$ from the h-BDMI procedure (its CF based version) discussed in Section \ref{standard_bayesian_approach}. We compare their estimation efficiencies based on the empirical mean squared error ({\bf MSE}) and report their relative efficiencies ({\bf RE}) compared to the supervised estimator $\wh\theta_{\sup}$. Further, for evaluating the accuracy of inference, we report the empirical coverage probabilities (\textbf{CovP}) and lengths ({\bf Len}) of the $95\%$ credible intervals ({\bf CI}s) obtained from their respective posteriors. Finally, as a performance benchmark for estimation efficiency, we also report the maximum (oracle) asymptotic relative efficiency ({\bf ORE}) relative to $\wh\theta_{\sup}$, given by $\Var(\wh\theta_{\sup})/\tau^2_{n, N}(m^*)$, where $\tau^2_{n, N}(m^*) = \Var\{Y - m^*(\boldX)\}/n + \Var\{m^*(\boldX)\}/N$ with $m^*(\cdot) = m_0(\cdot)$. For the choice of the number of folds $K$, we consider $K = 5$ and $10$. The reported simulation results are all based on 500 replications. We examine various true data generating mechanisms and different methods for nuisance parameter estimation, leading to both correctly specified and misspecified models for $m_0(\cdot)$. We discuss the correctly specified and misspecified model settings and their corresponding results in Sections \ref{sim_correctly_specified}-- \ref{sim_misspecified}.

\subsection[Simulation studies: Correctly specified models]{Simulation studies: Correctly specified models}\label{sim_correctly_specified}

Throughout, we set $n = 500$ and $N = 10000$, and considered $p = 50$ and $p = 166$ ($\approx n/3)$, representing moderate and high dimensional settings (relative to $n$), respectively. We generated $\boldX \sim \calN_p(\bzero_p, I_p)$, and given $\boldX = \bx$, we generated $Y \sim \calN(m_0(\bx), \sigma_0^2)$, where $m_0(\bx) = \alpha_0 + \bx'\bbeta_0$
and
$\sigma_0^2 = \Var\{m_0(\boldX)\}/5$, and we used $\alpha_0 = 5$ and $\bbeta_0 = (\mathbf{1}_{s/2}',\ \mathbf{0.5}_{s/2}',\ \bzero_{p-s}')'$ (for different choices of $s$ discussed below). Here, $\calN_d(\boldsymbol{\mu}, \boldsymbol{\Sigma}$) denotes the $d$-variate ($d \geq 2$) Gaussian distribution with mean $\boldsymbol{\mu}_{d \times 1}$ and covariance matrix $\boldsymbol{\Sigma}_{d \times d}$, $I_d$ denotes the identity matrix of order $d$, and the notation $\mathbf{a}_{l}$, for any positive integer $l$ (e.g., $l = p$, $s/2$ or $p - s$, as above), denotes the vector $(a, \dots, a)'_{l \times 1}$ for any $a \in \bbR$ (e.g., $a = 0$, $0.5$ or $1$, as above). The parameter $s$ in $\bbeta_0$ above denotes the {\it sparsity} of $\bbeta_0$. For $p = 50$, we set $s = 7$ $(\approx \sqrt{p})$, or $s = 50 \equiv p$; while for $p = 166$, we set $s = 13$ ($\approx \sqrt{p}$), $ s= 55$ ($\approx p/3$), $s = 83$ ($\approx p/2$), or $s = 166 \equiv p$.
These choices of $s$ span a variety of settings, including {\it sparse} $(s= \sqrt{p}$), {\it moderately dense} $(s = p/2$ or $p/3$), or {\it fully dense} ($s =p$) cases. Note that, except for the sparse case, none of these choices correspond to settings where $s$ (or $p$) may be considered small or fixed relative to $n$, and therefore appropriate sparsity-friendly nuisance estimation methods may {\it still} fail to consistently estimate $m_0$. For illustrative purposes, we consider {\it three choices} (all parametric model based) for obtaining the nuisance posterior $\Pi_{\mbm}$: Bayesian ordinary linear regression ({\tt Bols}), Bayesian ridge regression ({\tt Bridge}), and a sparse Bayesian linear regression method ({\tt Bsparse}) based on non-local priors (NLP) \citep{johnson2012bayesian}. In all cases, we consider the Gaussian linear regression model $Y_i \mid \boldX_i, \alpha, \bbeta, \sigma \iid \calN(\alpha +\boldX_i'\bbeta, \sigma^2)$ for $i = 1, \dots, n$. For {\tt Bols}, we use a prior on $(\alpha,\bbeta, \sigma^2)$ given by: $\pi(\alpha,\bbeta \mid \sigma^2) \propto 1$ and $\pi(\sigma^2) \propto (\sigma^2)^{-1}$; and for {\tt Bridge}, the prior employed on $(\alpha, \bbeta,\sigma^2)$ is: $\pi(\alpha \mid \sigma^2) \propto 1$, $\bbeta \mid \lambda, \sigma^2 \sim \calN_p(\bzero_p, \lambda^{-1}\sigma^2 I_p)$, with $\alpha$ and $\beta$ being independent, and $\pi(\sigma^2) \propto (\sigma^2)^{-1}$. We use an empirical Bayes approach to plug in a point estimate $\wh\lambda$ for the prior precision (or ridge) parameter $\lambda$. The estimate $\wh\lambda$ is obtained from the {\tt R} package {\tt glmnet} so that the posterior mean of $(\alpha, \bbeta')' \in \bbR^{(p+1)}$ coincides with the cross-validated point estimate obtained from {\tt cv.glmnet} in the {\tt glmnet} package. For both these methods, we obtain that the posteriors of $(\alpha, \bbeta)$ are multivariate $t$-distributions. For the {\tt Bsparse} method, we use the R package \texttt{mombf} to obtain posterior samples for $(\alpha, \bbeta)$. The implementation details of the \texttt{mombf} and {\tt glmnet} packages are provided in Section \ref{supp_implementation_details} of the \hyperref[sec:supplementary]{Supplementary Material}.

\begin{table}[!htb]
\centering
\caption{Relative efficiency (RE) of $\wh\theta_{\BDM,i}$ and $\wh\theta_{\hBDM,i}$ relative to $\wh\theta_{\sup}$, w.r.t. their empirical MSEs,
for the settings in Section \ref{sim_correctly_specified}, where the methods (the subscript ``$i$'') used to obtain the nuisance posterior $\Pi_{\mbm}$ for BDMI
are denoted as: $l = $ {\tt Bols}, $r = $ {\tt Bridge} and $s = $ {\tt Bsparse}. {\bf Settings:}
$n = 500$, $N = 10000$, and: (i) $p = 50$, with $s = 7$ or $50$; or (ii) $p = 166$, with $s = 13, 55, 83$ or $166$. (As a performance benchmark, we also report the maximum (oracle) asymptotic relative efficiency (ORE) relative to $\wh\theta_{\sup}$.)}
\label{table_n500_mse_try}
\begin{tabular}{lccccccccccc}
\toprule
 \multicolumn{3}{c}{} &
 \multicolumn{2}{c}{$\wh\theta_{\sup}$} &
 \multicolumn{1}{c}{$\wh\theta_{\BDM,l}$} &
 \multicolumn{1}{c}{$\wh\theta_{\BDM,r}$} &
 \multicolumn{1}{c}{$\wh\theta_{\BDM,s}$} &
 \multicolumn{1}{c}{$\wh\theta_{\hBDM,l}$} & \multicolumn{1}{c}{$\wh\theta_{\hBDM,r}$} &
\multicolumn{1}{c}{$\wh\theta_{\hBDM,s}$} &
 \multicolumn{1}{c}{} \\
 \midrule
 $p$ & $s$ & $K$ & MSE & RE & RE & RE & RE  & RE & RE  & RE & \textbf{ORE} \\
 \midrule
50  & 7  &  5 &  0.01 & 1.00  & 3.99   & 4.38 & 5.00 &  4.61  & 4.69  & 5.06  &  4.80 \\
 &   & 10 &  0.01 & 1.00  & 4.12  & 4.73 & 5.04  & 4.67   &  4.72 & 5.08  &  4.80 \\
 \hline
50  & 50  &  5 &  0.08 & 1.00  & 4.31   & 4.38   & 3.88    & 4.33   &  4.35  &  4.22  & 4.80 \\
 &   & 10 &  0.08 & 1.00  & 4.35  & 4.41   &  4.02   & 4.37   &  4.42   &  4.30  & 4.80 \\
   \midrule
   \midrule
166  & 13  &  5 &  0.02 & 1.00  & 2.84 & 3.46 & 4.56 & 3.30 & 3.62  & 4.75 &   4.80 \\
 &   & 10 & 0.02 & 1.00  & 3.17 & 3.61 & 4.70 & 3.64 & 3.88 & 4.81 & 4.80 \\
   \hline
166  & 55  &  5 & 0.09  & 1.00  & 3.02 & 3.48 & 3.15 & 3.08 & 3.47 & 3.42 &  4.80 \\
 &   &  10 & 0.09  & 1.00  & 3.45 & 3.83 & 3.41 & 3.49 & 3.81 & 3.78 &  4.80 \\
 \hline
166  & 83  &  5 & 0.13  & 1.00  & 3.01 & 3.28 & 1.40 & 3.03 & 3.31 & 1.49 &  4.80 \\
 &   &  10 & 0.13  & 1.00  & 3.33 & 3.59 & 1.96 & 3.35 & 3.56 & 2.15 &  4.80 \\
   \hline
166  & 166  &  5 &  0.26 & 1.00  & 3.30 & 3.64 & 0.98 & 3.32 & 3.66 & 1.00 &  4.80 \\
 &   &  10 &  0.26 & 1.00  & 3.60  & 3.81  & 0.98 & 3.58 & 3.82 & 1.00 &  4.80 \\
   \bottomrule
   \end{tabular}
\end{table}

\begin{figure}[ht!]
    \centering
    \begin{subfigure}{0.45\textwidth}
        \centering
        \includegraphics[width=\linewidth]{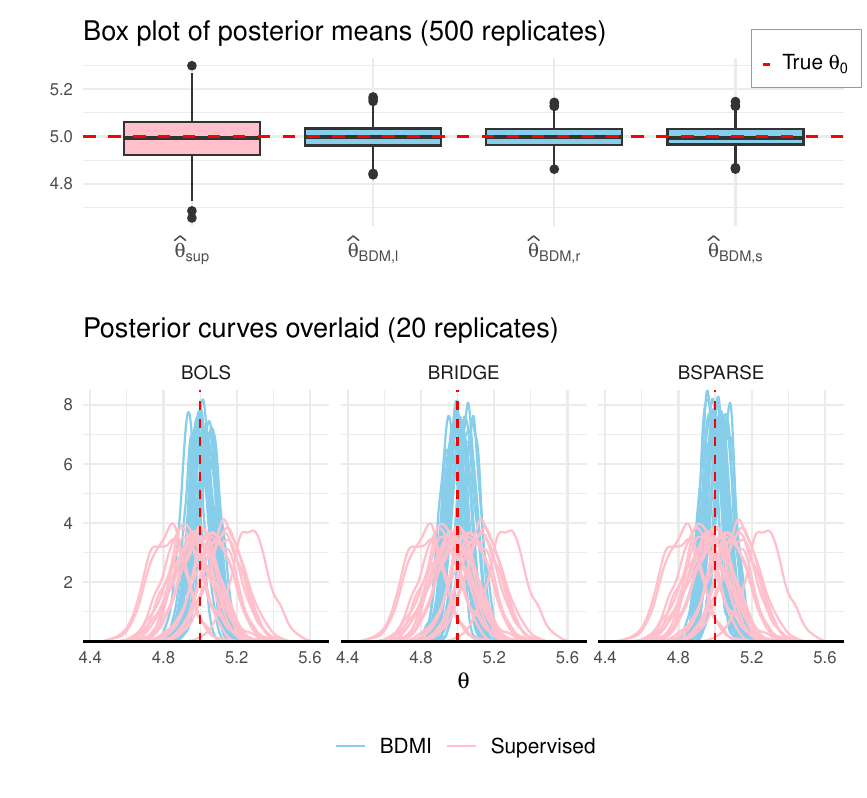}
        \caption{Setting: $p = 50$ with $\mathbf{s = 7}$.}
        \label{fig:subfig1:p50s7}
    \end{subfigure}
    \hfill
    \begin{subfigure}{0.45\textwidth}
        \centering
        \includegraphics[width=\linewidth]{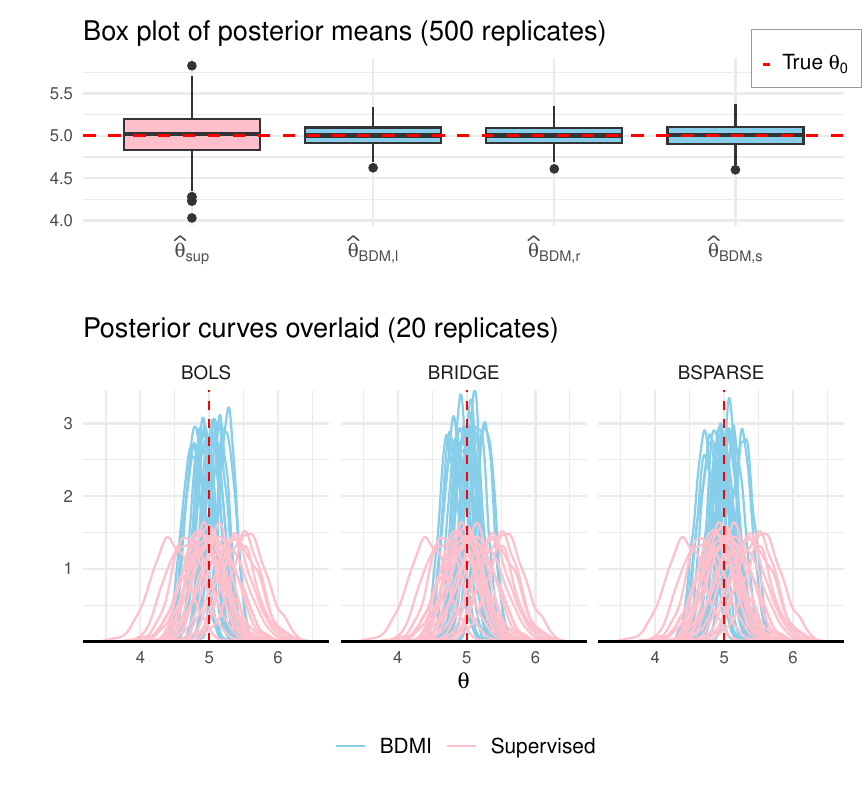}
        \caption{Setting: $p = 50$ with $\mathbf{s = 50}$.}
        \label{fig:subfig2:p50s50}
    \end{subfigure}
    \caption{Box plots of posterior means  (based on 500 replications) and plots of overlaid density curves (based on 20 iterations) for the posteriors $\Pi_{\sup}$ (pink) and $\Pi_{\btheta}$ (blue) of $\theta$,
    with three different methods ({\tt Bols}, {\tt Bridge} and {\tt Bsparse}) to obtain the nuisance posterior $\Pi_{\mbm}$ for BDMI. {\bf Setting:} $n = 500$, $N = 10000$, $p = 50$, and $s = 7$ or $10$. Each density curve is generated using 1000 posterior samples of $\theta$. The red dashed vertical line indicates the true parameter of interest $\theta_0$ and equals 5 for all settings.}
    \label{fig:p50s7_50}
\end{figure}

\begin{figure}[ht!]
    \centering
    \begin{subfigure}{0.45\textwidth}
        \centering
        \includegraphics[width=\linewidth]{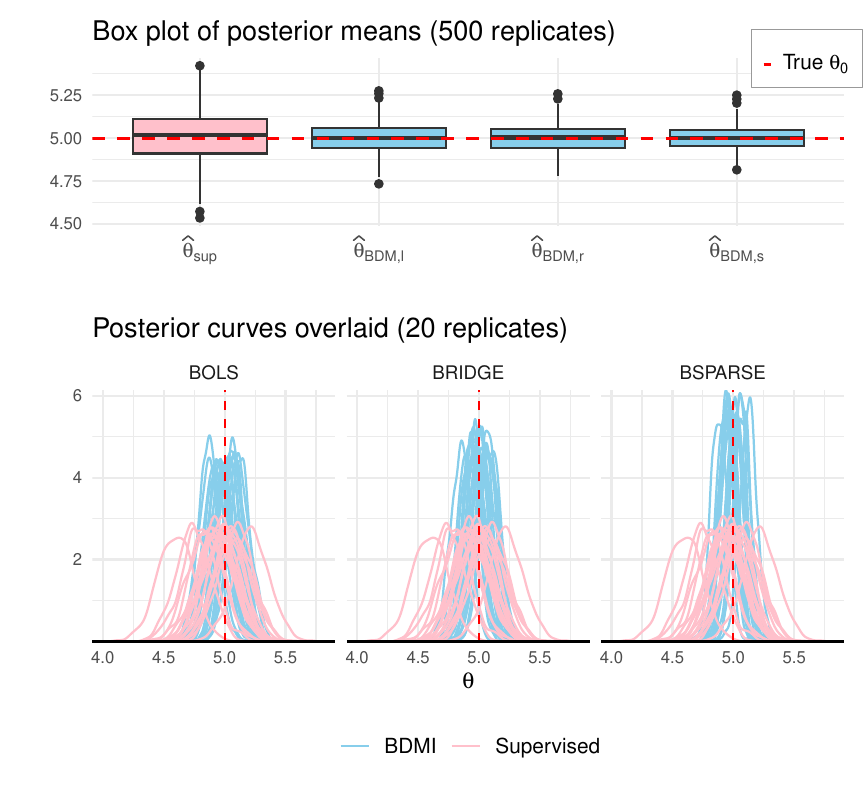}
        \caption{Setting: $p = 166$ with $\mathbf{s = 13}$.}
        \label{fig:subfig1:p166s13}
    \end{subfigure}
    \hfill
    \begin{subfigure}{0.45\textwidth}
        \centering
        \includegraphics[width=\linewidth]{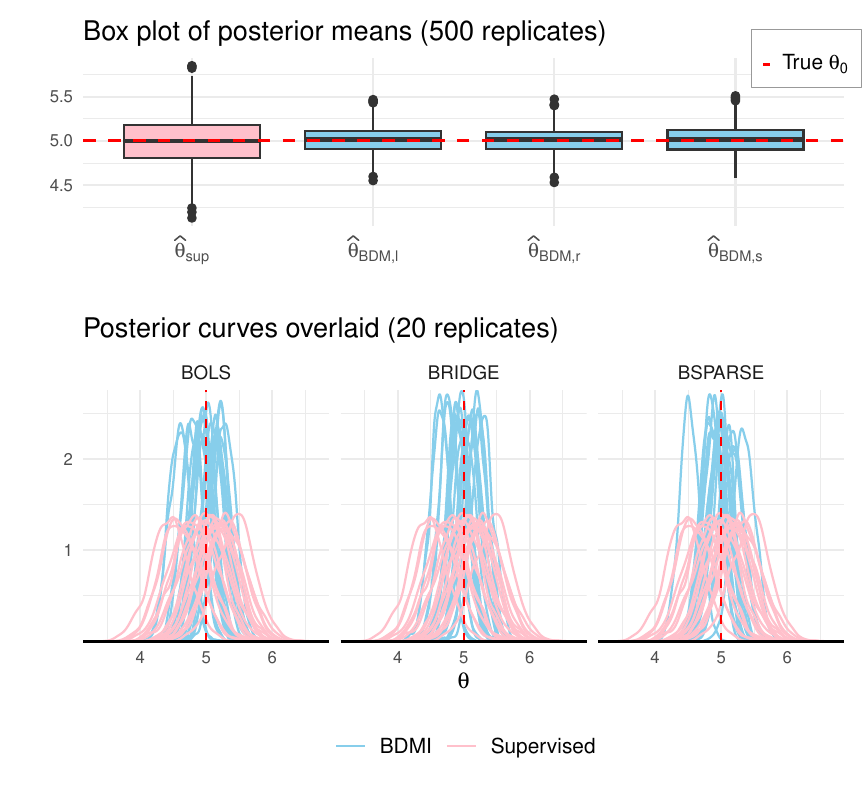}
        \caption{Setting: $p = 166$ with $\mathbf{s = 55}$.}
        \label{fig:subfig2:p166s55}
    \end{subfigure}
    \vspace{1em}

    \begin{subfigure}{0.45\textwidth}
        \centering
        \includegraphics[width=\linewidth]{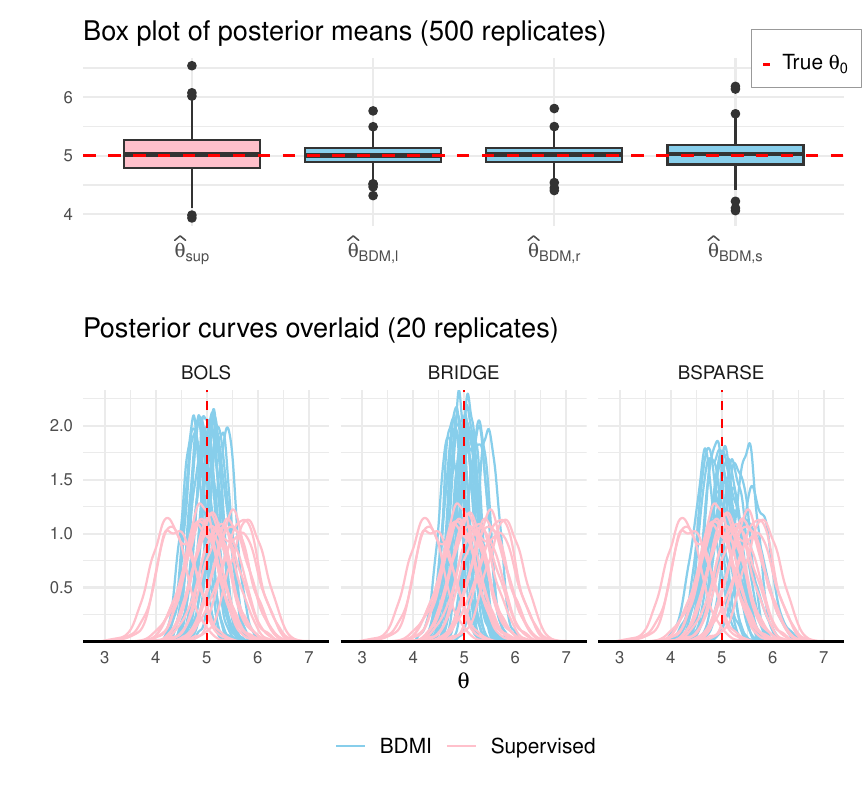}
        \caption{Setting: $p = 166$ with $\mathbf{s = 83}$.}
        \label{fig:subfig1:p166s83}
    \end{subfigure}
    \hfill
    \begin{subfigure}{0.45\textwidth}
        \centering
        \includegraphics[width=\linewidth]{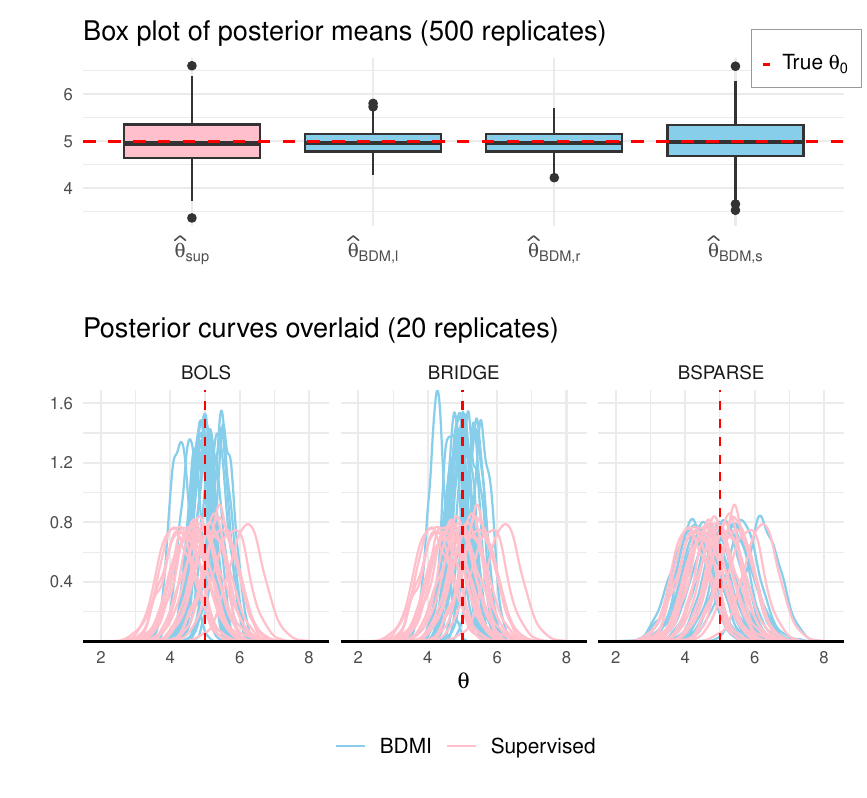}
        \caption{Setting: $p = 166$ with $\mathbf{s = 166}$.}
        \label{fig:subfig2:p166s166}
    \end{subfigure}
    \caption{Box plots of posterior means and plots of overlaid density curves for the posteriors $\Pi_{\sup}$ (pink) and $\Pi_{\btheta}$ (blue) of $\theta$. {\bf Setting:} $n = 500$, $N = 10000$, $p = 166$, and $s=13, 55, 83$ or $166$. The rest of the caption details remain the same as in Figure \ref{fig:p50s7_50}.}
    \label{supp:fig:2x2_matrix}
\end{figure}

Table \ref{table_n500_mse_try} and Tables \ref{table_n500_p50_cov}--\ref{table_n500_p166_cov} present the results on estimation efficiency and inference, respectively, along with illustrations of the posteriors and their overall behaviors in Figures \ref{fig:p50s7_50}--\ref{supp:fig:2x2_matrix}. As seen from Table \ref{table_n500_mse_try} (as well as the box plots in Figures \ref{fig:p50s7_50}--\ref{supp:fig:2x2_matrix}),
the REs of $\wh\theta_{\BDM}$ and $\wh\theta_{\hBDM}$ w.r.t. $\wh\theta_{\sup}$, i.e., $\text{MSE}(\wh\theta_{\sup})/ \text{MSE}(\wh\theta_{\BDM})$ and $\text{MSE}(\wh\theta_{\sup})/ \text{MSE}(\wh\theta_{\hBDM})$, are consistently greater than 1, ranging roughly between 2 to 5 across most settings. This highlights the substantial efficiency improvement achieved by BDMI over the supervised approach. In addition, as illustrated in  Figures \ref{fig:p50s7_50}--\ref{supp:fig:2x2_matrix}, apart from point estimators, the {\it posteriors themselves} are consistently and significantly {\it tighter} than the supervised posteriors, while throughout resembling a Gaussian behavior centered at the true $\theta_0$. These patterns hold generally {\it regardless} of the setting and/or the nuisance posterior.

Furthermore, Table \ref{table_n500_mse_try} illustrates that the efficiency improvement depends primarily on the dimensionality $p$ and the sparsity level $s$. In moderate-dimensional settings ($p = 50$), BDMI achieves (near-)optimal efficiency gains, with RE values close to each other and approaching the ORE value, regardless of the sparsity levels (sparse $s = \sqrt{p}$ and fully dense $s=p$). This confirms that BDMI performs {\it optimally} when the model is correctly specified {\it and} estimated well enough. The impact of the sparsity level becomes particularly apparent in high-dimensional scenarios ($n = 500$ with $p = 166$), where finite-sample nuisance estimation {\it bias} introduces a {\it soft} form of misspecification. Specifically, sparsity-friendly nuisance models (e.g., \texttt{Bsparse}) struggle to consistently estimate $m_0(\cdot)$ in moderately dense ($s = p/2$) or fully dense ($s = p$) settings, leading to somewhat lower RE values. However, in sparse settings $s = \sqrt{p}$ ($s = 13$), \texttt{Bsparse} achieves RE values that are close to ORE by leveraging the underlying sparse structure, outperforming non-sparse methods
such as \texttt{Bols} and \texttt{Bridge}. Conversely, in fully dense settings ($p = s$, $s = 166$), \texttt{Bsparse} struggles to adapt and estimates a nearly constant function, resulting in RE values close to 1. In contrast, the non-sparse methods \texttt{Bols} and \texttt{Bridge} still target non-trivial approximations of $m_0$, yielding reasonably high RE values (approximately 3.70; see Table~\ref{table_n500_mse_try}). These observations highlight that while BDMI remains robust under soft misspecification, the choice of a nuisance model can influence the {\it extent} of efficiency gain in dense settings, emphasizing the interplay among both $p$ and $s$. Notably, even in high-dimensional (fully dense) cases, RE values remain still acceptable (RE $> 1$), albeit not optimal, and BDMI consistently provides correct coverage around $95\%$ regardless of the nuisance parameter estimation methods. Moreover, Table \ref{table_n500_mse_try} shows that the RE values of the SS estimators tend to be slightly higher for $K = 10$. But in general, the results -- both for estimation and inference -- seem to be fairly robust across both choices of $K$. We thus recommend either choice in practice.

\begin{table}[!htb]
\centering
\caption{Inference results for $\theta_0$ based on the 95\% CIs from the posteriors $\Pi_{\sup}, \Pi_{\btheta}$ (BDM) and $\wt \Pi_{\btheta}$ (hBDM), for the settings
in Section \ref{sim_correctly_specified}, with $n = 500$, $N = 10000$, $p = 50$, and $s = 7$ or $50$. The methods used to obtain the nuisance posterior $\Pi_{\mbm}$ for BDM (or hBDM) are denoted as: $l = $ {\tt Bols}, $r = $ {\tt Bridge} and $s = $ {\tt Bsparse}. The columns `\textbf{CovP}' and `Len' respectively denote the average empirical coverage probability and the average length of the 95\% CIs across the iterations.} \label{table_n500_p50_cov}
\tabcolsep=0.13cm
\begin{tabular}{lccccccccccccccc}
\toprule
  \multicolumn{2}{c}{} & \multicolumn{2}{c}{$\CI_{\sup}$} & \multicolumn{2}{c}{$\CI_{\BDM,l}$} & \multicolumn{2}{c}{$\CI_{\BDM,r}$} & \multicolumn{2}{c}{$\CI_{\BDM,s}$} & \multicolumn{2}{c}{$\CI_{\hBDM,l}$} & \multicolumn{2}{c}{$\CI_{\hBDM,r}$} & \multicolumn{2}{c}{$\CI_{\hBDM,s}$}\\
     \midrule
   $s$ & $K$ & \textbf{\small CovP} & Len & \textbf{\small CovP} & Len & \textbf{\small CovP} & Len  & \textbf{\small CovP}& Len & \textbf{\small CovP} & Len & \textbf{\small CovP} & Len & \textbf{\small CovP} & Len  \\
 \midrule
7 & 5 & 0.95 & 0.42 & 0.94 & 0.21 & 0.95 & 0.21 & 0.95 & 0.20 & 0.96 & 0.23 & 0.96 &  0.21 & 0.95 & 0.20 \\
 & 10 & 0.95 & 0.42 & 0.95 & 0.21 & 0.96 & 0.21 & 0.96 & 0.20 & 0.97 & 0.23 & 0.97 & 0.21 & 0.96 & 0.20 \\
 \hline
50 & 5& 0.95 & 1.07 & 0.96 & 0.55 & 0.96 & 0.54 & 0.95 & 0.55 & 0.96 & 0.58 & 0.97 & 0.57 & 0.97 & 0.57  \\
 & 10 & 0.95 & 1.07 & 0.96 & 0.55 & 0.96 & 0.54 & 0.95 & 0.55 & 0.97 & 0.57 & 0.97 & 0.57 & 0.97 & 0.57   \\
 \bottomrule
 \end{tabular}
\end{table}

\begin{table}[!htb]
\centering
\caption{Inference results for $\theta_0$ for the settings in Section \ref{sim_correctly_specified}, with $n = 500$, $N = 10000$, $p = 166$, and $s = 13, 55, 83$ or $166$. The rest of the caption details remain the same as in Table \ref{table_n500_p50_cov}.}
\label{table_n500_p166_cov}
\tabcolsep=0.124cm
\begin{tabular}{lccccccccccccccc}
\toprule
  \multicolumn{2}{c}{} & \multicolumn{2}{c}{$\CI_{\sup}$} & \multicolumn{2}{c}{$\CI_{\BDM,l}$} & \multicolumn{2}{c}{$\CI_{\BDM,r}$} & \multicolumn{2}{c}{$\CI_{\BDM,s}$} & \multicolumn{2}{c}{$\CI_{\hBDM,l}$} & \multicolumn{2}{c}{$\CI_{\hBDM,r}$} & \multicolumn{2}{c}{$\CI_{\hBDM,s}$}\\
     \midrule
   $s$ & $K$ & \textbf{\small CovP} & Len & \textbf{\small CovP} & Len & \textbf{\small CovP} & Len  & \textbf{\small CovP}& Len & \textbf{\small CovP} & Len & \textbf{\small CovP} & Len & \textbf{\small CovP} & Len  \\
 \midrule
13 & 5 & 0.95 & 0.56 & 0.94 & 0.35 & 0.96 & 0.32   & 0.94 & 0.26 & 0.98 & 0.36 & 0.96 & 0.32 & 0.95 & 0.27  \\

 & 10 & 0.95 & 0.56 & 0.96 & 0.33 & 0.96 & 0.31 & 0.95 & 0.27 & 0.98 & 0.35 & 0.97 & 0.32 & 0.95 & 0.27 \\
 \hline
55 & 5  &  0.94 & 1.13 & 0.95 & 0.66  & 0.95 & 0.62 & 0.95  & 0.66  &0.94 & 0.66  & 0.95 & 0.62 & 0.97 & 0.69 \\
    & 10 &  0.94 & 1.13 & 0.95 & 0.64 & 0.96 &0.62  &0.95  & 0.64 & 0.95 & 0.64 & 0.95 & 0.62  & 0.97  & 0.67 \\
    \hline
83 & 5  &  0.95 & 1.38 & 0.96 & 0.80  & 0.95 & 0.77 & 0.95  & 1.16  & 0.96 & 0.81  & 0.95 & 0.76 & 0.97 & 1.12 \\
    & 10 &  0.95 & 1.38 & 0.96 & 0.79 & 0.95 & 0.75  & 0.95  & 0.97 & 0.95 & 0.78 & 0.95 & 0.75  & 0.96  & 1.01 \\
\hline
166 & 5 & 0.95 & 1.95 & 0.96 & 1.13 & 0.96 & 1.08 & 0.95 & 1.98 & 0.96 & 1.13 & 0.96 & 1.08 & 0.95 & 2.02 \\

 & 10 & 0.95 & 1.95 & 0.96 & 1.10 & 0.96 & 1.06 & 0.93 & 2.00 & 0.96 & 1.10 & 0.96 & 1.06 & 0.95 & 2.04 \\
 \bottomrule
 \end{tabular}
\end{table}

Tables \ref{table_n500_p50_cov}--\ref{table_n500_p166_cov} exhibit that BDMI {\it consistently} achieves {\it correct} coverage probabilities for $\theta_0$, maintaining approximately $95\%$ coverage across all settings with various choices of $p, s, K$, as well as different methods for obtaining the nuisance posterior $\Pi_{\mbm}$. This highlights the {\it robustness} of BDMI in providing valid and accurate inference (correct coverage), as well as substantial {\it improvement} over supervised inference with {\it tighter} CIs (typically around 50\% tighter) across settings -- thereby validating its construction and our claimed theoretical properties. Figures \ref{fig:p50s7_50}--\ref{supp:fig:2x2_matrix} provide visual confirmation of these findings, showing that BDMI-based posteriors consistently exhibit always tighter spread than the supervised posterior, regardless of the setting or the nuisance posterior method. Additionally, the variability of the BDMI posteriors {\it remains} consistent within each setting, further emphasizing its robustness and nuisance-insensitivity across the different scenarios.

Lastly, comparing the BDMI and h-BDMI approaches, we observe that despite the former requiring only one sample from $\Pi_{\mbm}$, both methods perform similarly across most settings, which: (i) validates our earlier claims on their common theoretical properties, {and} (ii) also {\it reinforces} the crucial role of {\it debiasing} common to both, that {\it negates} any distinction between the use of one vs. many $\wt m$ samples. The point estimators $\wh\theta_{\BDM}$ and $\wh\theta_{\hBDM}$ show very similar efficiencies with h-BDMI marginally higher in some cases, while for inference, h-BDMI often tends to give slightly conservative coverages $> 95\%$ (likely due to more noise from its hierarchical nature). Overall, given its computational simplicity, we recommend the original BDMI approach.

\subsection[Simulation studies: Misspecified models]{Simulation studies: Misspecified models} \label{sim_misspecified}

Section \ref{sim_correctly_specified} considered scenarios where the true model is linear, with Bayesian linear methods used to obtain the nuisance posterior $\Pi_{\mbm}$ of $m$. Although the models were technically ``correctly" specified, high dimensional (and dense) settings do {\it not} necessarily guarantee consistent estimation of the true $m_0(\cdot)$, leading to a `soft' form of misspecification. We now examine the functional form of misspecification where the limiting function $m^*(\cdot)$ around which $\Pi_{\mbm}$ contracts, is {\it not} equal to $m_0(\cdot)$, i.e., $m_0(\cdot)$ is nonlinear but the {\it fitted} model for learning $\Pi_{\mbm}$ remains linear. Even in such cases, Theorems \ref{main_thm}--\ref{bvm_standard} ensure BDMI's validity with efficiency improvement persisting (see Table \ref{table_efficiency}), though the improvements may not reach the optimal ORE.

Throughout, we set $N = 10000$ and $n = 500$. To illustrate our points, we specifically study non-linear, but low or moderate dimensional models with $p = 10$ (and sparsity $s = 10$ or $3$) or $p = 50$ (and sparsity $s = 50$ or $7$). We generated $\boldX \sim \calN_p(\bzero, I_p)$ as in Section \ref{sim_correctly_specified} and given $\boldX = \bx$, we generate $Y \sim \calN(m_0(\bx), \sigma_0^2)$ with $m_0(\bx) = \alpha_0 + \bx'\bbeta_0 + (\bx'\bgamma_0)^2$ and $\sigma_0^2 = \Var\{m_0(\boldX)\}/5$. Here, $\alpha_0 = 5$, $\bbeta_0 = (\mathbf{1}_{s/2}',\ \mathbf{0.5}_{s/2}',\ \bzero_{p-s}')'$ and $\bgamma_0$ is constructed to ensure $\sqrt{\bbE\{(\bbeta_0'\boldX)^2\}/\bbE\{(\bgamma_0'\boldX)^4\}} = 3$, a reasonable balance between linear and quadratic signal parts. Despite the true $m_0(\cdot)$ being non-linear, we employed linear working models to update the nuisance posterior $\Pi_{\mbm}$ like {\tt Bols, Bridge} and {\tt Bsparse} methods as detailed in Section \ref{sim_correctly_specified}. Note that due to potential misspecification, $\Pi_{\mbm}$ now contracts around a non-random limiting function $m^*(\boldX) := \wt\boldX'\bbeta^*$, where $\wt\boldX = (1, \boldX')'$ and $\bbeta^* := \arg \min_{\bbeta \in \bbR^{p+1}} \bbE ( Y - \wt\boldX' \bbeta )^2$, i.e., $\bbeta^* = \{\bbE(\wt\boldX\wt\boldX')\}^{-1}\bbE(\wt\boldX Y)$, refer to Remark \ref{remark_nuisance_contraction_rate}.

Unlike the settings in Section \ref{sim_correctly_specified}, where the theoretical ORE is attainable, it is {\it not} achievable here due to model misspecification. Instead, we calculated the {\it achievable} oracle asymptotic RE $(\ORE^*)$, defined as $\ORE^* := \Var(\wh\theta_{\sup})/\tau^2_{n, N}(m^*)$, where $\tau^2_{n, N}(m^*) = \Var\{Y - m^*(\boldX)\}/n + \Var\{m^*(\boldX)\}/N$ and $m^*(\cdot)$ is the possibly misspecified limit of $\Pi_{\mbm}$. However, both $\ORE$ and $\ORE^*$ are reported as performance benchmarks.

Table \ref{table_n500_mse_p1050} and Tables \ref{table_n500_p10_cov_miss}--\ref{table_n500_p50_cov_miss} present the results on estimation and inference, respectively. Table \ref{table_n500_mse_p1050} shows that the REs of the SS estimators $\wh\theta_{\BDM}$ and $\wh\theta_{\hBDM}$, compared to the supervised estimator $\wh\theta_{\sup}$, are substantially greater than 1, ranging roughly from 2.4 to 2.8 (matching the $\ORE^*$ closely) across most settings. Further, Tables \ref{table_n500_p10_cov_miss}--\ref{table_n500_p50_cov_miss} show that BDMI consistently achieves CovPs close to the nominal 95\% level, and with significantly tighter CIs (typically 25--40\% tighter) compared to the supervised approach across all settings and methods for $\Pi_{\mbm}$. All these findings highlight: (i) the {\it efficiency improvement} and (ii) {\it global robustness} that BDMI {continues} to enjoy {\it even under misspecification of $\Pi_{\mbm}$}, and further reinforces its first-order {\it insensitivity} to nuisance estimation bias. For more
 visual illustrations, see Figures \ref{fig:miss_p10s3-10}--\ref{fig:miss_p500s7-50} in the \hyperref[sec:supplementary]{Supplementary Material}.

\begin{table}[!ht]
\centering
\caption{Relative efficiency (RE) of $\wh\theta_{\BDM,i}$ and $\wh\theta_{\hBDM,i}$ relative to $\wh\theta_{\sup}$, w.r.t. their empirical MSEs,
for the settings in Section \ref{sim_misspecified}, with $n = 500$, $N = 10000$, and: (i) $p = 10$, with $s = 3$ or $10$; or (ii) $p = 50$, with $s = 7$ or $50$. The rest of the caption details remain the same as in Table \ref{table_n500_mse_try}. Further, apart from the ORE, as an additional benchmark appropriate for these misspecified settings considered in Section \ref{sim_misspecified}, we also report the oracle {\it achievable} asymptotic relative efficiency ($\ORE^*$) relative to $\wh\theta_{\sup}$.}\label{table_n500_mse_p1050}
\tabcolsep=0.17cm
\begin{tabular}{lcccccccccccc}
\toprule
 \multicolumn{3}{c}{} &
 \multicolumn{2}{c}{$\wh\theta_{\sup}$} &
 \multicolumn{1}{c}{$\wh\theta_{\BDM,l}$} &
 \multicolumn{1}{c}{$\wh\theta_{\BDM,r}$} &
 \multicolumn{1}{c}{$\wh\theta_{\BDM,s}$} &
 \multicolumn{1}{c}{$\wh\theta_{\hBDM,l}$} & \multicolumn{1}{c}{$\wh\theta_{\hBDM,r}$} &
\multicolumn{1}{c}{$\wh\theta_{\hBDM,s}$} & \multicolumn{1}{c}{} & \multicolumn{1}{c}{}  \\
 \midrule
$p$ & $s$ & $K$ & MSE & RE & RE & RE & RE  & RE & RE  & RE & \textbf{ORE*} & \textbf{ORE} \\
 \midrule
10  & 3  &  5 &  0.002 & 1.00  & 2.64   & 2.71 & 2.72 &  2.71  & 2.74  & 2.75 &  2.89  & 4.8 \\
 &   & 10 &  0.002 & 1.00  & 2.65  & 2.71 & 2.82  & 2.72 &  2.74 & 2.82  &  2.89  & 4.8 \\
 \hline
 10 & 10  &  5 &  0.017 & 1.00 & 2.62 & 2.67 & 2.60  & 2.63   &  2.66  &  2.63 &  2.81  &   4.8 \\
 &   & 10 &  0.017 & 1.00  & 2.62 & 2.65  &  2.57  & 2.63   &  2.65 & 2.63  &  2.81 & 4.8 \\
    \midrule
    \midrule
 50  & 7  &  5 & 0.014 & 1.00  & 2.47 & 2.51 & 2.67 & 2.48  & 2.54  & 2.73  & 3.29  &   4.8 \\
 &   & 10 &  0.014 & 1.00  & 2.47 & 2.54 & 2.72  & 2.52 &  2.57 & 2.75  &  3.29 &  4.8 \\
 \hline
 50  & 50  &  5 &  0.093 & 1.00  & 2.44   & 2.49   & 1.70  & 2.45 &  2.49  &  1.92  &  2.78 &  4.8 \\
 &   & 10 &  0.093 & 1.00  & 2.50 & 2.54  & 1.83 &  2.51   & 2.54 &  2.04 & 2.78 &   4.8 \\
   \bottomrule
   \end{tabular}
\end{table}

\begin{table}[!ht]
\centering
\caption{Inference results for $\theta_0$ for the settings in Section \ref{sim_misspecified}, with $n = 500$, $N = 10000$, $p = 10$, and $s = 3$ or $10$. The rest of the caption details remain the same as in Table \ref{table_n500_p50_cov}. }\label{table_n500_p10_cov_miss}
\tabcolsep=0.13cm
\begin{tabular}{lccccccccccccccc}
\toprule
    \multicolumn{2}{c}{} & \multicolumn{2}{c}{$\CI_{\sup}$} & \multicolumn{2}{c}{$\CI_{\BDM,l}$} & \multicolumn{2}{c}{$\CI_{\BDM,r}$} & \multicolumn{2}{c}{$\CI_{\BDM,s}$} & \multicolumn{2}{c}{$\CI_{\hBDM,l}$} & \multicolumn{2}{c}{$\CI_{\hBDM,r}$} & \multicolumn{2}{c}{$\CI_{\hBDM,s}$}\\
     \midrule
   $s$ & $K$ & \textbf{\small CovP} & Len & \textbf{\small CovP} & Len & \textbf{\small CovP} & Len  & \textbf{\small CovP}& Len & \textbf{\small CovP} & Len & \textbf{\small CovP} & Len & \textbf{\small CovP} & Len  \\
 \midrule
 3 & 5 & 0.94 & 0.32 & 0.95 & 0.19 & 0.94 & 0.19 & 0.95  & 0.19 & 0.95 & 0.20 & 0.95 & 0.20 & 0.95 & 0.19  \\

 & 10 & 0.94 & 0.32 & 0.95 & 0.19 & 0.94 & 0.19 & 0.96 & 0.19 & 0.96 & 0.20 & 0.95 & 0.19 & 0.96 & 0.19 \\
 \hline

 10 & 5 & 0.95 & 0.53 & 0.95 & 0.32 & 0.94 & 0.32 & 0.95 & 0.32 & 0.94 & 0.32 & 0.95 & 0.32 & 0.95 & 0.33 \\

 & 10 & 0.95 & 0.53 & 0.96 & 0.32 & 0.96 & 0.32 & 0.95 & 0.33 & 0.95 & 0.32 & 0.95 & 0.32 & 0.96 & 0.33 \\
 \bottomrule
 \end{tabular}
\end{table}

\begin{table}[!ht]
\centering
\caption{Inference results for $\theta_0$ for the settings in Section \ref{sim_misspecified}, with $n = 500$, $N = 10000$, $p = 50$, and $s = 7$ or $50$. The rest of the caption details remain the same as in Table \ref{table_n500_p50_cov}.}\label{table_n500_p50_cov_miss}
\tabcolsep=0.13cm
\begin{tabular}{lccccccccccccccc}
\toprule
    \multicolumn{2}{c}{} & \multicolumn{2}{c}{$\CI_{\sup}$} & \multicolumn{2}{c}{$\CI_{\BDM,l}$} & \multicolumn{2}{c}{$\CI_{\BDM,r}$} & \multicolumn{2}{c}{$\CI_{\BDM,s}$} & \multicolumn{2}{c}{$\CI_{\hBDM,l}$} & \multicolumn{2}{c}{$\CI_{\hBDM,r}$} & \multicolumn{2}{c}{$\CI_{\hBDM,s}$}\\
     \midrule
   $s$ & $K$ & \textbf{\small CovP} & Len & \textbf{\small CovP} & Len & \textbf{\small CovP} & Len  & \textbf{\small CovP}& Len & \textbf{\small CovP} & Len & \textbf{\small CovP} & Len & \textbf{\small CovP} & Len  \\
 \midrule
 7 & 5 & 0.95 & 0.48 & 0.95 & 0.34 & 0.96 & 0.34  & 0.94  & 0.34 & 0.98 & 0.36 & 0.98 & 0.35 & 0.97 & 0.36 \\

 & 10 & 0.95 & 0.48 & 0.94 & 0.34 & 0.96 & 0.34 & 0.95 & 0.34 & 0.97 & 0.36 & 0.98 & 0.35 & 0.97 & 0.36 \\
 \hline
 50 & 5 & 0.94 & 1.18 & 0.94 & 0.75 & 0.95 & 0.75 & 0.94 & 0.88 & 0.94 & 0.77 & 0.94 & 0.75 & 0.96 & 0.92 \\

 & 10 & 0.94 & 1.18 & 0.95 & 0.75 & 0.94 & 0.75 & 0.94 & 0.86 & 0.95 & 0.76 & 0.94 & 0.75 & 0.96 & 0.90 \\
 \bottomrule
 \end{tabular}
\end{table}

A notable aspect of the RE values in Table \ref{table_n500_mse_p1050} is that the extent of the efficiency improvement is fairly {\it uniform} across the settings, and quite close to the {\it achievable}
$\ORE^*$ in most cases -- with a slight lowering, in general, for the higher $p = 50$ case, as expected. This indicates no substantial additional finite sample losses in estimating $m^*(\cdot)$ under the low/moderate dimensional settings here. On the other hand, the difference between the achievable $\ORE^*$ and the optimal $\ORE$ indicate the (unrecoverable) difference due to the $O(1)$ bias stemming from $\Pi_{\mbm}$ targeting $m^*(\cdot)$ and not the true $m_0(\cdot)$. One notable exception to the general uniform pattern in the REs is the case of \texttt{Bsparse} for $p = s = 50$, where the REs, while still high, are closer to 2. This arises since the dense setting introduces an \textit{additional} layer of (soft) misspecification, making consistent estimation of even the $m^*(\cdot)$ more challenging for a sparsity-friendly method at such a choice of $(p,s,n)$. Conversely, \texttt{Bols} and \texttt{Bridge}, which do not depend on sparsity, continue to provide higher REs around 2.5.

Finally, consistent with our findings in Section~\ref{sim_correctly_specified}, BDMI and h-BDMI {\it still} perform similarly across all settings, with h-BDMI having slightly higher REs, while also exhibiting some conservativeness in CovPs, at least in some cases.
Furthermore, similar to
Section~\ref{sim_correctly_specified}, the results (both for estimation and inference) remain fairly robust across $K = 5$ and $K=10$. Thus, we continue to recommend either choice in practice.

Overall, as shown in Sections \ref{sim_correctly_specified}--\ref{sim_misspecified}, BDMI {\it always} achieves significant efficiency improvements and valid inference, under both correctly specified and misspecified models, thus validating our theoretical results.

\subsection[Real data analysis: Application to NHEFS data]{Real data analysis: Application to NHEFS data}\label{sec_real_data_analysis}

In this section, we apply the proposed BDMI approach to a subset of data from the National Health and Nutrition Examination Survey Epidemiologic Follow-up Study (NHEFS), a longitudinal study jointly initiated by the National Center for Health Statistics and the National Institute on Aging in collaboration with other agencies of the United States Public Health Service \citep{Hernan2020}. The NHEFS was designed to investigate the effects of clinical, nutritional, demographic, and behavioral factors on various health outcomes, including morbidity and mortality. Data were collected during a baseline visit in 1971 and a follow-up visit in 1982. For our analysis, we focus on a cohort of 1425 individuals from this study. A detailed description of the dataset is available at \hyperlink{https://hsph.harvard.edu/miguel-hernan/causal-inference-book}{https://hsph.harvard.edu/miguel-hernan/causal-inference-book}. This dataset has been widely used in other studies for different purposes. For instance, \citet{Ertefaie2022} used this dataset to estimate causal parameters like the average treatment effect of quitting smoking on weight gain, and \citet{chakrabortty2022semi} focused on quantile estimation under an SS framework.

Our primary goal is to estimate the {\it mean body weight}, $\theta_0$, of the entire cohort in 1982 under a semi-supervised framework. Additionally, we aim to investigate whether there is a significant change in body weight within the cohort between 1971 and 1982. To achieve this, we compared the analysis results to the baseline measurement from 1971, which had a mean of 70.99 and a standard error of 0.41 for the 1425
individuals. To benchmark our results, we also consider a {\it gold standard} scenario where all 1425 observations (for the response) are available in 1982 (which is the case for this data). We take the mean weight $\wh \theta_{\GS} = 73.6$ of all 1425 individuals in the 1982 cohort as the gold standard ($\GS$) estimator (i.e., a close `proxy' of the truth).

To evaluate the performance of the proposed BDMI approach, we randomly select $n = 200$ observations as the labeled dataset $\calL$, where body weight (response variable) is observed. For the unlabeled data $\calU$, we randomly designate $N \in \{ 400, 800, 1220\}$ observations from the remaining data. This setup allows us to explore and compare the performance of BDMI under varying ratios of labeled and unlabeled data ($n/N$), particularly as this ratio approaches 0. In addition to body weight as the response variable, we considered a set of 20 important covariates in our analysis, including demographic, clinical, and behavioral factors (see Table \ref{supp:tab:nhefs_data} in the \hyperref[sec:supplementary]{Supplementary Material} for their names and descriptions). These variables were also considered in other studies on this dataset,
e.g., \citet{chakrabortty2022semi} used them for SS quantile estimation.

The gold standard estimator $\wh \theta_{\GS}$ provides a benchmark for evaluating and comparing the performance of BDMI versus the supervised approach (based on the labeled data only). Given the labeled and unlabeled data, we calculated the supervised posteriors $\Pi_{\sup}$ (see Section \ref{sec_sup}) and $\Pi_{\btheta} \equiv \Pi_{\BDM}$ based on the BDMI-CF approach (Algorithm \ref{algo}) with $K =5$. From each posterior distribution, 1000 samples were obtained to compute the point estimators $\wh \theta_{\sup}$ and $\wh \theta_{\BDM}$, along with the respective $95\%$ credible intervals (CIs). Additionally, we calculated the {\it ratio} of the lengths ({\bf RL}) of the $95\%$ CIs from the supervised approach to those from BDMI. This RL serves as a natural measure of the relative efficiency of BDMI, where an RL greater than 1 indicates that BDMI provides tighter (and hence more efficient) CIs. Similar to our simulation study, 3 different methods are used to update the posterior $\Pi_{\mbm}$ of the nuisance parameter $m$, resulting in 3 distinct posteriors $\Pi_{\btheta} \equiv \Pi_{\BDM}$ for the BDMI approach. Table \ref{nhefss_data} summarizes our findings from the data analysis.

\begin{table}[!ht]
    \centering
    \caption{Results for the data analysis in Section \ref{sec_real_data_analysis}. Estimation and inference for the mean weight of the cohort in 1982 based on the supervised ($\Pi_{\sup}$) and BDMI ($\Pi_{\BDM}$) approaches. Description of notations: $\mathbf{n}$, the labeled data size; $\mathbf{N}$, the unlabeled data size; {\bf $\mathbf{95\%}$ CI}, the $95\%$ credible interval (CI); {\bf RL}, the ratios of the lengths of the $95\%$ CIs based on supervised approach versus BDMI; $\wh \theta_{\GS}$, the gold standard estimator (based on the entire cohort); $\wh \theta_{\sup}$, the supervised estimator; $\wh \theta_{\BDM, i}$, the BDMI estimator where the subscript ``i" denotes the method used to obtain the posterior of $m$: $l = $ {\tt Bols}, $r = $ {\tt Bridge}, $s = $ {\tt Bsparse}.}
    \tabcolsep=0.023cm
    \begin{tabular}{llcccccccccccr}
    \toprule
  \multicolumn{2}{c}{} & \multicolumn{1}{c}{~} &  \multicolumn{2}{c}{$\Pi_{\sup}$}  &  \multicolumn{3}{c}{ $\Pi_{\BDM, l}$} &  \multicolumn{3}{c}{ $\Pi_{\BDM, r}$} & \multicolumn{3}{c}{$\Pi_{\BDM, s}$} \\
  \midrule
     $n$  & $N$ & $\wh \theta_{\GS}$ ~ & $\wh \theta_{\sup}$  & $95\%$ CI & $\wh \theta_{\BDM, l}$ & $95\%$ CI & { \bf RL}~ & $\wh \theta_{\BDM, r}$ & $95\%$ CI & {\bf RL}~ & $\wh \theta_{\BDM, s}$ & $95\%$ CI & {\bf RL} \\
       \midrule
       & 400 & 73.6~ & 72.9 &  [70.6, 75.0] & 73.6 &  [71.8, 75.4] & 1.20~ & 73.7 & [72.0, 75.4] & 1.28~  & 73.8 & [72.1, 75.6] & 1.24 \\
      200 & 800 & 73.6~ & 72.9 & [70.6, 75.0] & 73.7 &  [72.2, 75.2] & 1.45~ & 73.6 & [72.1, 75.0] & 1.53~ & 73.7 & [72.3, 75.1] & 1.56  \\
       & 1220 & 73.6~ & 72.9 & [70.6, 75.0] & 73.2 & [71.9, 74.5] & 1.64\; & 73.2 & [71.9, 74.5] & 1.74\; & 73.3 &  [72.1, 74.7] & 1.74\\
        \bottomrule
    \end{tabular}
    \label{nhefss_data}
\end{table}

Table \ref{nhefss_data} highlights that BDMI demonstrates two key advantages over the supervised approach: {\it improved} accuracy and efficiency. First, the SS point estimates based on BDMI (across all versions) are consistently closer to the gold standard estimate, $\wh \theta_{\GS} = 73.6$, compared to the supervised estimate $\wh \theta_{\sup} = 72.6$ for all settings of $N$. Second, BDMI (across all versions) consistently produces significantly tighter $95\%$ CIs than the supervised approach, with efficiency gains quantified by the ratio of CI lengths (RL), ranging from $1.2$ to $1.7$ across all settings. This corresponds to $20-70\%$ tighter intervals for BDMI. Notably, for a fixed number of labeled data, as the ratio $n/N$ decreases (i.e., increasing the size of the unlabeled data), BDMI achieves substantial efficiency improvements by further reducing CI lengths compared to the supervised approach. For example, with $n=200$, increasing $N$ from 400 to 1220 improves the RL from around 1.24 to 1.74, reflecting a $40\%$ reduction in CI length for BDMI. These results indicate that the posterior spread under BDMI becomes increasingly tighter as more unlabeled data are incorporated. Hence, these findings highlight BDMI's ability to {\it efficiently} leverage unlabeled data, providing strong empirical support for our theoretical framework regarding the importance of $\lim_{n, N \to \infty} n/N = c \in [0,1)$. These results show that the BDMI procedure delivers both accurate point estimates (near identical to the GS version) and enhanced efficiency through shorter/tighter credible intervals, underscoring its advantage over the supervised approach. Finally, a notable feature of the BDMI based CIs for the mean weight of the 1982 cohort is that they consistently {\it exclude} the 1971 mean weight (70.99), indicating a significant {weight gain}, likely due to aging or quitting smoking \citep{Ertefaie2022}. In contrast, the supervised approach {\it fails to detect} this change, as its $95\%$ CI $(70.6, 75.0)$ includes the 1971 mean.
These results highlight the improved efficiency and higher {\it power} of BDMI for detecting significant (and scientifically meaningful) differences in weights between the two cohorts.

\section{Concluding discussions} \label{conclusion_discussion}

We proposed the BDMI procedure for estimating the population mean $\theta_0 = \bbE(Y)$ under the SS setting. To the best of our knowledge, this is the first attempt to establish a Bayesian method that achieves desirable SS inference goals, including {\it efficiency improvement} and {\it global robustness}, while providing {\it rigorous} theoretical guarantees. Our methodology ensures that the posterior $\Pi_{\btheta}$ of the parameter of interest $\theta$ contracts around the true parameter $\theta_0$ at the parametric rate $n^{-1/2}$ and is asymptotically Normal, {\it regardless} of the choice of method used to obtain a posterior for the nuisance parameter $m$, its contraction rate, or even potential misspecification of $m$. Moreover, the posterior mean of $\Pi_{\btheta}$, as an SS estimator of $\theta_0$, {\it always} possesses $\sqrt{n}$-consistency, asymptotic normality, and first-order {\it insensitivity}, in addition to being at least as {\it efficient} as the supervised estimator (sample mean of $Y$). These theoretical results have been rigorously established in Section \ref{sec_theory}. One of the key contributions of BDMI lies in its ability to disentangle nuisance parameter estimation from inference on the target parameter by developing a novel debiasing approach under the Bayesian paradigm. It enables joint learning of the nuisance bias and the main parameter through targeted modeling of summary statistics, along with careful usage of sample splitting. We hope this research brings attention to the rarely used idea of modeling summary statistics within Bayesian inference and demonstrates its potential to address other Bayesian semi-parametric inference problems. While this work focuses on SS mean estimation, the underlying principles of BDMI can be extended to a broad range of problems, including missing data analysis, causal inference, and SS inference for other functionals. For instance, BDMI could be adapted to handle selection bias or distribution shifts between labeled and unlabeled data; this was recently explored in the frequentist SS literature \citep{zhang2021double} but not yet addressed within a Bayesian framework. Further, extending BDMI to Bayesian SS inference for high dimensional target parameters (e.g., regression coefficients) poses additional theoretical and computational challenges, but also represents an important direction for future research. Finally, adapting BDMI’s debiasing framework to causal inference or missing data settings offers exciting opportunities for advancing Bayesian semi-parametric methodologies. We hope this work generates interest in considering such Bayesian problems in the future.

\section*{Supplementary Material}\label{sec:supplementary}
\addcontentsline{toc}{section}{Supplementary Material}

\noindent {\bf Supplement to `Bayesian Semi-supervised Inference via a Debiased Modeling Approach'}. The supplement (Sections \ref{supp_sec_simulation}--\ref{supp:prelim_proof}) includes additional discussions, numerical results, and all technical materials (e.g., proofs) that could not be accommodated in the main paper: {\bf (i)} additional figures and a table for the simulations and data analysis in Sections \ref{sim_misspecified}--\ref{sec_real_data_analysis} (Section \ref{supp_sec_simulation}); {\bf (ii)} additional discussion on the imputation approach introduced in Section \ref{motivation}, along with a detailed numerical study for comparison with BDMI (Section \ref{supp:imputation_approach}); {\bf (iii)} implementation details of the \texttt{Bridge} and \texttt{Bsparse} methods
used to obtain the nuisance posterior $\Pi_{\mbm}$ in our numerical studies (Section \ref{supp_implementation_details}); {\bf (iv)} proofs of all the main theoretical results (Section \ref{appendix}); and {\bf (v)} proofs of preliminary results and intermediate lemmas utilized in the proofs of the main results (Section \ref{supp:prelim_proof}).

\section*{Acknowledgements}
\addcontentsline{toc}{section}{Acknowledgements}
\noindent The authors would like to thank the Editor, the Associate Editor, and the three Reviewers for their constructive comments and suggestions that significantly helped improve the presentation and the content of the article.

\noindent This research was partially supported by the National Science Foundation grants: NSF-DMS 2113768 (to Abhishek Chakrabortty), and NSF-DMS 2210689 and NSF-DMS 1916371 (to Anirban Bhattacharya).

\phantomsection
\addcontentsline{toc}{section}{References}
\bibliography{reference}

\clearpage

\section*{\centering \Large Supplement to ``Bayesian Semi-supervised Inference via a Debiased Modeling Approach"}\label{supplementary_material}

\renewcommand{\thefootnote}{\arabic{footnote}}

{\centering
\large Gözde Sert, Abhishek Chakrabortty, and Anirban Bhattacharya \par }

{\centering \large \it Department of Statistics, Texas A\&M University \footnote{{\it Email addresses}: {\tt gozdesert@stat.tamu.edu} (Gözde Sert), {\tt abhishek@stat.tamu.edu} (Abhishek Chakrabortty),\\
{\tt anirbanb@stat.tamu.edu} (Anirban Bhattacharya)} \par }

\renewcommand*{\theHsection}{\thesection}
\renewcommand*{\theHsubsection}{\thesubsection}
\setcounter{section}{0}
\setcounter{equation}{0}
\setcounter{subsection}{0}

\renewcommand \thesection{S\arabic{section}}
\renewcommand \thesubsection{\thesection.\arabic{subsection}}
\renewcommand \thesubsubsection{\thesubsection.\arabic{subsubsection}}
\renewcommand\thetable{S.\arabic{table}}
\renewcommand \thefigure{S.\arabic{figure}}
\renewcommand{\theequation}{S.\arabic{equation}}

\setcounter{table}{0}
\setcounter{figure}{0}

\par\medskip
This supplementary document (Sections~\ref{supp_sec_simulation}--\ref{supp:prelim_proof}) includes additional discussions and numerical analyses, as well as technical details such as proofs and extended discussions that could not be accommodated in the main paper. Section~\ref{supp_sec_simulation} includes additional figures for the simulation results in Section~\ref{sim_misspecified} and a supplementary table for the data analysis in Section~\ref{sec_real_data_analysis}. In Section~\ref{supp:imputation_approach}, we provide a detailed construction of the imputation approach, initially introduced in Section~\ref{motivation}, and then present numerical studies to highlight its limitations, along with a comparative analysis with the BDMI approach. Section~\ref{supp_implementation_details} outlines the implementation details of the methods used to obtain the nuisance posteriors in the numerical studies of Section~\ref{simulations}. Section~\ref{appendix} presents the proofs of all the results in the main paper. Finally, Section~\ref{supp:prelim_proof} provides the proofs for all supporting lemmas or intermediate lemmas introduced in the course of the main proofs in Section~\ref{appendix}.

\section[Additional figures and tables for numerical studies]{Additional figures and tables for numerical studies}\label{supp_sec_simulation}

\begin{figure}[ht!]
    \centering
    \begin{subfigure}{0.45\textwidth}
        \centering
        \includegraphics[width=\linewidth]{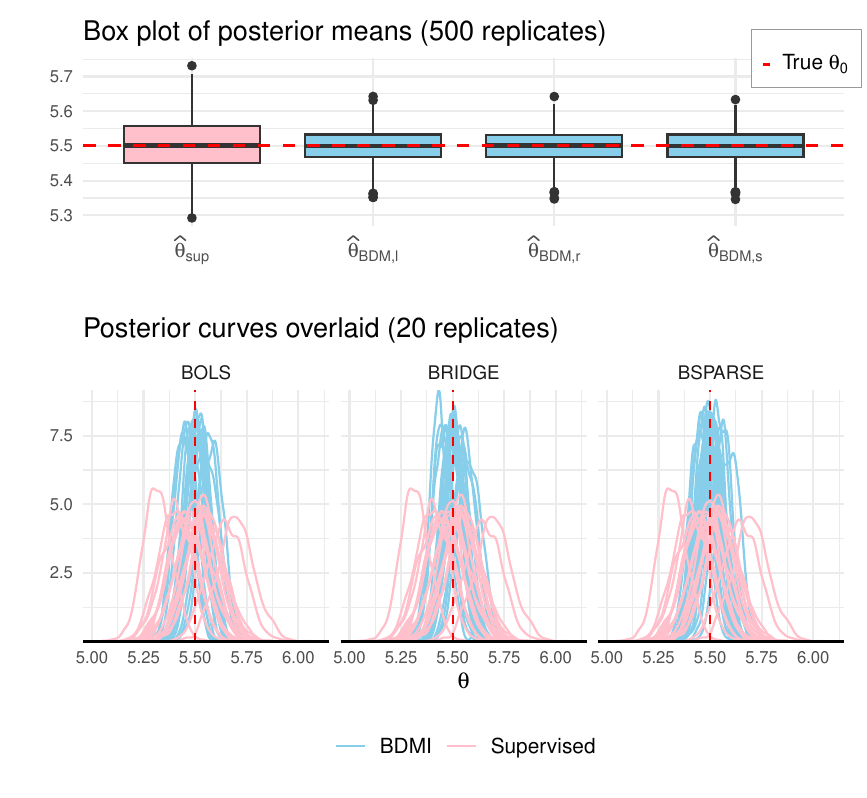}
        \caption{Setting for misspecified model: $p = 10$ with $\mathbf{s = 3}$}
        \label{fig:subfig1:p10s3}
    \end{subfigure}
    \hfill
    \begin{subfigure}{0.45\textwidth}
        \centering
        \includegraphics[width=\linewidth]{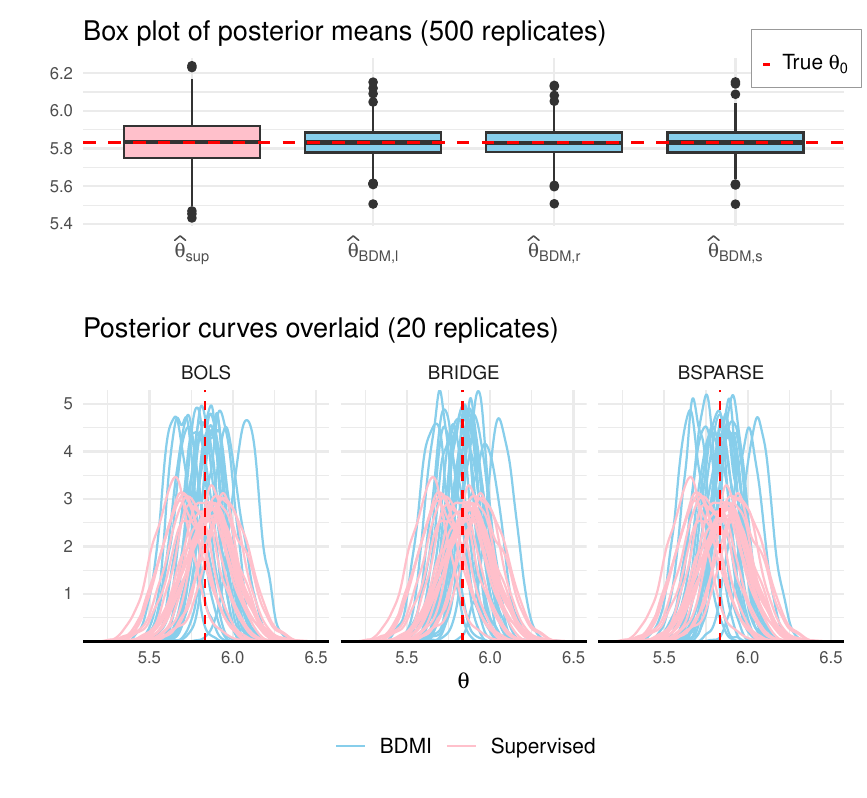}
        \caption{Setting for misspecified model: $p = 10$ with $\mathbf{s = 10}$}
        \label{fig:subfig2:p10s10}
    \end{subfigure}
    \caption{Box plots of posterior means (based on 500 replications) and plots of overlaid density curves (based on 20 iterations) for the posteriors $\Pi_{\sup}$ (pink) and $\Pi_{\btheta}$ (blue) of $\theta$, with three different methods ({\tt Bols}, {\tt Bridge} and {\tt Bsparse}) to obtain the nuisance posterior $\Pi_{\mbm}$ for BDMI. {\bf Setting} (from Section~\ref{sim_misspecified}): $n = 500$, $N = 10000$, $p = 10$, and $s = 3$ or $s = 10$. (Each density curve is generated using 1000 posterior samples of $\theta$. The red dashed vertical line indicates the true parameter of interest $\theta_0$.)}
    \label{fig:miss_p10s3-10}
\end{figure}

\begin{figure}[ht!]
    \centering
    \begin{subfigure}{0.45\textwidth}
        \centering
        \includegraphics[width=\linewidth]{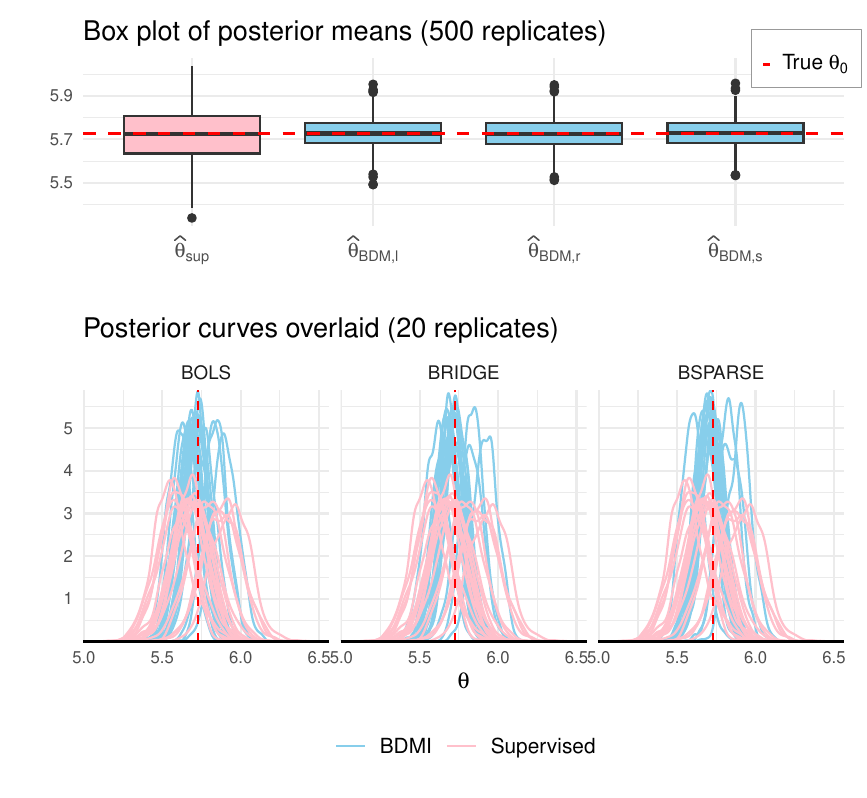}
        \caption{Setting for misspecified model: $p = 50$ with $\mathbf{s = 7}$.}
        \label{fig:subfig1:miss_p50s7}
    \end{subfigure}
    \hfill
    \begin{subfigure}{0.45\textwidth}
        \centering
        \includegraphics[width=\linewidth]{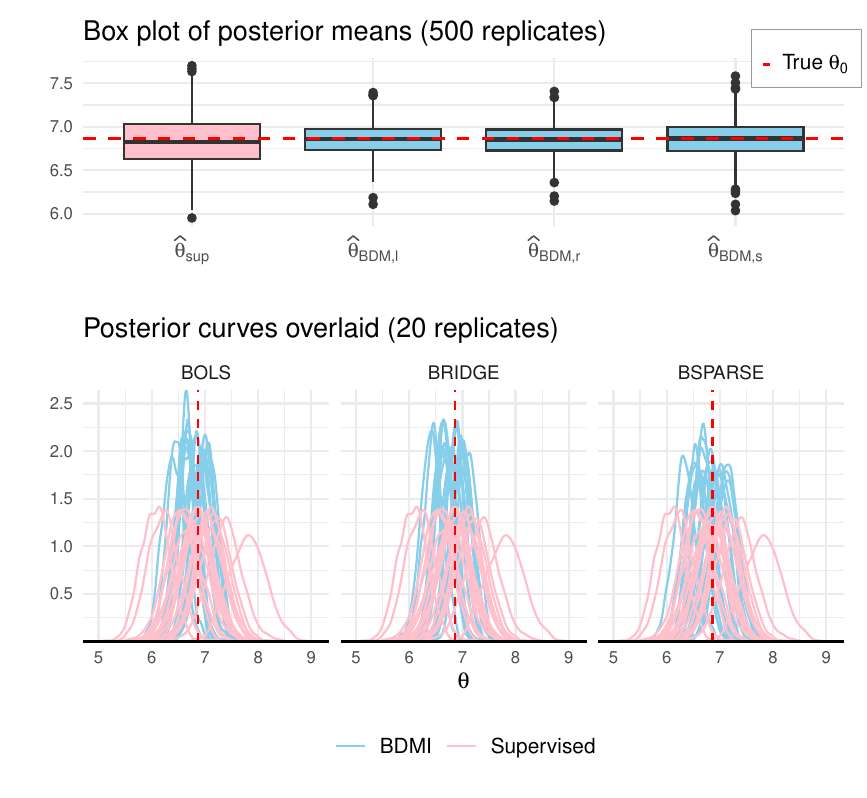}
        \caption{Setting for misspecified model: $p = 50$ with $\mathbf{s = 50}$.}
        \label{fig:subfig2:miss_p50s50}
    \end{subfigure}
    \caption{Box plots of posterior means and plots of overlaid density curves for the posteriors $\Pi_{\sup}$ (pink) and $\Pi_{\btheta}$ (blue) of $\theta$. {\bf Setting (Section~\ref{sim_misspecified}):} $n = 500$, $N = 10000$, $p = 50$, and $s = 7$ or $s = 50$. The rest of the caption details
    are the same as Figure
   ~\ref{fig:miss_p10s3-10}.}
    \label{fig:miss_p500s7-50}
\end{figure}

Figures~\ref{fig:miss_p10s3-10}--\ref{fig:miss_p500s7-50} present additional plots for the simulation results in Section~\ref{sim_misspecified} for misspecified models.

\noindent Table~\ref{supp:tab:nhefs_data} lists the names and descriptions of the covariates used for the NHEFS data
analysis in Section~\ref{sec_real_data_analysis}.
\begin{table}[!ht]
\def~{\hphantom{0}}
\caption{Covariates included for the NHEFS data analysis in Section~\ref{sec_real_data_analysis}.}
\centering
	\begin{tabular}{ll}
\toprule
		\textbf{Variable name} & \textbf{Description} \\
		\hline
		{\tt active} & On your usual day, how active were you in 1971? \\
		{\tt age} & Age in 1971 \\
		{\tt alcoholfreq} & How often did you drink in 1971?    \\
		{\tt allergies} & Use allergies medication in 1971 \\
		{\tt asthma} &  DX asthma in 1971 \\
		{\tt cholesterol} & Serum cholesterol (mg/100ml) in 1971 \\
		{\tt dbp} & Diastolic blood pressure in 1982 \\
		{\tt education} & Amount of education by 1971 \\
		{\tt exercise} & In recreation, how much exercise in 1971? \\
		{\tt ht} & Height in centimeters in 1971 \\
		{\tt price71} & Average tobacco price in the state of residence 1971 (US\$2008) \\
		{\tt price82} & Average tobacco price in the state of residence 1982 (US\$2008) \\
		{\tt race} & White, black or other in 1971 \\
		{\tt sbp} & Systolic blood pressure in 1982 \\
		{\tt sex} & Male or female  \\
		{\tt smokeintensity} & Number of cigarettes smoked per day in 1971 \\
		{\tt smokeyrs}  & Years of smoking            \\
		{\tt tax71} & Tobacco tax in the state of residence 1971 (US\$2008)              \\
		{\tt tax82} & Tobacco tax in the state of residence 1971 (US\$2008) \\
		{\tt wt71} & Weight in kilograms  in 1971 \\
        \bottomrule
\end{tabular}
\label{supp:tab:nhefs_data}
\end{table}

 \section[The imputation approach and its limitations: A comparative analysis with BDMI]{The imputation approach and its limitations: A comparative analysis with BDMI}\label{supp:imputation_approach}

This section provides an extensive numerical comparison of the imputation-type approach (henceforth IMP) introduced in Section~\ref{motivation} with BDMI. For IMP, recall that one selects a Bayesian regression method to construct the nuisance posterior $\Pi_{\mbm}$ for $m$ from the labeled data $\calL$. Using the imputation (regression) representation $\theta_0 = \bbE_{\boldX}\{m_0(\boldX)\}$, one can compute the induced posterior by approximating $\bbE_{\boldX}$ with an empirical average over $\calU$. Specifically, given $\tm \sim \Pi_{\mbm}$, we define a new random variable:
\vspace{-0.2cm}
\begin{align}\label{eq:imp_post_supp}
  \theta_{\imp} ~\equiv ~\theta_{\imp}(\tm) ~=~ \frac{1}{N}\sum_{i=n+1}^{n+N}\tm(\boldX_i), \ \text{ and let $\Pi_{\imp}$ be the (induced) posterior of $\theta_{\imp}$.}
\end{align}

The posterior mean $\wh \theta_{\imp}$ of $\Pi_{\imp}$, a point estimate of $\theta_0$ under IMP, by linearity of expectation, is given by:
\begin{align*}
\wh \theta_{\imp} \, \equiv \wh \theta_{\imp}(\wh m) := \frac{1}{N}\sum_{i=n+1}^{n+N}\wh m(\boldX_i), ~ \mbox{where $\wh m(\cdot) := \bbE_{m \sim \Pi_{\mbm}}\big\{m(\cdot) \hspace{0.05cm}|\hspace{0.05cm} \calL \big\}$ is the posterior mean of $\Pi_{\mbm}$.}
\end{align*}
It is straightforward to sample from $\Pi_{\imp}$; to generate $B$ samples of $\theta$ from $\Pi_{\imp}$, one first draws $B$ samples $\tm^{(1)}, \dots, \tm^{(B)}$ of $m$ from the nuisance posterior $\Pi_{m}$, and then uses the construction in \eqref{eq:imp_post_supp} to obtain the corresponding posterior samples $\theta^{(1)}, \dots, \theta^{(B)}$. The posterior mean $\wh \theta_{\imp}$ is approximated by $B^{-1} \sum_{b=1}^B \theta^{(b)}$.

To enable a fair comparison, we compare IMP with h-BDMI, as both methods share a hierarchical structure, and thus differences in performance can be attributed to {\it debiasing}, which is the key distinction between these approaches. Specifically, we use the CF version of h-BDMI with $K = 10$ throughout. As shown in Section~\ref{sim_correctly_specified}, the performance of our original BDMI (single-sample version) is nearly indistinguishable from h-BDMI, and we have observed the same trends we report below when comparing IMP with BDMI.
\begin{figure}[!ht]
    \centering
    \begin{subfigure}[b]{0.45\textwidth}
        \centering
        \includegraphics[width=\textwidth]{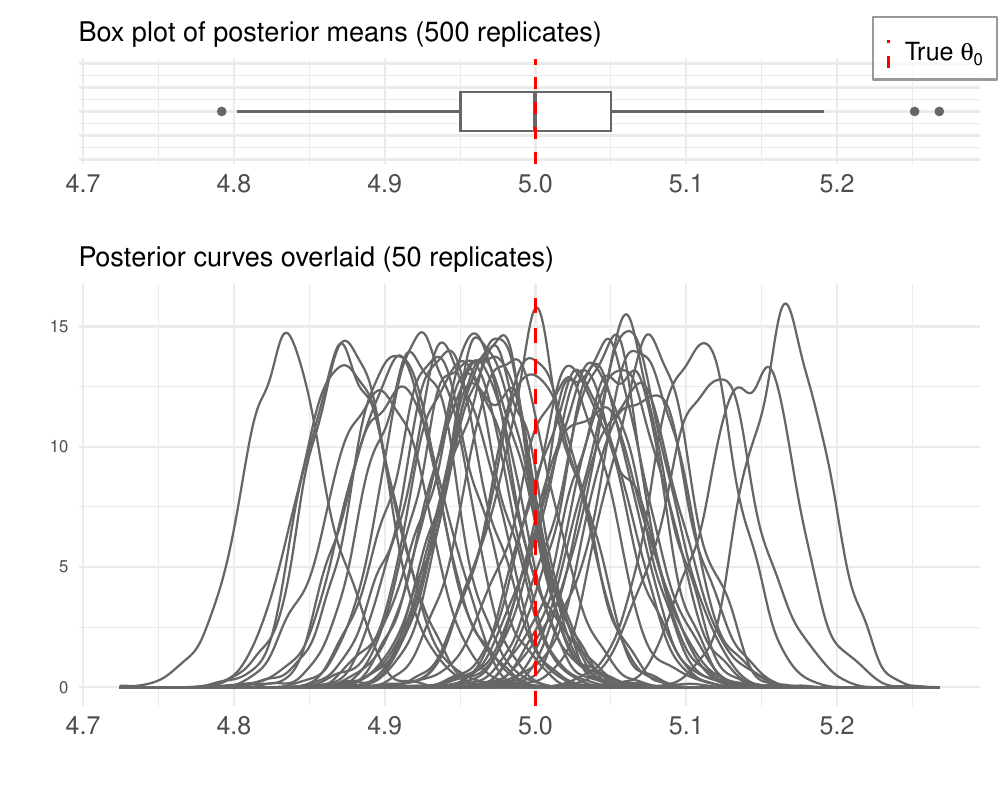}
        \caption{{\tt Bridge}}
    \end{subfigure}
    \hfill
    \begin{subfigure}[b]{0.45\textwidth}
        \centering
        \includegraphics[width=\textwidth]{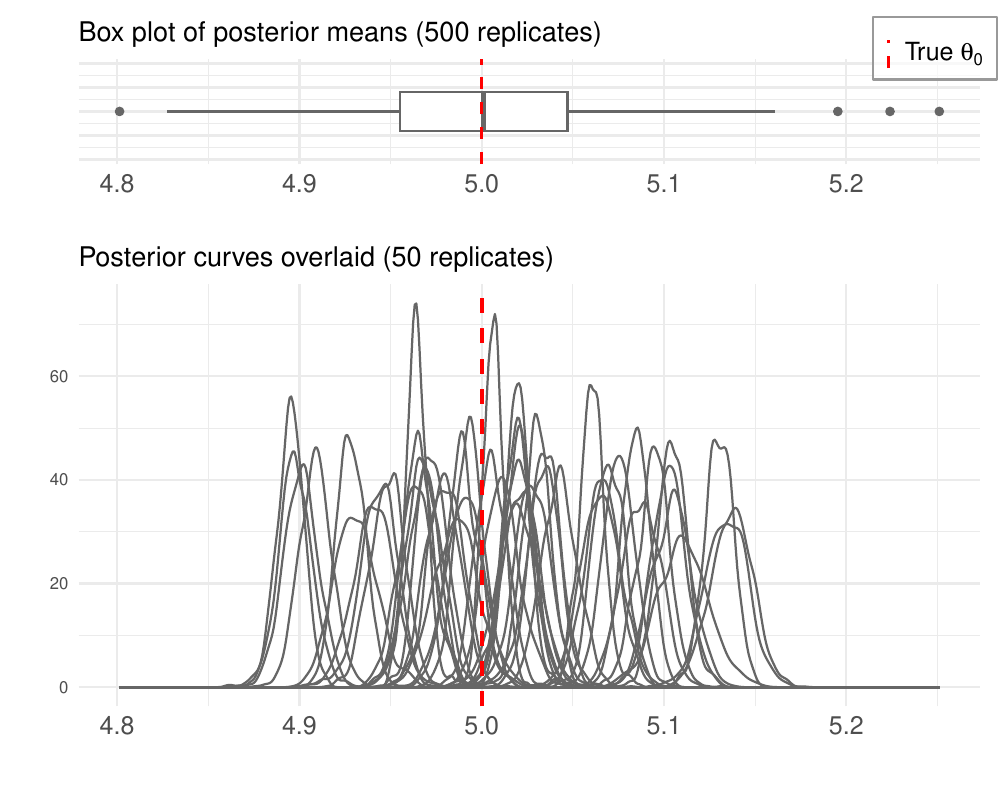}
        \caption{{\tt Bsparse}}
    \end{subfigure}
    \caption{Box plot of posterior means (based on 500 replications) and the overlaid density curves (based on 50 iterations) of the posterior $\Pi_{\imp}$ of $\theta$ for the {\bf imputation} approach (IMP) with two different methods (left: {\tt Bridge}; right: {\tt Bsparse}) to obtain the posterior $\Pi_{\mbm}$ of the nuisance parameter $m$. Each density curve is generated using 1000 posterior samples of $\theta \sim \Pi_{\imp}$.
    {\bf Setting:} $n = 500$, $N = 10000$, $p = 166$, and $\mathbf{s = 13}$. The coverage probabilities based on IMP are $\mathbf{56\%}$ for the {\tt Bridge} method and $\mathbf{23\%}$ for the {\tt Bsparse} method. (The red dashed vertical line indicates the true parameter of interest $\theta_0$ and equals 5 for all settings.)}
    \label{fig:imp_p166s13}
\end{figure}

\begin{figure}[!ht]
    \centering
    \begin{subfigure}[b]{0.45\textwidth}
        \centering
        \includegraphics[width=\textwidth]{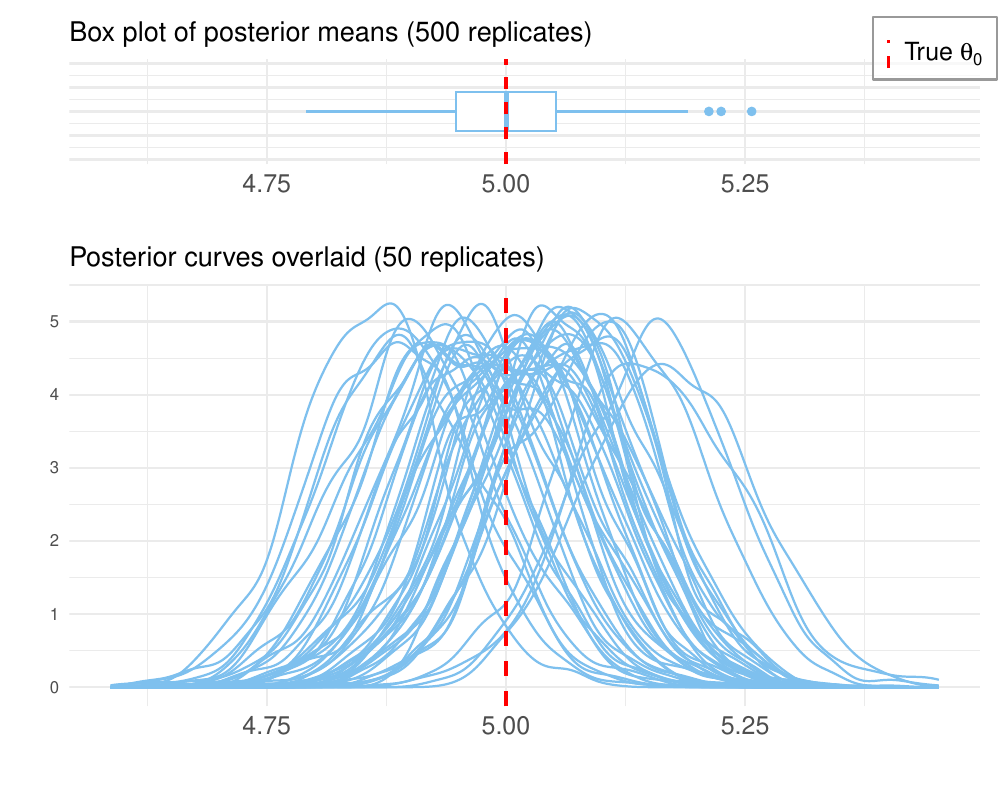}
        \caption{{\tt Bridge}}
    \end{subfigure}
    \hfill
    \begin{subfigure}[b]{0.45\textwidth}
        \centering
        \includegraphics[width=\textwidth]{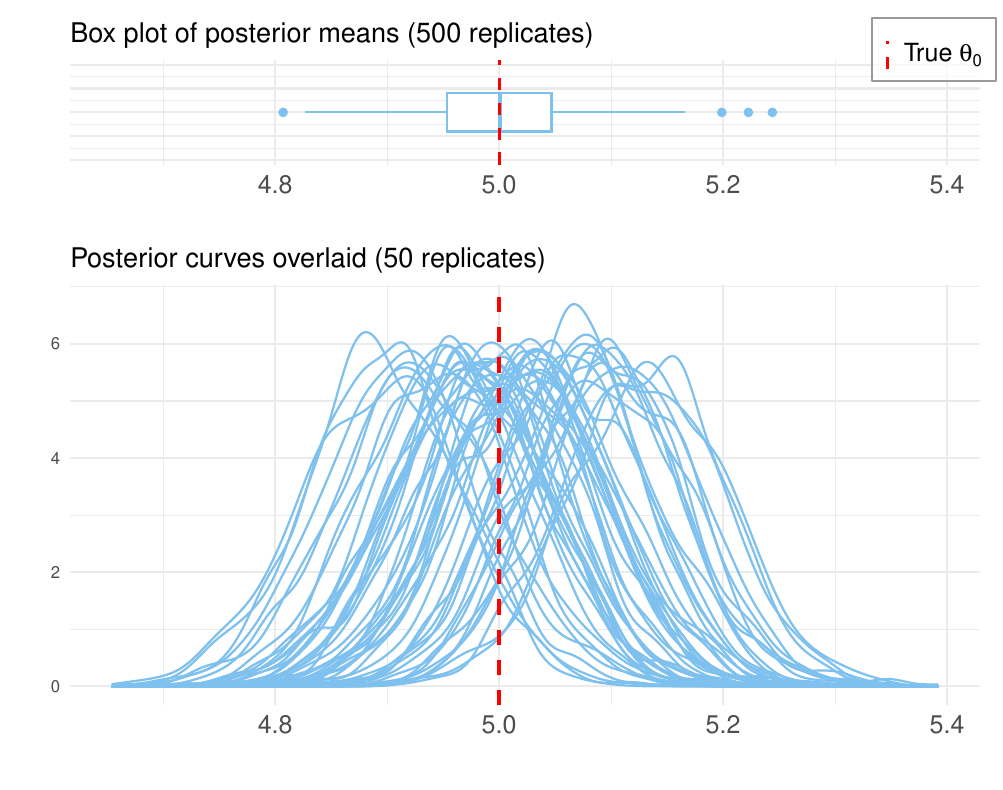}
        \caption{{\tt Bsparse}}
    \end{subfigure}
    \caption{Box plot of posterior means (based on 500 replications) and the overlaid density curves (based on 50 iterations) of the posterior $\Pi_{\btheta}$ of $\theta$ for the {\bf BDMI} approach with two different methods (left: {\tt Bridge}; right: {\tt Bsparse}) to obtain the nuisance posterior $\Pi_{\mbm}$ of $m$. Each density curve is generated using 1000 posterior samples of $\theta \sim \Pi_{\btheta}$. {\bf Setting:} $n = 500$, $N = 10000$, $p = 166$, and $\mathbf{s = 13}$ (and $K = 10$). The coverage probabilities based on BDMI are $96\%$ for {\tt Bridge} and $95\%$ for {\tt Bsparse}.}
    \label{fig:BDMI_p166s13}
\end{figure}

\begin{figure}[!ht]
    \centering
    \begin{subfigure}[b]{0.45\textwidth}
        \centering
        \includegraphics[width=\textwidth]{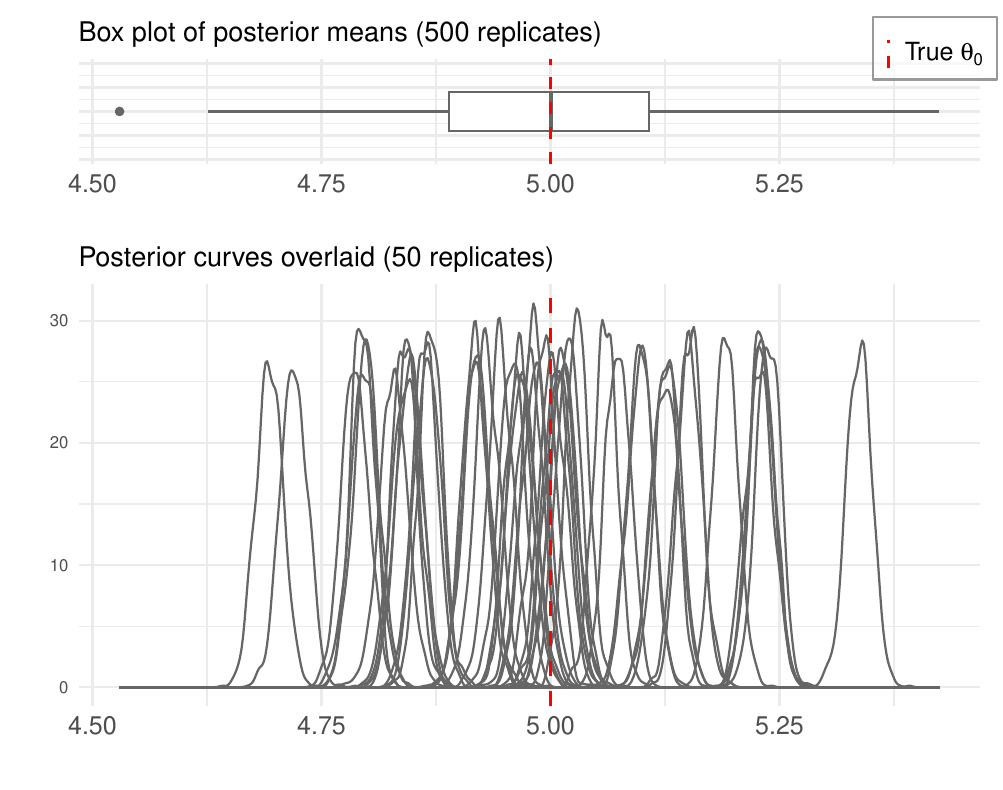}
        \caption{{\tt Bridge}}
    \end{subfigure}
    \hfill
    \begin{subfigure}[b]{0.45\textwidth}
        \centering
        \includegraphics[width=\textwidth]{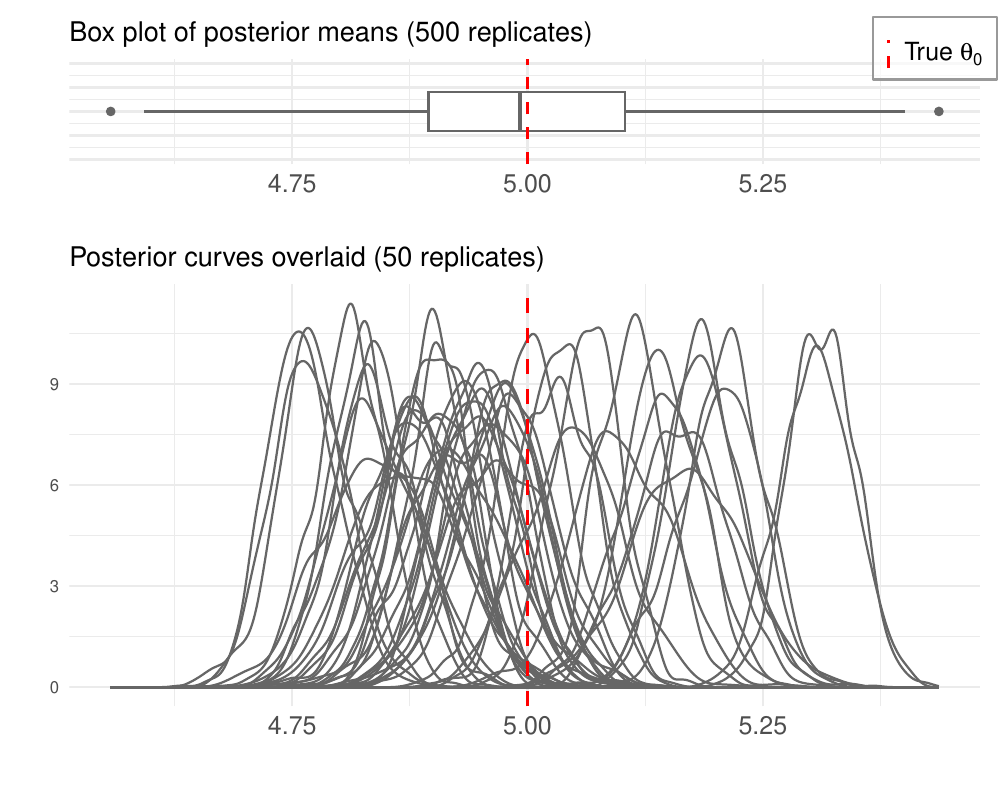}
        \caption{{\tt Bsparse}}
    \end{subfigure}
    \caption{Box plot of posterior means and the overlaid density curves of $\Pi_{\imp}$ for $\theta$ based on {\bf IMP} with {\tt Bridge} and {\tt Bsparse} methods for $\mathbf{s = 55}$. The corresponding coverages are $\mathbf{12\%}$ and $\mathbf{43\%}$. The rest of the caption remains the same as in Figure~\ref{fig:imp_p166s13}.}
    \label{fig:imp_p166s55}
\end{figure}

\begin{figure}[!ht]
    \centering
    \begin{subfigure}[b]{0.45\textwidth}
        \centering
        \includegraphics[width=\textwidth]{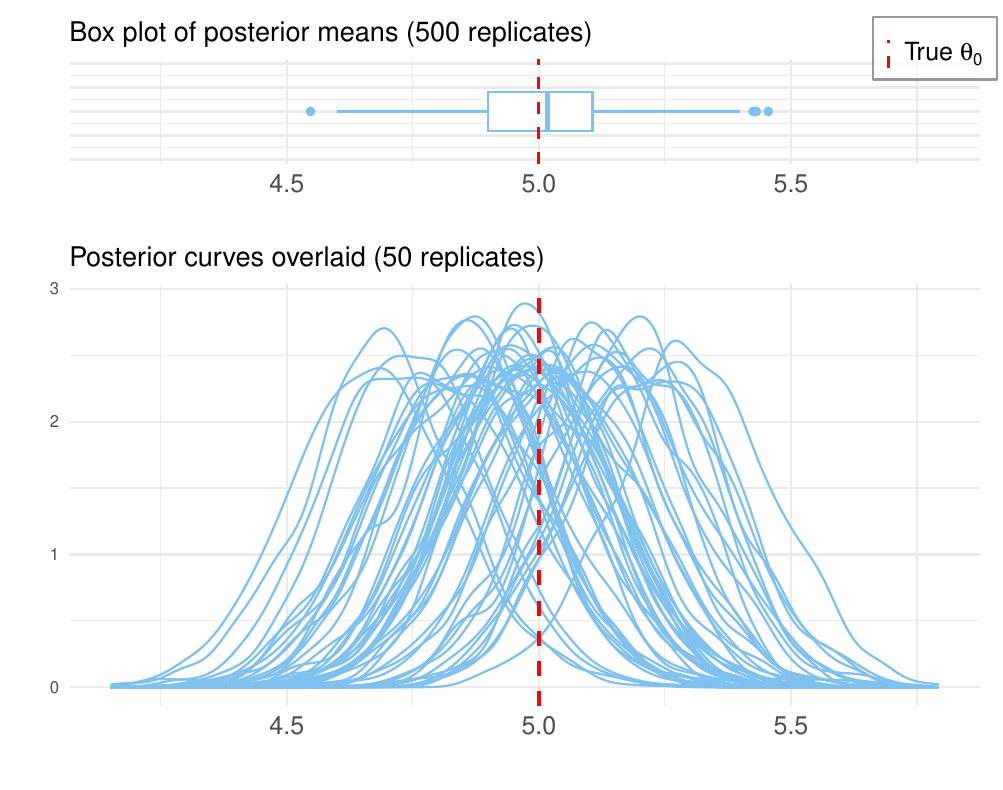}
        \caption{{\tt Bridge}}
    \end{subfigure}
    \hfill
    \begin{subfigure}[b]{0.45\textwidth}
        \centering
        \includegraphics[width=\textwidth]{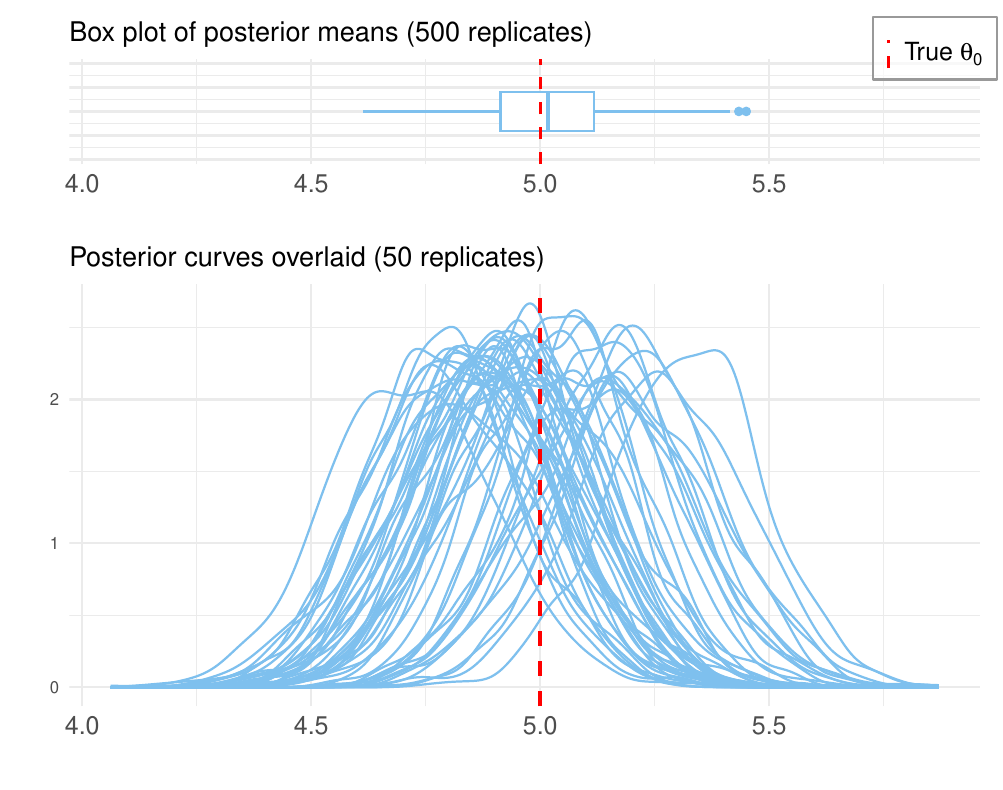}
        \caption{{\tt Bsparse}}
    \end{subfigure}
    \caption{Box plot of posterior means and overlaid density curves of $\Pi_{\btheta}$ for $\theta$ based on {\bf BDMI} with the {\tt Bridge} and {\tt Bsparse} methods for $\mathbf{s = 55}$. The corresponding coverages are $96\%$ and $95\%$. The rest of the caption remains the same as in Figure~\ref{fig:BDMI_p166s13}.}
    \label{fig:BDMI_p166s55}
\end{figure}

\begin{figure}[!ht]
    \centering
    \begin{subfigure}[b]{0.45\textwidth}
        \centering
        \includegraphics[width=\textwidth]{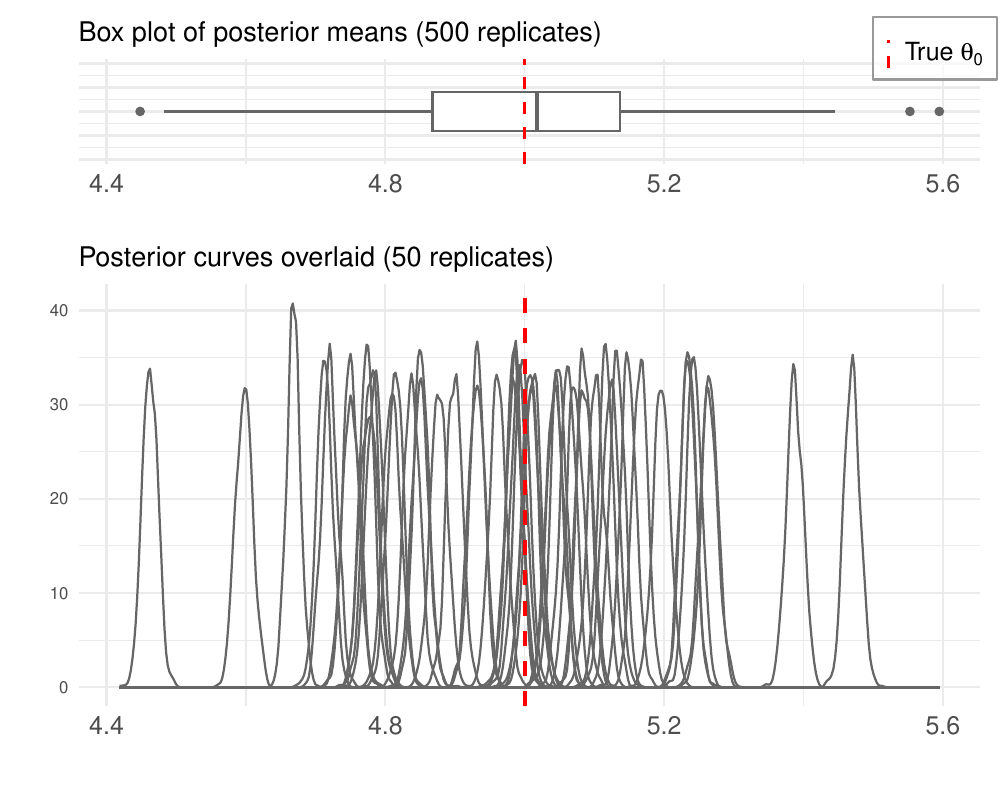}
        \caption{{\tt Bridge}}
    \end{subfigure}
    \hfill
    \begin{subfigure}[b]{0.45\textwidth}
        \centering
        \includegraphics[width=\textwidth]{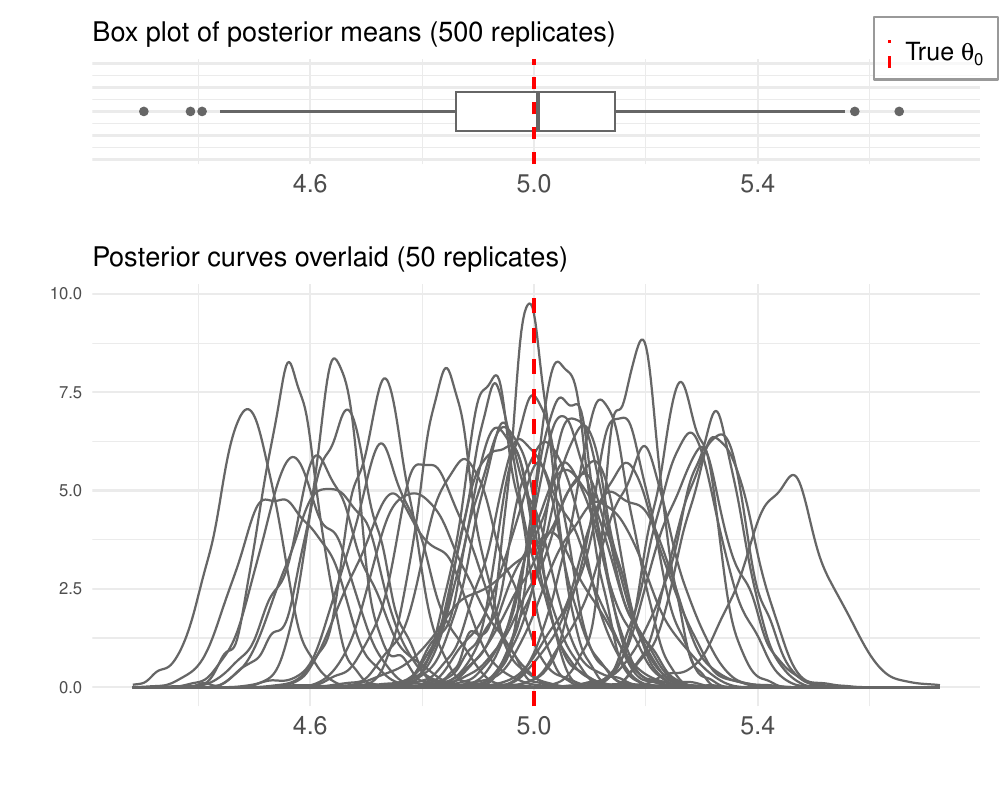}
        \caption{{\tt Bsparse}}
    \end{subfigure}
   \caption{Box plot of posterior means and the overlaid density curves of $\Pi_{\imp}$ for $\theta$ based on {\bf IMP} with {\tt Bridge} and {\tt Bsparse} methods for $\mathbf{s = 83}$. The corresponding coverages are $\mathbf{7\%}$ and $\mathbf{45\%}$. The rest of the caption remains the same as in Figure~\ref{fig:imp_p166s13}.}
    \label{fig:imp_p166s83}
\end{figure}

\begin{figure}[!ht]
    \centering
    \begin{subfigure}[b]{0.45\textwidth}
        \centering
        \includegraphics[width=\textwidth]{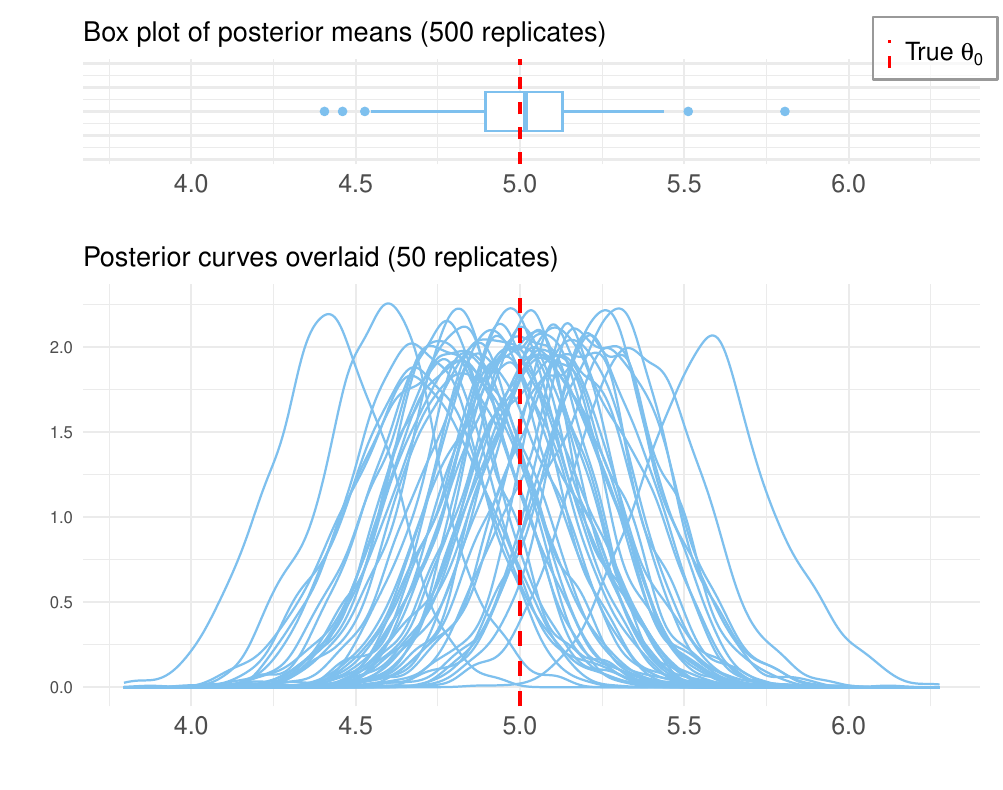}
        \caption{{\tt Bridge}}
    \end{subfigure}
    \hfill
    \begin{subfigure}[b]{0.45\textwidth}
        \centering
        \includegraphics[width=\textwidth]{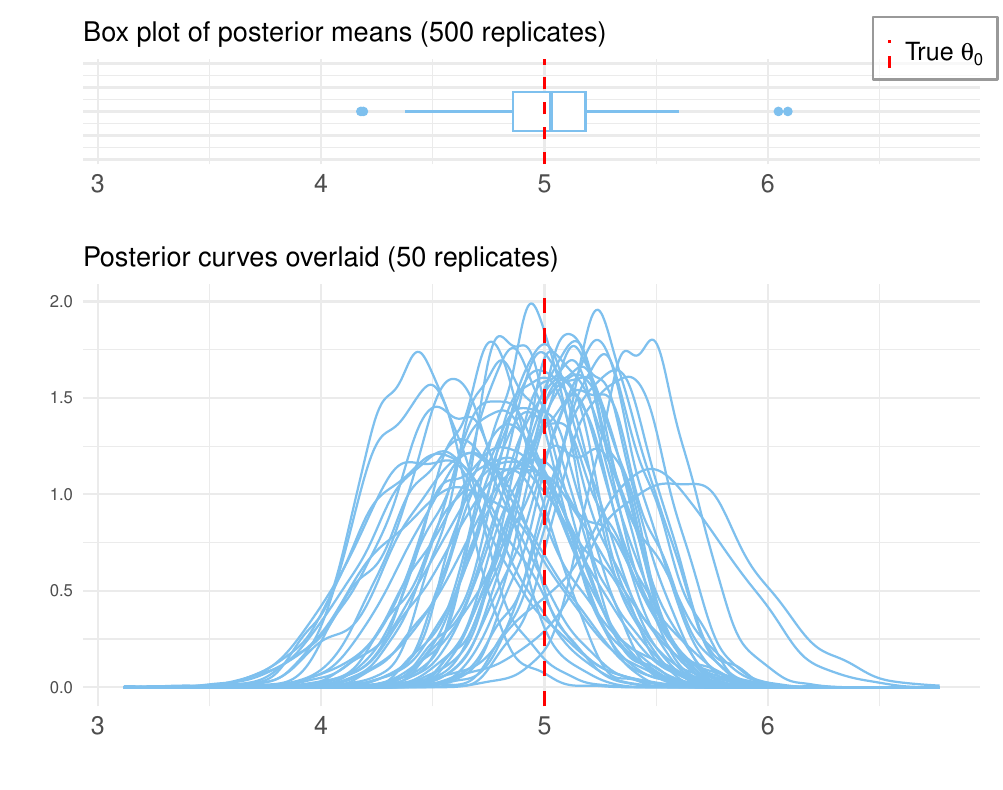}
        \caption{{\tt Bsparse}}
    \end{subfigure}
    \caption{Box plot of posterior means and overlaid density curves of $\Pi_{\btheta}$ for $\theta$ based on {\bf BDMI} with the {\tt Bridge} and {\tt Bsparse} methods for $\mathbf{s = 83}$. The corresponding coverages are $95\%$ and $95\%$. The rest of the caption remains the same as in Figure~\ref{fig:BDMI_p166s13}.}
    \label{fig:BDMI_p166s83}
\end{figure}

\begin{figure}[!ht]
    \centering
    \begin{subfigure}[b]{0.45\textwidth}
        \centering
        \includegraphics[width=\textwidth]{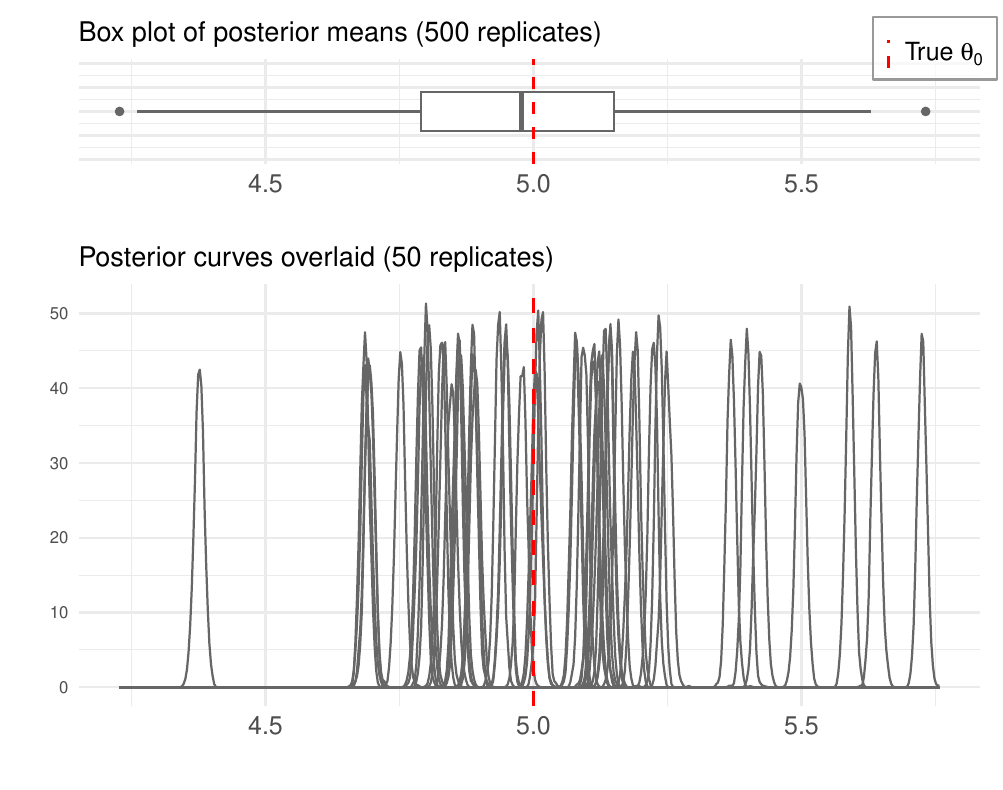}
        \caption{{\tt Bridge}}
    \end{subfigure}
    \hfill
    \begin{subfigure}[b]{0.45\textwidth}
        \centering
        \includegraphics[width=\textwidth]{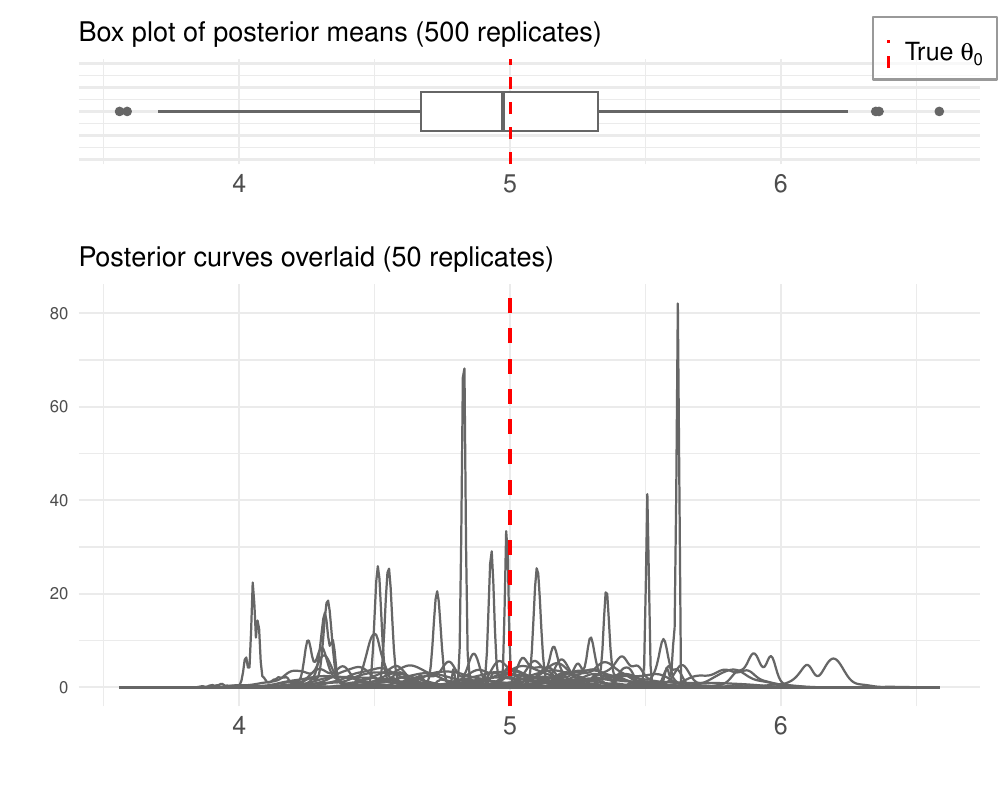}
        \caption{{\tt Bsparse}}
    \end{subfigure}
    \caption{Box plot of posterior means and the overlaid density curves of $\Pi_{\imp}$ for $\theta$ based on {\bf IMP} with {\tt Bridge} and {\tt Bsparse} methods for $\mathbf{s = 166}$. The corresponding coverages are $\mathbf{5\%}$ and $\mathbf{25\%}$. The rest of the caption remains the same as in Figure~\ref{fig:imp_p166s13}.}
    \label{fig:imp_p166s166}
\end{figure}

\begin{figure}[!ht]
    \centering
    \begin{subfigure}[b]{0.45\textwidth}
        \centering
        \includegraphics[width=\textwidth]{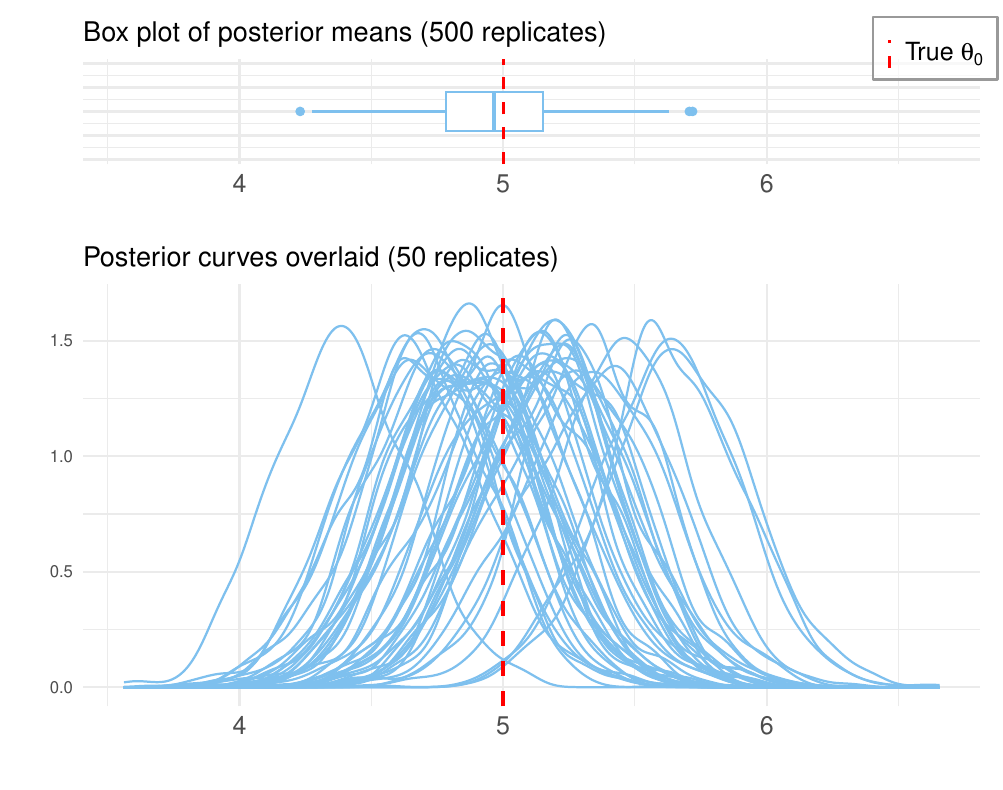}
        \caption{{\tt Bridge}}
    \end{subfigure}
    \hfill
    \begin{subfigure}[b]{0.45\textwidth}
        \centering
        \includegraphics[width=\textwidth]{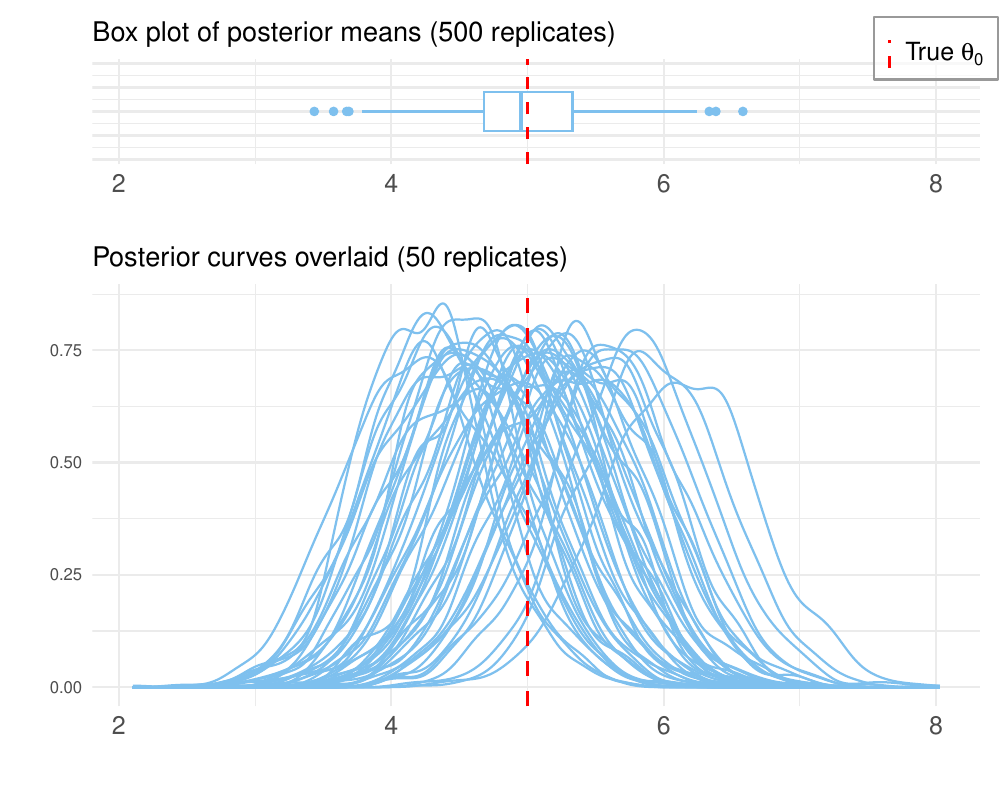}
        \caption{{\tt Bsparse}}
    \end{subfigure}
    \caption{Box plot of posterior means and overlaid density curves  of $\Pi_{\btheta}$ for $\theta$ based on {\bf BDMI} with the {\tt Bridge} and {\tt Bsparse} methods for $\mathbf{s = 166}$. The corresponding coverages are $96\%$ and $93\%$. The rest of the caption remains the same as in Figure~\ref{fig:BDMI_p166s13}.}
    \label{fig:BDMI_p166s166}
\end{figure}

We adhere to the data generation setting described in Section~\ref{sim_correctly_specified}. Specifically, we examine the case where $p = 166$ with four different sparsity levels: $s = 13$ (sparse), $s = 55$ or $s = 83$ (moderately dense), and $s = 166$ (fully dense). For $\Pi_{\mbm}$, we consider two different methods: {\tt Bridge} and {\tt Bsparse}, as described in Section~\ref{sim_correctly_specified}. This yields two versions each for the induced posterior $\Pi_{\imp}$ under IMP and the aggregated posterior $\Pi_{\btheta}$ under BDMI, along with their respective posterior means ($\wh \theta_{\imp}$ and $\wh \theta_{\BDM}$). Figures~\ref{fig:imp_p166s13}--\ref{fig:BDMI_p166s166} display boxplots of the point estimators (based on 500 replications) and density plots of the posteriors across a random subset of 50 replications to improve visual clarity. The {\it odd-numbered} figures correspond to IMP and the {\it even-numbered} ones to BDMI. The posterior curves are based on 1000 posterior samples each. The left and right panels in each figure correspond to {\tt Bridge} and {\tt Bsparse}, respectively.

We first comment on the point estimators $\wh \theta_{\imp}$ and $\wh \theta_{\BDM}$. Figures~\ref{fig:imp_p166s13}--\ref{fig:BDMI_p166s166} show that both point estimators appear unbiased across all settings (sparse, moderately dense, and dense), regardless of the method used ({\tt Bridge} or {\tt Bsparse}) to obtain the nuisance posterior $\Pi_{\mbm}$. Their medians are consistently centered around $\theta_0$ with similar variability. While IMP performs comparably to BDMI in terms of point estimation, important differences emerge when examining the entire {\it posteriors}, $\Pi_{\imp}$ and $\Pi_{\btheta}$, themselves.

The posteriors from the imputation approach exhibit substantial variability {\it across} the two methods as well as the different sparsity settings, showcasing its {\it sensitivity} (in the first order) to nuisance estimation (both in method choice and the setting). Moreover, the imputation posteriors are often very narrow, especially in the more dense cases, and show considerable variation across simulation replicates, with their supports increasingly becoming disjoint. As a result, the imputation posteriors often fail to cover $\theta_0$, leading to {\it severe undercoverage}. Across the two methods and the four different settings, the imputation posterior's coverage of the symmetric $95\%$ credible interval ranges between $5\%- 56\%$. In stark contrast, the BDMI posteriors {\it remain} stable across methods and settings, maintaining a Gaussian shape, and vary smoothly across simulation replicates, with coverage {\it consistently} close to the nominal level, showcasing the superiority of BDMI over IMP. Its ability to provide provably valid inference and its stability (more generally, the overall posterior's smooth behavior) across settings and choices of nuisance models reinforces the importance of its {\it debiased} nature and {\it insensitivity} to nuisance estimation -- an aspect that may be useful more generally in other settings as well.

\section[Implementation details of the Bridge and Bsparse methods used to obtain Pim in Section 5]{Implementation details of the \texttt{Bridge} and \texttt{Bsparse} methods to obtain $\Pi_{\mbm}$ in Section~\ref{simulations}}\label{supp_implementation_details}

In this section, we collect some technical details regarding implementations of two of the methods used to obtain the nuisance posterior $\Pi_{\mbm}$ in our numerical studies in Section~\ref{simulations}: Bayesian ridge regression (\texttt{Bridge}) (in Section~\ref{supp:empirical_bayes_ridge_reg}), and sparse Bayesian linear regression via non-local priors (\texttt{Bsparse}) (in Section~\ref{supp:mombf}).

\subsection[Implementation details for Bridge: Empirical Bayes approach for tuning parameter selection]{Implementation details for \texttt{Bridge}: Empirical Bayes approach for tuning parameter selection}\label{supp:empirical_bayes_ridge_reg}

For the \texttt{Bridge} method, we adopt an empirical Bayes approach to estimate the prior precision parameter (or the ridge tuning parameter, in frequentist terminology) $\lambda$, effectively bridging frequentist and Bayesian methodologies. The estimate $\wh \lambda$ is obtained using the {\tt R} package {\tt glmnet}, specifically its {\tt cv.glmnet} function, along with a scale transformation thereafter. This approach ensures that the posterior mean of $(\alpha, \beta')' \in \bbR^{(p+1)}$ from our approach aligns with the cross-validated point estimate from \texttt{glmnet}, offering a data-driven approach for hyper-parameter selection. A notable aspect of {\tt cv.glmnet} is its {\it standardization} (column-by-column)
of the design matrix $\mathbb{X}_{n \times p} := (\bX_1, \ldots, \bX_n)'$, as well as {\it scaling} of the response vector $\boldY_{n \times 1} := (Y_1,\ldots, Y_n)'$ to have unit standard deviation. Since penalized methods, in general, are not scale invariant, these adjustments are critical to ensure equal penalization of all predictors and mitigate any potential biases due to differences in scale (for both $\mathbb{X}$ and $\boldY$). Writing $\mathbb{X}$ column-wise as $\mathbb{X}_{n \times p} \equiv (\bx_1,\ldots, \bx_p)$, let $\bbZ_{n \times p} \equiv (\mathbf{z}_1,\ldots, \mathbf{z}_p)$ denote the corresponding (column-by-column) standardized version of $\mathbb{X}$, i.e., $\mathbf{z}_j := (\bx_j - \overline{x}_j\mathbf{J}_n)/s_j \in \bbR^n$, where $\overline{x}_j$ and $s_j$ respectively denote the sample mean and sample standard deviation of $\bx_j$, for $j =1,\ldots, p$; and the vector $\mathbf{J}_n := (1,\ldots, 1) \in \bbR^n$. Further, let $s_Y$ be the sample standard deviation of $\boldY$.

Then, the objective function minimized by {\tt cv.glmnet} is given by:
\begin{align*}
l(a, \mathbf{b}) \, := \, \frac{1}{2n} \bigg\|\frac{\boldY}{s_Y}  - a \mathbf{J}_n - \mathbb{Z}\mathbf{b} \bigg\|_2^2 \, + \, \frac{\lambda}{2} \hspace{0.09cm} \|\mathbf{b}\|_2^2, 
\end{align*}
where $\mathbf{J}_n := (1,\ldots,1)' \in \bbR^n$ and $\| \cdot\|_2$ is the $L_2$-vector norm. Note that the intercept parameter $a$ is {\it not} penalized, as is the usual practice. It is also important to note that while \texttt{glmnet} performs the standardization and scaling {\it internally} (by default), the {\it final estimator} it returns is in the {\it original scale} of the data (for both $\boldY$ and $\mathbb{X}$). That is, if $(\wh a, \wh{\mathbf{b}}')'$ denotes the minimizer from above (with optimally chosen $\lambda$) for fitting $\boldY/s_Y \sim a \mathbf{J}_n + \mathbb{Z} \mathbf{b}$, then the final (ridge) estimator it returns is $(\wh \alpha, \wh \bbeta')'$ (for fitting the model $\boldY \sim \alpha \mathbf{J}_n + \mathbb{X} \bbeta$), with $\wh \alpha$ and $\wh{\bbeta}$ obtained from appropriately transforming back $\wh a$ and $\wh{\mathbf{b}}$.

The optimal $\lambda$ value, denoted as $\widetilde\lambda = $ {\tt lambda.min}, from \texttt{cv.glmnet} is selected to minimize the cross-validation error of the above optimization (involving the scaled version of $\boldY$). Therefore, to integrate this into our own Bayesian modeling framework (involving the original $\boldY$), we apply the following transformation:
\[
\wh\lambda ~=~ \widetilde\lambda \cdot (n/s_y), ~~\mbox{and we use this transformed $\wh \lambda$ as our Gaussian prior's precision parameter.}
\]
This transformation ensures consistency between the frequentist and Bayesian approaches by aligning the ridge point estimator $(\wh \alpha, \wh \bbeta')'$ from \texttt{cv.glmnet} with the posterior mean of $(\alpha, \bbeta')'$ from our Bayesian modeling.

For the Bayesian component, we model the data $\widetilde D_n := (\bY_{n \times 1}, \bbZ_{n\times p})$, where $\bY$ is the original response vector and $\bbZ_{n\times p}$ is the standardized design matrix, using a Gaussian likelihood and specify priors as follows:
\begin{align*}
    & \bY \mid \bbZ, \wt\alpha, \wt\bbeta, \sigma^2 \, \iid \, \calN_n\big(\wt\alpha \mathbf{J}_n + \bbZ \wt \bbeta, \sigma^2 I_n\big), ~~\text{with}~ \\
& \pi(\wt\alpha \mid \sigma^2)
    \propto 1, \ \wt\bbeta \mid \sigma^2 \sim \calN_p(\bzero_p, \wh\lambda^{-1} \sigma^2 I_p) ~\, \text{and}~\, \pi(\sigma^2) \propto (\sigma^2)^{-1},
\end{align*}
and we assume that $\wt\alpha$ and $\wt\bbeta$ are independent. After obtaining samples from the posterior distribution $\Pi_{\wt \gamma}$ (a multivariate $t$-distribution) for the parameter $\wt \gamma:= (\wt \alpha, \wt \bbeta')' \in \bbR^{(p+1)}$, we transform these samples back to the {\it original scale,}
i.e., the same scale of $(\bY, \mathbb{X})$, to obtain posterior samples of $(\alpha, \bbeta')'$ as follows:
\[
\alpha \, = \, \wt \alpha - \sum_{j=1}^p \frac{(\wt \beta)_j \cdot \overline{x}_j}{s_j} ~~\mbox{and}~~ (\bbeta)_j \, = \, \frac{(\wt \bbeta)_j}{s_j}, \quad \text{for } j = 1, \dots, p, ~~\mbox{[and $(\mathbf{v})_j$ denotes the $j^{th}$ entry of $\mathbf{v}$]},
\]
where $\overline{x}_j$ and $s_j$ respectively denote the sample mean and sample standard deviation of $\bx_j$, for $j =1,\ldots, p$.

\subsection[Implementation details for Bsparse: The R package mombf]{Implementation details for \texttt{Bsparse}: The \texttt{R} package \texttt{mombf}}
\label{supp:mombf}

For the \texttt{Bsparse} method of obtaining $\Pi_{\mbm}$, we used a sparse Bayesian linear regression based on non-local priors (NLP) \citep{johnson2012bayesian}. We implemented it using the R package \texttt{mombf}, which provides tools for Bayesian model selection and parameter estimation with a focus on NLP. In our implementation, we used the package's default options to ensure consistency and simplicity. The main function used from the package is \texttt{modelSelection}, which performs Bayesian variable selection for linear models using NLP.

This function has two key arguments that allow specification of prior distributions as follows:
\begin{itemize}
    \item \texttt{priorCoef}: Determines the prior for the regression coefficients.
    \item \texttt{priorDelta}: Specifies a prior distribution for the model space.
\end{itemize}

For our implementation, we selected the default choices for these arguments:
\begin{itemize}
    \item \texttt{priorCoef = momprior(tau = 0.348)}, where $\tau$ represents the prior dispersion parameter, controlling the strength of penalization applied to small regression coefficients.
    \item \texttt{priorDelta = modelbbprior(alpha.p = 1, beta.p = 1)}, which sets a Beta-Binomial prior for the model space.
\end{itemize}
Using these settings, we performed Bayesian model selection with \texttt{modelSelection}. To obtain posterior samples for the regression coefficients, we employed the \texttt{rnlp} function from the \texttt{mombf} package. The \texttt{rnlp} function includes two important parameters: \texttt{center} and \texttt{scale} that control whether pre-processing is applied to the response variable ($Y$) and covariates ($\boldX$). We used the following choices for each of these:
\begin{itemize}
    \item \texttt{center = TRUE}: Centers $Y$ and $\boldX$ by subtracting their means to remove potential biases caused by non-zero means in predictors.
    \item \texttt{scale = TRUE}: Scales $\boldX$ by dividing each covariate by its standard deviation to ensure fair penalization across all predictors.
\end{itemize}
By default, both parameters are set to \texttt{FALSE}, meaning no pre-processing is applied unless explicitly specified. However, since our data was not pre-standardized, we set both \texttt{center = TRUE} and \texttt{scale = TRUE} in our implementation. It is important to note that even when centering and scaling are applied (i.e., \texttt{center = TRUE} and \texttt{scale = TRUE}), the \texttt{rnlp} function provides posterior samples for regression coefficients on the {\it original scale} of the data. This is achieved since the \texttt{mombf} package internally stores information on the centering and scaling transformations applied during pre-processing. These stored values are used to transform posterior samples back to their original scale, ensuring that results remain interpretable in terms of the original data.

\section[Proofs of the main results]{Proofs of the main results}\label{appendix}
In this section, we present the proofs of all the results from the main paper. We begin by introducing some additional notations and some preliminary lemmas that will be used throughout the proofs of the main results. The rest of the section is organized as follows: {\bf (i)} Section~\ref{section_preliminaries} enlists the preliminary lemmas; {\bf (ii)} Section~\ref{proof_prop_ind_data} presents the proof of Proposition~\ref{prop_ind_data}; {\bf (iii)} Section~\ref{proof_prop_half_data} presents the proof of Proposition~\ref{prop_half_data}; {\bf (iv)} Section~\ref{proof_bvm_half_data} provides the proof of Theorem~\ref{bvm_on_first_half_data}; {\bf (v)} Section~\ref{proof_main_thm} presents the proof of Theorem~\ref{main_thm}; {\bf (vi)} Section~\ref{proof_of_corr_asymp_equiv_pmean} presents the proof of Corollary~\ref{corollory_asymp_equiv_postmean}; and {\bf (vii)} Section~\ref{proof_bvm_standard} provides the proof of Theorem~\ref{bvm_standard}.

Throughout this section, we will use the following {\it additional notations}, in addition to those introduced in the main paper. For any functions $f(\cdot) \in \bbL_2(\bbP_{\boldZ})$ and $g(\cdot)
\in \bbL_2(\bbP_{\boldX})$, and for any $k \in \{ 1, \dots, K \}$, define:
\begin{align}
&\mbox{(i)}~~ \bbE_{\nK}^{(k)}\{f(\boldZ)\} ~:=~ \nK^{-1} \sum_{i \in \calI_k} f(\boldZ_i)~~\mbox{and}~~
\bbG_{\nK}^{(k)}\{f(\boldZ)\} ~:=~ \nK^{1/2} \big[\hspace{0.5mm} \bbE_{\nK}^{(k)}\{f(\boldZ)\} - \bbE_{\boldZ}\{f(\boldZ)\} \hspace{0.5mm}\big]; \quad \mbox{and} \nonumber \\
& \mbox{(ii)}~~ \bbE_{\NK}^{(k)}\{g(\boldX)\} ~:=~ \NK^{-1} \sum_{i \in \calJ_k} g(\boldX_i)  ~~\mbox{and}~~
\bbG_{\NK}^{(k)}\{g(\boldX)\} ~:=~ \NK^{1/2} \big[\hspace{0.5mm} \bbE_{\NK}^{(k)}\{g(\boldX)\} - \bbE_{\boldX}\{g(\boldX)\} \hspace{0.5mm} \big]. \label{emp_notations}
\end{align}

Further, for any two absolutely continuous probability measures
$P$ and $Q$ on $\bbR$ with corresponding densities $p(\cdot)$ and $q(\cdot)$, we will use the well-known identity for their TV distance: $\| P - Q \|_{\TV} = (1/2) \int |p(x) - q(x)| \dd x$ {\citep[Lemma 2.1 (Scheff\'es' theorem)]{tsybakov2009}}. Additionally, a Gamma distribution with shape and rate parameters $(\alpha, \beta) > 0$ and density $\frac{\beta^\alpha}{\Gamma(\alpha)} e^{-\beta x} x^{\alpha-1} \mathbbm{1}_{(0, \infty)}(x)$ is denoted $\text{Gamma}(\alpha, \beta)$; and if $W \sim \text{Gamma}(\alpha, \beta)$, then we denote $1/W \sim \text{IG}(\alpha, \beta)$, i.e., the inverse Gamma (IG) distribution with parameters $(\alpha, \beta) > 0$.

\begin{remark}[Empirical process notations]\label{rem_empirical_process_notation}
The notations $\bbE_{\nK}^{(k)}(\cdot)$ and $\bbE_{\NK}^{(k)}(\cdot)$ in \eqref{emp_notations} simply denote the empirical mean operators on the data folds $\calL_k$ (indexed by $\calI_k$ and of size $\nK = n/K$) and $\calU_k$ (indexed by $\calJ_k$ and of size $\NK = N/K$), respectively. Similarly, $\bbG_{\nK}^{(k)}(\cdot)$ denotes the corresponding $\nK^{1/2}$-scaled (and centered) empirical process on $\calL_k$ indexed by $f(\cdot)$, and and $\bbG_{\NK}^{(k)}(\cdot)$ denotes the $\NK^{1/2}$-scaled (and centered) empirical process on $\calU_k$ indexed by $g(\cdot)$. Notations of this flavor are fairly common in the modern semi-parametric inference literature, as well as SS inference literature, that require empirical process and/or sample-splitting techniques \citep{chernozhukov2018double, van2000asymptotic, zhang2000value, chakrabortty2022semi}.
\end{remark}

\subsection{Preliminaries}\label{section_preliminaries}
The following results will be used to prove the main results of the paper.

\begin{lemma}[Invariance property of the TV distance {\citep[Chapter 3]{pollard2002user}}]\label{TV_invariance}
Let $P$ and $Q$ be absolutely continuous probability measures on $\bbR$ with the corresponding densities $p(\cdot)$ and $q(\cdot)$. For fixed $\mu \in \bbR$ and $\sigma > 0$, define $p^{\mu, \sigma}(t) := \sigma^{-1}\hspace{0.5mm} p\{(t - \mu)/\sigma\}$ and $q^{\mu, \sigma}(t) := \sigma^{-1}\hspace{0.5mm}q\{(t - \mu)/\sigma\}$ as the corresponding location-shifted and scaled version of $p(\cdot)$ and $q(\cdot)$,
with the respective probability measures $P^{\mu, \sigma}$ and $Q^{\mu, \sigma}$. Then, $\|P - Q \|_\TV = \|P^{\mu, \sigma} - Q^{\mu, \sigma} \|_\TV$.
\end{lemma}

\begin{lemma}[An upper bound for the TV distance between two Gaussian distributions
with the same variance ]\label{TV_2normal}
Let $P = \calN(\mu_1, \sigma^2)$ and $Q = \calN(\mu_2, \sigma^2)$ be two Normal distributions. Then, $\|P - Q \|_{\TV} ~ = ~ 2 \hspace{0.5mm} \Phi\{ |\mu_1 - \mu_2|/(2\sigma) \} - 1$, where $\Phi(\cdot)$ is the cumulative distribution function (CDF) of the standard Normal distribution $\calN(0, 1)$. This further implies that $
\|P - Q \|_{\TV} \leq (2 \pi)^{-1/2} |\mu_1 - \mu_2|/\sigma$.
\end{lemma}

\begin{lemma}[Adopted from Lemma 4.9 in \citet{Klartag2006ACL}]\label{klartag} Let $P = \calN(0, \sigma^2_1)$ and $Q = \calN(0, \sigma^2_2)$.
Then, for some universal constant $C > 0$,
$
\|P - Q\|_{\TV} \leq~ C \hspace{0.1mm} | (\sigma^2_2/\sigma^2_1) - 1 |$.
\end{lemma}

\begin{lemma}[TV distance between a
$t$--distribution and a Normal distribution] \label{TV_tnormal}
Let $P = t_\nu(\mu, \sigma^2)$ and $Q = \calN(\mu, \sigma^2)$. Then, for some constant $C_0 > 0$, we have $
\|P - Q \|_{\TV} ~\leq~ C_0/\sqrt \nu$.
\end{lemma}

\begin{lemma}[TV distance for
convolutions]\label{TVD_convolution}
Let $P$, $Q$ be two probability distributions with the pdfs $p(\cdot)$, $q (\cdot)$, respectively. Suppose $
p(x)=(p_1 * p_2)(x)$ and $q(x)= (q_1 * q_2)(x)$, where $p_i(\cdot), q_i(\cdot)$ are the pdfs of the corresponding distributions $P_i, Q_i$ for $i = 1, 2$, and $*$ is the convolution operator which is defined in Proposition~\ref{prop_ind_data}. Then, $
\| P - Q \|_{\TV} ~\leq~ \| P_1 - Q_1 \|_{\TV} ~+~ \| P_2 - Q_2 \|_{\TV}. $
\end{lemma}

\begin{lemma}[Conditional convergence $\Rightarrow$ unconditional convergence; adopted from {\citet[Lemma 6.1]{chernozhukov2018double}}]\label{chernozhukov_lemma6_1} Let $U_n$ and $V_n$ be sequences of random variables (with a joint distribution). (a) If, for $\varepsilon_n \to 0$, $\bbP(|U_n| > \varepsilon_n \hspace{0.5mm} | \hspace{0.5mm} V_n ) \cvP 0$, then $\bbP( |U_n| > \varepsilon_n) \to 0$. (b) Let $b_n$ be a sequence of positive constants. If $|U_n| = O_{\bbP}(b_n)$ conditional on $V_n$, namely, that for any $t_n \to \infty$, $\bbP(|U_n| > t_n \hspace{0.3mm} b_n \hspace{0.5mm} | \hspace{0.5mm} V_n) \cvP 0$, then $|U_n| = O_{\bbP}(b_n)$ unconditionally, namely, that for any $t_n \to \infty$, $\bbP(|U_n| > t_n \hspace{0.3mm} b_n) \to 0$.
\end{lemma}

Proofs of the preliminary results are presented in  Section~\ref{supp:prelim_proof} of the \hyperref[sec:supplementary]{Supplementary Material}. We are now ready to present the proof of the main results in Sections~\ref{bayesian_SS} and~\ref{sec_theory}.

\subsection{Proof of Proposition~\ref{prop_ind_data}}\label{proof_prop_ind_data}
For notational simplicity, we define $W(\boldZ;\tm) := Y - \tm(\boldX)$ and $\delta(\tm) := \theta - b(\tm)$. Then, the likelihood function in \eqref{likelihood_function_indp_data} becomes
\begin{align*}
L\big\{\delta(\tm), b(\tm), \sigma^2_{1}(\tm), \sigma^2_{2}(\tm) \big\} ~\propto~ &  \ \frac{1}{\{\sigma^2_{1}(\tm)\}^{n/2}} \exp\left[-\frac{1}{2\sigma^2_{1}(\tm)} \sum_{i = 1}^n \big\{W(\boldZ_i;\tm)- b(\tm) \big\}^2\right]  \times \\
& \ \frac{1}{\{\sigma^2_{2}(\tm)\}^{N/2}} \exp\left[-\frac{1}{2\sigma^2_{2}(\tm)} \sum_{i = n+1}^{n+N}\big\{\tm(\boldX_i)- \delta(\tm) \big\}^2\right] \\
~:=~ & \ L\{ b(\tm), \sigma^2_{1}(\tm)\} \ L\{ \delta(\tm), \sigma^2_{2}(\tm)\}.
\end{align*}
Since the determinant of the Jacobian matrix is 1, the prior on the model parameters becomes
\begin{equation}\label{scaled_prior}
\pi\big\{\delta(\tm), b(\tm), \sigma^2_{1}(\tm), \sigma^2_{2}(\tm) \big\} ~\propto~ \big\{\sigma^2_{1}(\tm) \sigma^2_{2}(\tm)\big\}^{-1}.
\end{equation}
By Bayes' theorem, the joint posterior density of $\{\delta(\tm), b(\tm)\}$ can be calculated by integrating out the parameters $\{\sigma^2_{1}(\tm), \sigma^2_{2}(\tm)\}$. Then, we obtain that
\begin{align*}
\pi\{\delta(\tm), b(\tm) \mid \calD\} ~\propto~ & \int \frac{L\{\delta(\tm), \sigma^2_{2}(\tm)\}}{\sigma^2_{2}(\tm)} \ \dd\sigma^2_{2}(\tm)  \times \int \frac{L\{ b(\tm), \sigma^2_{1}(\tm)\}}{\sigma^2_{1}(\tm)} \ \dd \sigma^2_{1}(\tm) \\
~=~ & \pi\{\delta(\tm) \mid \calD\}  \pi\{b(\tm) \mid \calD\}.
\end{align*}
Then the joint posterior density $\pi\{\delta(\tm), b(\tm) \mid \calD\}$ is the product of the marginal posterior densities of $\delta(\tm)$ and $b(\tm)$. This implies that posterior distributions of $\delta(\tm)$ and $b(\tm)$ are independent. Since $\theta = \delta(\tm) + b(\tm)$ by the construction of $\delta(\tm)$, the marginal posterior distribution of $\theta$ can be calculated as a convolution of posterior distributions of $\delta(\tm)$ and $b(\tm)$.

To conclude the proof of Proposition~\ref{prop_ind_data}, it is enough to show the marginal posterior distribution of $b(\tm)$ is a $t$--distribution with degrees of freedom $v_n$, center $\mu_n(\tm)$ and scale $\wh \sigma^2_{1, n}(\tm)/n$ as defined in \eqref{notations_ind_data}, since the calculation of the posterior distribution of $\delta(\tm)$ follows the same steps.

Towards establishing this, we first observe that
\begin{align*}
b(\tm) \mid \sigma^2_{1}(\tm), \calD  \sim  \calN\!(\mu_{n}(\tm), \sigma^2_{1}(\tm)/n) ~\mbox{and}~
\sigma^2_{1}(\tm) \mid \calD  \sim  \text{IG}\!\left( \frac{n - 1}{2}, \frac{\sum_{i = 1}^n \{W(\boldZ_i;\tm) - \mu_{n}(\tm)\}^2}{2}\right)\!,
\end{align*}
where $\mu_{n}(\tm) = n^{-1} \sum_{i = 1}^n W(\boldZ_i;\tm)$.
Since the $t$--distribution can be expressed as a scale mixture of a Normal distribution, we obtain that $
b(\tm) \mid \calD \sim t_{\nu_{n}}(\mu_{n}(\tm), \wh \sigma^2_{1, n}(\tm)/n)$ where $\nu_{n} := n - 1$ and
\begin{align*}
\frac{\wh \sigma^2_{1, n}(\tm)}{n} \ := \ \frac{\sum_{i = 1}^n \{ W(\boldZ_i;\tm) - \mu_{n}(\tm)\}^2}{n(n - 1)}.
\end{align*}
By following the same steps, we obtain that
$\delta(\tm) \sim t_{\nu_{N}}(\mu_{N}(\tm), \wh \sigma^2_{2, N}(\tm)/N)$ where $\nu_{N} := N - 1$,
\begin{align*}
\mu_{N}(\tm) \, :=  \, \frac{1}{N}\sum_{i = n+1}^{n+N} \tm(\boldX_i) ~\, \text{ and } ~\,
\frac{\wh \sigma^2_{2, N}(\tm)}{N} \, := \, \frac{\sum_{i = n+1}^{n+N}\{\tm(\boldX_i) - \mu_{N}(\tm)\}^2}{N(N - 1)}.
\end{align*}
Hence, the marginal posterior of $\theta$ is a convolution of two $t$-distributions: $t_{\nu_{n}}(\mu_{n}(\tm), \wh \sigma^2_{1, n}(\tm)/n)$ and $t_{\nu_{N}}(\mu_{N}(\tm), \wh \sigma^2_{2, N}(\tm)/N)$. This completes the proof. \qeds

\subsection{Proof of Proposition~\ref{prop_half_data}}\label{proof_prop_half_data}
To avoid repetition, we refer to the proof of Proposition~\ref{prop_ind_data} in Section~\ref{proof_prop_ind_data}. Consider the likelihood function in \eqref{supp:likelihood_func_half_data} and the prior density in \eqref{scaled_prior}. Then, by following the same steps as in the proof of Proposition~\ref{prop_ind_data} this time applied to the data fold $\wt\calD_k$ instead of $\calD$ as in Proposition~\ref{prop_ind_data}, we obtain that the marginal posterior distribution $\Pi_{\btheta}^{(k)}$ is a convolution of two $t$-distributions with the desired parameters as given in the statement of Proposition~\ref{prop_half_data}. This concludes the proof. \qeds

\subsection{Proof of Theorem~\ref{bvm_on_first_half_data}} \label{proof_bvm_half_data}
For notational simplicity, we set $k = 1$ w.l.o.g. and present the proof for $k = 1$. By the triangle inequality, we first observe that
\begin{align}\label{TV_triangular_eq}
& \|\Pi_{\btheta}^{(1)} - \calN(\wh\theta_{\BDM}^{(1)}(m^*), \tau^2_{\nK,\NK}(m^*)) \|_{\TV} \nonumber \\
~\leq~ & \ \|\Pi_{\btheta}^{(1)} - \calN(\wh\theta_{\BDM}^{(1)}(\tm_1), \tau^2_{\nK,\NK}(\tm_1)) \|_{\TV} \nonumber \\
& ~+~ \|\calN(\wh\theta_{\BDM}^{(1)}(\tm_1), \tau^2_{\nK, \NK}(\tm_1)) - \calN(\wh\theta_{\BDM}^{(1)}(m^*), \tau^2_{\nK, \NK}(m^*))\|_{\TV} ~:=~ T_1 + T_2,
\end{align}
where 
$\wh\theta_{\BDM}^{(k)}(\tm_k):= \mu_{\nK}(\tm_k)+ \mu_{\NK}(\tm_k)= \nK^{-1} \sum_{i \in \calI_k}\{Y_i - \tm_k(\boldX_i)\} + \NK^{-1} \sum_{i \in \calJ_k}\tm_1(\boldX_i)$ and $\tau^2_{\nK, \NK}(\tm_k):= \sigma^2_{1}(\tm_k)/\nK + \sigma^2_{2}(\tm_k)/\NK
$ for any $k = 1, \dots , K$ (we specifically set $k =1$ here).
Then, the problem reduces to showing both $T_1$ and $T_2$ converge to 0 in probability w.r.t. $\bbP_{\wt\calD_1}$.

We first consider $T_1$ in \eqref{TV_triangular_eq}. By Proposition~\ref{prop_half_data} (refer to the \hyperref[sec:supplementary]{Supplementary Material}), we have that the posterior $ \Pi_{\btheta}^{(1)}$ of $\theta$ is the convolution of two $t$-distributions $
t_{\nu_{\nK}}(\mu_{\nK}(\tm_1), \wh \sigma^2_{1, \nK}(\tm_1)/\nK)$ and $t_{\nu_{\NK}}(\mu_{\NK}(\tm_1), \wh \sigma^2_{2, \NK}(\tm_1)/\NK)$, where the parameters are as defined in \eqref{supp:notations_half_data} (refer to the \hyperref[sec:supplementary]{Supplementary Material}) by setting $k= 1$. Also, we can always write a Normal distribution as a convolution of two independent Normal distributions. Further, by Lemma~\ref{TVD_convolution}, we observe that
\begin{align}\label{TV_posterior_normal}
\hspace{-0.3cm}T_1 =~ & \| \Pi_{\btheta}^{(1)} - \calN(\wh\theta_{\BDM}^{(1)}(\tm_1), \tau^2_{\nK, \NK}(\tm_1)) \|_{\TV} \nonumber \\
\leq~ & \|t_{\nu_{\nK}}(\mu_{\nK}(\tm_1), \wh \sigma^2_{1, \nK}(\tm_1)/\nK) - \calN(\mu_{\nK}(\tm_1), \sigma^2_{1}(\tm_1)/\nK)\|_{\TV} \nonumber \\
& +~ \|t_{\nu_{\NK}}(\mu_{\NK}(\tm_1), \wh \sigma^2_{2, \NK}(\tm_1)/\NK) - \calN(\mu_{\NK}(\tm_1), \sigma^2_{2}(\tm_1)/\NK)\|_{\TV} \nonumber\\
= ~ & \|t_{\nu_{\nK}}(0,\wh\sigma^2_{1, \nK}(\tm_1)) - \calN(0, \sigma^2_{1}(\tm_1))\|_{\TV} + \|t_{\nu_{\NK}}(0, \wh\sigma^2_{2, \NK}(\tm_1)) - \calN(0, \sigma^2_{2}(\tm_1))\|_{\TV} \nonumber \\
:=~ & T_{11} + T_{12},
\end{align}
where \eqref{TV_posterior_normal} is obtained from the invariance property of the TV distance from Lemma~\ref{TV_invariance}.

Next, we consider $T_{11}$ in \eqref{TV_posterior_normal}. By the triangle inequality and the construction of $\wh \sigma^2_{1, \nK}(\tm_1)$, we get
\begin{align*}
T_{11} & ~=~ \|t_{\nu_{\nK}}(0, \wh \sigma^2_{1, \nK}(\tm_1)) - \calN(0, \sigma^2_{1}(\tm_1))\|_{\TV}  \\
& ~\leq~ \|t_{\nu_{\nK}}(0, \wh \sigma^2_{1, \nK}(\tm_1)) - \calN(0, \wh \sigma^2_{1, \nK}(\tm_1))\|_{\TV} + \|\calN(0,\wh \sigma^2_{1, \nK}(\tm_1)) - \calN(0, \sigma^2_{1}(\tm_1)) \|_{\TV} \\
& ~\leq~ \frac{C_0}{\nu_{\nK}} + \left| \frac{ \wh \sigma^2_{1, \nK}(\tm_1) - \sigma^2_{1}(\tm_1)}{\sigma^2_{1}(\tm_1)}  \right|,
\end{align*}
where the last step follows from Lemma~\ref{TV_tnormal} (applied to the first TV distance in the second line above) and Lemma~\ref{klartag} (applied to the second TV distance in the second line above). By the definition of $\nu_{\nK}$ (i.e., $\nu_{\nK} = \nK - 1$), $1/\nu_{\nK} \to 0$ as $n \to \infty$ (since $\nK = n/K$ and $K$ is fixed, refer to Section~\ref{BCF} for further notational clarification). Thus to show
$T_{11} \cvP 0$ under
$\bbP_{\calD_1,\tm_1}$, it is enough to show
\begin{align*}
    \Sigma_{\epsilon, \nK}(\calD_1,\tm_1) ~ := ~ \left| \frac{ \wh \sigma^2_{1, \nK}(\tm_1) - \sigma^2_{1}(\tm_1)}{\sigma^2_{1}(\tm_1)}  \right| ~ \to ~ 0, \ \text{in probability w.r.t. } \bbP_{\calD_1, \tm_1}.
\end{align*}
This means that for any $t > 0$, $ \bbP_{\calD_1, \tm_1}(\Sigma_{\epsilon, \nK}(\calD_1,\tm_1) > t)   \to 0
$. We further observe that
\begin{align}\label{equivalence_for_DCT}
\bbP_{\calD_1, \tm_1}\{\Sigma_{\epsilon, \nK}(\calD_1,\tm_1) > t\} &  = \bbE_{\calD_1, \tm_1} [ \mathbf{1}\{\Sigma_{\epsilon, \nK}(\calD_1,\tm_1) > t \}] \nonumber\\
  & =  \bbE_{\tm_1} (\bbE_{\calD_1}[\mathbf{1}\{\Sigma_{\epsilon, \nK}(\calD_1,\tm_1) > t \}\mid \tm_1]) \\
   & = \bbE_{\tm_1} [\bbP_{\calD_1}\{ \Sigma_{\epsilon, \nK}(\calD_1,\tm_1) > t \mid \tm_1 \} ], \nonumber
\end{align}
where $\mathbf{1}\{\cdot\}$ denotes the indicator function and the second step uses the fact that $\tm_1 \ind \calD_1$ (refer to the construction of BDMI in Section~\ref{BCF} by taking $k =1$). We note that $0 \leq \bbP_{\calD_1}\{\Sigma_{\epsilon, \nK}(\calD_1,\tm_1) > t \mid \tm_1 \} \leq 1$, and it is random through $\tm_1$. Then, by the dominated convergence theorem (DCT) (alternatively, refer to Lemma~\ref{chernozhukov_lemma6_1} here), it is sufficient to show $\bbP_{\calD_1}\{\Sigma_{\epsilon, \nK}(\calD_1,\tm_1) > t \mid \tm_1 \} \cvP 0$ under $\bbP_{\tm_1}$ to conclude that $\bbP_{\calD_1, \tm_1}\{\Sigma_{\epsilon, \nK}(\calD_1,\tm_1) > t\} \to 0$. Next, we observe that for any $t > 0$,
\begin{align*}
\bbP_{\calD_1} \{ \Sigma_{\epsilon, \nK}(\calD_1,\tm_1) > t \mid \tm_1 \} \ = \ \bbP_{\calD_1} \{ | \wh \sigma^2_{1, \nK}(\tm_1) - \sigma^2_{1}(\tm_1) | > \tilde t \mid \tm_1\},
\end{align*}
for $\tilde t ~:=~ t~|\sigma^2_{1}(\tm_1)| > 0$, where given $\tm_1$, we can think of $\sigma^2_{1}(\tm_1)$ as a fixed non-random quantity. Then, by Chebyshev’s inequality, we obtain that
\begin{align*}
    Z_{\nK}(\tm_1) ~:=~ \bbP_{\calD_1} \{| \wh \sigma^2_{1, \nK}(\tm_1) - \sigma^2_{1}(\tm_1)| > \tilde t \mid \tm_1\} ~\leq~ (\hspace{0.3mm} \tilde t \hspace{0.3mm})^{-2}\hspace{0.7mm} \Var\{\wh \sigma^2_{1, \nK}(\tm_1) \mid \tm_1 \},
\end{align*}
where the last inequality uses $\bbE_{\calD_1}\{\wh \sigma^2_{1, \nK}(\tm_1) - \sigma^2_{1}(\tm_1) \hspace{0.4mm} | \hspace{0.4mm} \tm_1\} = 0$ that can be obtained by the construction of $\wh \sigma^2_{1, \nK}(\tm_1)$ (refer to \eqref{supp:notations_half_data} by setting $k = 1$). For notational simplicity, let $W(\boldZ;\tm_1) := Y - \tm_1(\boldX)$. Then, using Theorem 2 in Chapter VI of \citet{mood1974introduction}, we obtain that
\begin{equation}\label{variance_of_hat_sigma2_epsilon_mtilde}
    \Var\{\wh \sigma^2_{1, \nK}(\tm_1) \hspace{1mm} | \hspace{1mm} \tm_1 \}  ~=~  \frac{\mu_4(\tm_1)}{\nK} ~+~ \frac{(\nK - 3)\{\sigma^2_{1}(\tm_1)\}^2}{\nK (\nK - 1)},
\end{equation}
where $\mu_4(\tm_1):= \bbE_{\boldZ}([W(\boldZ;\tm_1)- \bbE_{\boldZ}\{W(\boldZ; \tm_1)\}]^4 \hspace{0.4mm} | \hspace{0.4mm} \tm_1)$ is the fourth central moment of $W(\boldZ; \tm_1)$ w.r.t. $\boldZ~(\ind \tm_1)$ given $\tm_1$. Now, by Assumption~\ref{assumption_for_half_fold} (i) and the construction/definition of $W(\boldZ; \tm_1)$ as above, i.e., $W(\boldZ; \tm_1) \equiv Y - \tm_1(\boldX)$, we have that $\mu_4(\tm_1) = \Op(1)$ under the joint probability distribution $\Pi_{\mbm}^{(1)}(\calS_1)$.

Consequently, we obtain that $Z_{\nK}(\tm_1) = o_{\bbP_{\tm_1}}(1)$. This equivalently gives that for some sequence $b_{\nK, \sK} \to 0$, $Z_{\nK}(\tm_1) = O_{\bbP_{\tm_1}}(b_{\nK,\sK})$, where $\sK = n - n/K$ (the size of $\calS_1$). We note that the double index used in $b_{\nK,\sK}$ indicates that the rate depends not only on the term $\wh \sigma^2_{1, \nK}(\tm_1)$ but also on the size of $\calS_1$ which is used to obtain the distribution $\Pi_\mbm^{(1)}$ of $\tm_1$. Then by applying Lemma~\ref{chernozhukov_lemma6_1} (b), we obtain that $\Sigma_{\epsilon, \nK}(\calD_1,\tm_1) = O_{\bbP_{\calD_1, \tm_1}}(b_{\nK, \sK})$ which implies that $\Sigma_{\epsilon, \nK}(\calD_1,\tm_1) = o_{\bbP_{\calD_1, \tm_1}}(1)$. Hence, we conclude that as $n, N \to \infty$, $T_{11} \cvP 0$ under $\bbP_{\calD_1, \tm_1}$. \qeds

Similarly, we follow the same steps for the term $T_{12}$ in \eqref{TV_posterior_normal} and finally obtain that
\begin{align*}
    T_{12} ~=~ & \| t_{\nu_{\NK}}(0, \wh \sigma^2_{2, \NK}(\tm_1)) - \calN(0, \sigma^2_{2}(\tm_1))\|_{\TV} ~\leq~ \frac{C_0}{\nu_{\NK}} + \left| \frac{\wh \sigma^2_{2, \NK}(\tm_1) - \sigma^2_{2}(\tm_1)}{\sigma^2_{2}(\tm_1)}  \right|.
\end{align*}
Then, it is enough to show that
\begin{align*}
    \Sigma_{r, \NK}(\calD_1,\tm_1) & ~ := ~ \left| \frac{\wh \sigma^2_{2, \NK}(\tm_1) - \sigma^2_{2}(\tm_1)}{\sigma^2_{2}(\tm_1)}  \right| \ \to \ 0 ~ \text{ in probability w.r.t. } \bbP_{\calD_1, \tm_1}.
\end{align*}
Next, by using $\calD_1 \ind \tm_1$ and by following the same idea and steps in \eqref{equivalence_for_DCT}, we observe that for any $t>0$,
\begin{align*}
\bbP_{\calD_1, \tm_1}\{\Sigma_{r, \NK}(\calD_1,\tm_1) > t\} \ = \ \bbE_{\tm_1}[\bbP_{\calD_1}\{ \Sigma_{r, \NK}(\calD_1,\tm_1) > t \mid \tm_1\} ].
\end{align*}
Then, by the DCT (or Lemma~\ref{chernozhukov_lemma6_1} (b)), it is enough to show $\bbP_{\calD_1}\{\Sigma_{r, \NK} (\calD_1,\tm_1) > t \mid \tm_1\} \cvP 0$ under $\bbP_{\tm_1}$ to conclude that $\bbP_{\calD_1, \tm_1} \{ \Sigma_{r, \NK}(\calD_1,\tm_1) > t \} \to 0.$

Further, given $\tm_1$, $\sigma^2_{2}(\tm_1)$ is a fixed non-random quantity, for any $t > 0$, we have that
\begin{align*}
\bbP_{\calD_1} \{ \Sigma_{r, \NK}(\calD_1,\tm_1)  > t \hspace{1mm} | \hspace{1mm} \tm_1 \} ~=~ \bbP_{\calD_1} \left\{ \left| \wh \sigma^2_{2, \NK}(\tm_1) - \sigma^2_{2}(\tm_1) \right| > \tilde t \hspace{1mm} | \hspace{1mm} \tm_1\right\},
\end{align*}
for $\tilde t = t~|\sigma^2_{2}(\tm_1)| > 0$. Let $Z_{\NK}(\tm_1): = \bbP_{\calD_1} \{ | \wh \sigma^2_{2, \NK}(\tm_1) - \sigma^2_{2}(\tm_1)| > \tilde t \mid \tm_1 \}$. We note that $Z_{\NK}(\tm_1)$ is a random variable where its randomness comes from $\tm_1$ and $0 \leq Z_{\NK}(\tm_1) \leq 1$ by its definition. Then, by applying the DCT (or directly using Lemma~\ref{chernozhukov_lemma6_1}), it is sufficient to prove that $Z_{\NK}(\tm_1) \to 0$ in probability w.r.t. $\bbP_{\tm_1}$ to conclude that $T_{12} \to 0$ in probability w.r.t. $\bbP_{\wt\calD_1}$. Towards that, since $\bbE_{\calD_1}\{ \wh \sigma^2_{2, \NK}(\tm_1) - \sigma^2_{2}(\tm_1) \hspace{1mm} | \hspace{1mm} \tm_1 \} = 0$ by the constriction of $\wh \sigma^2_{2, \NK}(\tm_1)$ (refer to \eqref{supp:notations_half_data} in the \hyperref[sec:supplementary]{Supplementary Material} by setting $k = 1$), by Chebyshev's inequality and following the same algebraic calculations as those used to obtain \eqref{variance_of_hat_sigma2_epsilon_mtilde} but applied to $\wh \sigma^2_{2, \NK}(\tm_1)$ this time, we have
\begin{align*}
    \bbP_{\calD_1} \{ \Sigma_{r, \NK}(\calD_1,\tm_1) > t \hspace{1mm} | \hspace{1mm} \tm_1 \} & \ \leq \ \frac{\Var \{ \wh \sigma^2_{2, \NK}(\tm_1)\} }{\tilde t^2} \ = \ \frac{\mu_4(\tm_1)}{\NK} + \frac{(\NK - 3)\{\sigma^2_{2}(\tm_1)\}^2}{\NK (\NK - 1)},
\end{align*}
where $\mu_4(\tm_1) := \bbE_{\boldX}([\tm_1(\boldX) - \bbE\{\tm_1(\boldX)\}]^4 \hspace{0.4mm} | \hspace{0.4mm} \tm_1)$ and the last step follows from \citet[Theorem 2 in Chapter VI]{mood1974introduction}. Then, by Assumption~\ref{assumption_for_half_fold} (i), we have $\mu_4(\tm_1) = \Op(1)$ under the joint probability distribution $\Pi_{\mbm}^{(1)}(\calS_1)$. Hence we obtain that $Z_{\NK}(\tm_1) = o_{\bbP_{\tm_1}}(1)$. This directly gives that for some sequence $d_{\NK, \sK} \to 0$, $Z_{\NK}(\tm_1) = O_{\bbP_{\tm_1}}(d_{\NK,\sK})$. We again note that the double index in $d_{\NK,\sK}$ indicates the dependency of the rate not only on the term $\wh \sigma^2_{2, \NK}(\tm_1)$ but also on the size of $\calS_1$ which is used to obtain the distribution $\Pi_\mbm^{(1)}$ of $\tm_1$. Then applying Lemma~\ref{chernozhukov_lemma6_1} (b), we obtain that $\Sigma_{r, \NK}(\calD_1, \tm_1) = O_{\bbP_{\calD_1, \tm_1}}(d_{\NK, \sK})$ which implies that
$\Sigma_{r, \NK}(\calD_1, \tm_1) = o_{\bbP_{\calD_1, \tm_1}}(1)$. This concludes that as $n, N \to \infty$, $T_{12} \cvP 0$ w.r.t. $\bbP_{\calD_1, \tm_1}$ and so $T_1$ in \eqref{TV_triangular_eq} $\cvP 0$ w.r.t. $\bbP_{\calD_1, \tm_1}$. \qeds

Next, we consider the term $T_2 = \|\calN(\wh\theta_{\BDM}^{(1)}(\tm_1), \tau^2_{\nK, \NK}(\tm_1)) - \calN(\wh\theta_{\BDM}^{(1)}(m^*), \tau^2_{\nK, \NK}(m^*))\|_{\TV}$ in \eqref{TV_triangular_eq} and show that $T_{2}$ goes to zero in probability w.r.t. $\bbP_{\wt\calD_1}$. To make the proof of this part clearer and streamlined, we first present the following lemma:
\begin{lemma}\label{lemma_half_of_data} Under Assumption~\ref{assumption_for_half_fold} and the setup
of Theorem~\ref{bvm_on_first_half_data}, we have
\begin{equation}\label{op_condition}
\| m^*(\boldX) - \tm_1(\boldX) \|_{\bbL_2(\bbP_{\boldX})} \ = \ o_{\bbP_{\tm_1}}(1).
\end{equation}
\end{lemma}
\noindent For brevity, the proof of Lemma~\ref{lemma_half_of_data} is presented in Section~\ref{proof_of_lemma_half_of_data} of the \hyperref[sec:supplementary]{Supplementary Material}.

Suppose Lemma~\ref{lemma_half_of_data} holds. Then, by the triangle inequality and the invariance property of the TV distance from Lemma~\ref{TV_invariance}, we obtain that
\begin{align*}
T_2 ~ = ~ & \ \|\calN(\wh\theta_{\BDM}^{(1)}(\tm_1), \tau^2_{\nK, \NK}(\tm_1)) - \calN(\wh\theta_{\BDM}^{(1)}(m^*), \tau^2_{\nK, \NK}(m^*))\|_{\TV} \\
~\leq~ &  \ \|\calN(\wh\theta_{\BDM}^{(1)}(\tm_1), \tau^2_{\nK, \NK}(\tm_1)) - \calN(\wh\theta_{\BDM}^{(1)}(\tm_1), \tau^2_{\nK, \NK}(m^*))\|_{\TV} \\
& ~+~ \| \calN(\wh\theta_{\BDM}^{(1)}(\tm_1), \tau^2_{\nK, \NK}(m^*)) - \calN(\wh\theta_{\BDM}^{(1)}(m^*), \tau^2_{\nK, \NK}(m^*)) \|_{\TV} \\
 ~=~ & \ \|\calN(0, \tau^2_{\nK, \NK}(\tm_1)) - \calN(0, \tau^2_{\nK, \NK}(m^*))\|_{\TV} + \| \calN(\alpha, 1) - \calN(0, 1) \|_{\TV} \\
  ~:=~ & \ T_{21} + T_{22}, \ \ \mbox{ where $\alpha = \ \{\wh\theta_{\BDM}^{(1)}(\tm_1) - \wh\theta_{\BDM}^{(1)}(m^*)\}/\tau_{\nK, \NK}(m^*)$.}
\end{align*}

We first consider the TV distance $ T_{21}$. By Lemma~\ref{klartag}, we obtain the following bound for $T_{21}$:
\begin{align}\label{T21_bound}
     T_{21} ~\equiv ~  \|\calN(0, \tau^2_{\nK, \NK}(\tm_1)) - \calN(0, \tau^2_{\nK, \NK}(m^*))\|_{\TV} ~\leq~ C \left|\frac{\tau^2_{\nK, \NK}(\tm_1) - \tau^2_{\nK, \NK}(m^*)}{\tau^2_{\nK, \NK}(m^*)} \right|,
\end{align}
for some constant $C < \infty$. By using the definitions of the terms $\tau^2_{\nK, \NK}(\tm_1)$ and $\tau^2_{\nK, \NK}(m^*)$ (refer to Theorem~\ref{bvm_on_first_half_data}), we obtain that
\begin{align*}
\|\calN\!(0, \tau^2_{\nK, \NK}(\tm_1)) - \calN\!(0, \tau^2_{\nK, \NK}(m^*))\|_{\TV} \leq \frac{C \left\{|\sigma^2_{1}(\tm_1) - \sigma^2_{1}(m^*)| + (n/N) |\sigma^2_{2}(\tm_1) - \sigma^2_{2}(m^*)| \right\}}{\sigma^2_{1}(m^*) + (n/N) \sigma^2_{2}(m^*)}.
\end{align*}
Since the denominator (on the right-hand side (RHS) above) is greater than and bounded away from zero, it is enough to show that both of the terms $
|\sigma^2_{1}(\tm_1) - \sigma^2_{1}(m^*)|$ and $|\sigma^2_{2}(\tm_1) - \sigma^2_{2}(m^*)|$ in the numerator (on the RHS above) converge to 0 in probability.
By using $\tm_1 \ind (Y, \boldX) \in \calD_1$ (refer to the construction of BDMI in Section~\ref{BCF} by taking $k = 1$), we observe that
\begin{align*}
|\sigma^2_{1}(\tm_1) - \sigma^2_{1}(m^*)| ~=~ & \ |\Var_{\boldZ}
[\{Y - \tm_1(\boldX) \hspace{0.4mm}\} | \hspace{0.4mm} \tm_1] - \Var_{\boldZ}
\{Y - m^*(\boldX)\}| \\
~\leq~ & \ |\bbE_{\boldZ}
[\{m^*(\boldX) - \tm_1(\boldX)\} \{ 2Y - \tm_1(\boldX) - m^*(\boldX)\} \hspace{1mm} | \hspace{1mm} \tm_1]| \\
& + |\bbE_{\boldZ}
[\{m^*(\boldX) - \tm_1(\boldX)\} \hspace{1mm} | \hspace{1mm} \tm_1] \hspace{0.5mm}\bbE_{\boldZ}
[\{2Y - \tm_1(\boldX) - m^*(\boldX)\} \hspace{1mm} | \hspace{1mm} \tm_1]|,
\end{align*}
where the last step uses the triangle inequality. By applying the Cauchy--Schwarz inequality and the triangle inequality for the $\bbL_2(\bbP_{\boldX})$-norm (and the underlying inner product), we have
\begin{align*}
|\sigma^2_{1}(\tm_1) - \sigma^2_{1}(m^*)| & ~\leq~ 2\| m^*(\boldX) - \tm_1(\boldX) \|_{\bbL_2(\bbP_{\boldX})}\{\| 2Y \|_{\bbL_2(\bbP_Y)} + \| 2m^*(\boldX)\|_{\bbL_2(\bbP_{\boldX})}\} \\
& ~~~~~+~ 2\| m^*(\boldX) - \tm_1(\boldX) \|^2_{\bbL_2(\bbP_{\boldX})}.
\end{align*}
Since \eqref{op_condition} holds by Lemma~\ref{lemma_half_of_data}, and $\| Y \|_{\bbL_2(\bbP_Y)} < \infty$ and $\| m^*(\boldX)\|_{\bbL_2(\bbP_{\boldX})} < \infty$ by Assumption~\ref{assumption_for_half_fold} (ii), we conclude that $
|\sigma^2_{1}(\tm_1) - \sigma^2_{1}(m^*)| \cvP 0$ under $\bbP_{\calD_1, \tm_1}$. Similarly, we have
 \begin{align*}
|\sigma^2_{2}(\tm_1) - \sigma^2_{2}(m^*)|& ~=~ |\Var\{\tm_1(\boldX) \hspace{0.4mm} | \hspace{0.4mm}  \tm_1\} - \Var\{ m^*(\boldX)\}| \\
& ~\leq~ 2 \|\tm_1(\boldX) - m^*(\boldX) \|_{\bbL_2(\bbP_{\boldX})}\{\| 2m^*(\boldX) \|_{\bbL_2(\bbP_{\boldX})} + \|\tm_1(\boldX) - m^*(\boldX) \|_{\bbL_2(\bbP_{\boldX})}\},
 \end{align*}
where the last step comes from the Cauchy-Schwarz inequality. Since $\| m^*(\boldX)\|_{\bbL_2(\bbP_{\boldX})} < \infty$ and  \eqref{op_condition} holds (by Assumption~\ref{assumption_for_half_fold} (ii)), we have $
|\sigma^2_{2}(\tm_1) - \sigma^2_{2}(m^*)| \ \cvP \ 0$ under $\bbP_{\calD_1, \tm_1}$. Referring back to the inequality \eqref{T21_bound}, and using all conclusions obtained above, and the fact that the denominator $\tau^2_{\nK, \NK}(m^*)$ in \eqref{T21_bound} is bounded away from zero, we now conclude that $T_{21} \to 0$ under $\bbP_{\calD_1, \tm_1}$. \qeds

Now, we consider $T_{22} = \|\calN(\{\wh\theta_{\BDM}^{(1)}(\tm_1) - \wh\theta_{\BDM}^{(1)}(m^*)\}/\tau_{\nK, \NK}(m^*), 1) - \calN(0, 1)\|_{\TV}$. Using Lemma~\ref{TV_2normal}, we observe that:
\begin{align}\label{T22_bound}
T_{22} ~\leq~ \frac{|\sqrt{\nK}\{\wh\theta_{\BDM}^{(1)}(\tm_1) - \wh\theta_{\BDM}^{(1)}(m^*)\}|}{\sqrt{2\pi \nK   \tau^2_{\nK, \NK}(m^*)}}.
\end{align}

Since the denominator $2 \hspace{0.5mm}\pi \hspace{0.5mm} \nK\hspace{0.5mm} \tau^2_{\nK, \NK}(m^*) = \sigma^2_{1}(m^*) + (n/N)\sigma^2_{2}(m^*)$ is bounded below and away from zero, as $n, N \to \infty$, it converges to a non--random quantity which is bounded below and away from zero. This reduces the problem to showing the numerator $|\sqrt{\nK}\{\wh\theta_{\BDM}^{(1)}(\tm_1) - \wh\theta_{\BDM}^{(1)}(m^*)\}| \cvP 0$ under $\bbP_{\wt \calD_1}$ (where recall that $\wt \calD_k \equiv \calD_k \cup \calS_k$) thanks
to the continuous mapping theorem (CMT) \citep[Theorem 2.3]{van2000asymptotic}.
We can define a continuous map $h(x, y) := xy^{-1}$ on $\bbR \times \bbR^+$ and then apply the CMT to argue that
\begin{align*}
    \frac{|\sqrt{\nK}\{\wh\theta_{\BDM}^{(1)}(\tm_1) - \wh\theta_{\BDM}^{(1)}(m^*)\}|}{\sqrt{2\pi \nK   \tau^2_{\nK, \NK}(m^*)}} \ \cvP \ 0,
\end{align*}
which implies that $
\|\calN(\{\wh\theta_{\BDM}^{(1)}(\tm_1) - \wh\theta_{\BDM}^{(1)}(m^*)\}/\tau^2_{\nK, \NK}(m^*), 1) - \calN(0, 1)\|_{\TV} \ \cvP \ 0$ under $\bbP_{\wt \calD_1}$.

Towards showing the numerator's convergence, by writing the terms explicitly, we observe that
\begin{align*}
&|\sqrt{\nK}\hspace{0.5mm}\{\wh \theta_{\BDM}^{(1)}(\tm_1) - \wh \theta_{\BDM}^{(1)}(m^*) \}| \\
& =~ \left|\sqrt{\nK} \!\left[ \frac{1}{\nK}\! \sum_{i \in \calI_1} \{Y_i - \tm_1(\boldX_i)\} + \frac{1}{\NK}\sum_{i \in \calJ_1} \tm_1(\boldX_i) -\frac{1}{\nK} \sum_{i \in \calI_1} \{Y_i - m^*(\boldX_i)\} - \frac{1}{\NK}\sum_{i \in \calJ_1} m^*(\boldX_i) \right] \right| \\
& =~ \left| \sqrt{\nK} \hspace{0.5mm} \left[ \frac{1}{\nK} \sum_{i \in \calI_1} \{m^*(\boldX_i) - \tm_1(\boldX_i)\} - \frac{1}{\NK} \sum_{i \in \calJ_1} \{m^*(\boldX_i) - \tm_1(\boldX_i)\} \right] \right|
\\
& =~ \left|\sqrt{\nK} \hspace{0.5mm} [\bbE_{\nK}^{(1)}\{m^*(\boldX) - \tm_1(\boldX)\} - \bbE_{\NK}^{(1)}\{m^*(\boldX) - \tm_1(\boldX)\} ] \right|
\\
& =~ \left| \bbG_{\nK}^{(1)}\{m^*(\boldX) - \tm_1(\boldX)\} - \frac{\sqrt{\nK}}{\sqrt{\NK}} \hspace{0.5mm} \bbG_{\NK}^{(1)} \{m^*(\boldX) - \tm_1(\boldX)\}\right|
\\
& \leq~ |\bbG_{\nK}^{(1)}\{m^*(\boldX) - \tm_1(\boldX)\}| + \frac{\sqrt{n}}{\sqrt{N}} \hspace{0.5mm} |\bbG_{\NK}^{(1)}\{m^*(\boldX) - \tm_1(\boldX)\}|,
\end{align*}
where recall the notations $\bbG_{\nK}^{(k)}(\cdot)$ and $\bbG_{\NK}^{(k)}(\cdot)$ as defined at the beginning of Section~\ref{appendix} (and here we set $k = 1$). We next want to show $\bbG_{\nK}^{(1)}\{m^*(\boldX) - \tm_1(\boldX)\} = o_{\bbP_{\calD_1, \tm_1}}(1)$. Since $\calD_1 \ind \tm_1$ by the construction of BDMI (see Section~\ref{BCF} by setting $k =1$), we observe that
\begin{align}\label{var_mtilde_op_1}
 \Var_{\boldX| \tm_1}[\{m^*(\boldX) - \tm_1(\boldX)\} \mid \tm_1] & = \Var_{\boldX}[\{m^*(\boldX) - \tm_1(\boldX)\} \hspace{1mm} | \hspace{1mm} \tm_1] ~~ \mbox{ [by }  \calD_1 \ind \tm_1] \nonumber \\
 & = \bbE_{\boldX}\!\left[\{m^*(\boldX) - \tm_1(\boldX) \}^2 \hspace{0.5mm} | \hspace{0.5mm} \tm_1 \right] - \left(\bbE_{\boldX}[\{m^*(\boldX) - \tm_1(\boldX)\} \hspace{0.5mm} | \hspace{0.5mm} \tm_1] \right)^2 \nonumber \\
& = o_{\bbP_{\tm_1}}(1),
\end{align}
where the last step
follows from \eqref{op_condition}.
Then by Chebyshev's inequality, for any $t > 0$, we have
\begin{align*}
V_{\nK}(\tm_1) & ~:=~ \bbP_{\calD_1}[\hspace{0.5mm} |\bbG_{\nK}^{(1)}\{m^*(\boldX) - \tm_1(\boldX)\}| > t \hspace{1mm} | \hspace{1mm} \tm_1 \hspace{0.5mm} ] \\
& ~\leq~ t^{-2} \hspace{0.5mm} \nK \hspace{0.5mm} \Var_{\boldX}[\hspace{0.5mm} \bbE_{\nK}^{(1)} \{m^*(\boldX) - \tm_1(\boldX)\} \hspace{1mm} | \hspace{1mm} \tm_1\hspace{0.5mm} ] \\
& ~=~ t^{-2} \hspace{0.5mm} \Var_{\boldX}[\hspace{0.5mm} \{m^*(\boldX) - \tm_1(\boldX)\} \hspace{1mm} | \hspace{1mm} \tm_1 \hspace{0.5mm} ] ~=~  o_{\bbP_{\tm_1}}(1),
\end{align*}
where the last step uses that $\bbE_{\nK}^{(1)} \{m^*(\boldX) - \tm_1(\boldX)\}$ is a sum of {\it independent} random variables given $\tm_1$, and the earlier conclusion obtained above in \eqref{var_mtilde_op_1}. Hence, we showed that
$V_{\nK}(\tm_1) \cvP 0$ under $\bbP_{\tm_1}$. This equivalently gives that $ V_{\nK}(\tm_1) = O_{\bbP_{\tm_1}}(c_{\nK, \sK})$, for some $c_{\nK, \sK} \to 0$.

We here also note that double index in $c_{\nK, \sK}$ signifies that the rate depends on both $\nK$ and the size $\sK$ of $\calS_1$ which is used to obtain $\Pi_{\mbm}^{(1)}$. Then by applying Lemma~\ref{chernozhukov_lemma6_1} (b), we obtain that $\bbG_{\nK}^{(1)}\{m^*(\boldX) - \tm_1(\boldX) \} = O_{\bbP_{\calD_1, \tm_1}}(c_{\nK, \sK})$ which implies that $\bbG_{\nK}^{(1)}\{m^*(\boldX) - \tm_1(\boldX) \} = o_{\bbP_{\calD_1, \tm_1}}(1)$.

By following similar steps as above (for $\bbG_{\nK}^{(1)}\{m^*(\boldX) - \tm_1(\boldX)\}$), we have the same conclusion for $\bbG_{\NK}^{(1)}\{m^*(\boldX) - \tm_1(\boldX)\}$, i.e., $\bbG_{\NK}^{(1)}\{m^*(\boldX) - \tm_1(\boldX)\} = o_{\bbP_{\calD_1, \tm_1}}(1)$. To be precise, since the condition (\ref{op_condition}) holds and $\calD_1 \ind \tm_1$ (in particular, $\boldX \in \calD_1$ $\ind \tm_1$), by using \eqref{var_mtilde_op_1}, we first obtain that
\begin{align*}
   W_{\NK}(\tm_1) & ~:=~ \bbP_{\calD_1}[\hspace{0.5mm} |\bbG_{\NK}^{(1)} \{m^*(\boldX) - \tm_1(\boldX)\}| > t \mid \tm_1 \hspace{0.5mm} ] ~~~\mbox{for any $t > 0$,}\\
   & ~ ~\leq~
   t^{-2} \hspace{0.7mm} \Var[ \{ m^*(\boldX) - \tm_1(\boldX) \} \mid \tm_1] ~=~ o_{\bbP_{\tm_1}}(1).
\end{align*}
This implies that for some $h_{\NK, \sK} \to 0$, $W_{\NK}(\tm_1) = O_{\bbP_{\tm_1}}(h_{\NK, \sK})$. We note that the double index in $h_{\NK, \sK}$ reveals the dependency of the rate on both $\NK$ and the size $\calS_1$ (used to obtain $\Pi_{\mbm}^{(1)}$ of $\tm_1$). Then by applying Lemma~\ref{chernozhukov_lemma6_1} (b), we obtain that $\bbG_{\NK}^{(1)} \{m^*(\boldX) - \tm_1(\boldX)\} = O_{\bbP_{\calD_1, \tm_1}}(h_{\NK, \sK})$ which implies that $\bbG_{\NK}^{(1)} \{m^*(\boldX) - \tm_1(\boldX)\} = o_{\bbP_{\calD_1, \tm_1}}(1)$. Referring back to the inequality \eqref{T22_bound}, and using all conclusions above along with the fact that the denominator $\nK \tau^2_{\nK, \NK}(m^*)$ in \eqref{T22_bound} is bounded away from zero, we now conclude that $T_{22} \cvP 0$ under $\bbP_{\calD_1, \tm_1}$. Hence, the entire proof of the first part of Theorem~\ref{bvm_on_first_half_data} is now completed (assuming Lemma~\ref{lemma_half_of_data} holds, as shown later in Section~\ref{proof_of_lemma_half_of_data}).  \qeds

Lastly, the second part of Theorem~\ref{bvm_on_first_half_data} immediately follows from the invariance property of the TV distance (refer to Lemma~\ref{TV_invariance}), by setting $h = \sqrt{\nK}(\theta - \theta_0)$. Specifically, let $\Pi^{(k)}_{\bh}$ be the posterior  of $h$. Then, using the invariance property of the TV distance from Lemma~\ref{TV_invariance}, we have
\begin{align*}
\big \| \Pi_{\btheta}^{(k)} - \calN(\wh \theta_{\BDM}^{(k)}(m^*), \tau^2_{\nK, \NK}(m^*)) \big\|_{\TV} =  \big\| \Pi_{\bh}^{(k)} - \calN(\sqrt{\nK} \{
\wh \theta_{\BDM}^{(k)}(m^*) - \theta_0 \}, \nK \tau^2_{\nK, \NK}(m^*)) \big\|_{\TV}.
\end{align*}
We already showed that the left-hand side of the equality above converges to 0 in probability under $\bbP_{\calD_k}$. This directly gives the second claim of Theorem~\ref{bvm_on_first_half_data} and completes the proof of the entire result. \qeds

\subsection{Proof of Theorem~\ref{main_thm}}\label{proof_main_thm}

Let $\wt\theta_{\BDM}$ be a new random variable defined as $\wt\theta_{\BDM} := K \hspace{0.5mm} \theta_{\BDM} = \sum_{k = 1}^K \theta_k$ where $\theta_1, \dots, \theta_k$ are independent random variables from the corresponding posteriors $\Pi_{\btheta}^{(1)}, \dots, \Pi_{\btheta}^{(K)}$ (refer to \eqref{eq:FINALposteriorBDMI}). Let $\wt\Pi_{\btheta}$ be the posterior distribution of $\wt\theta_{\BDM}$. Then, by using the invariance property of the TV distance from Lemma~\ref{TV_invariance}, to prove Theorem~\ref{main_thm}, it suffices to show the following:
\begin{align*}
\left\| \wt\Pi_{\btheta} - \calN(K\hspace{0.7mm}\wh \theta_{\BDM}(m^*), K^2\hspace{0.7mm}\tau^2_{n, N}(m^*)) \right\|_{\TV} \ \to \ 0 ~ \text{ in probability w.r.t.} \ \ \bbP_{\calD}.
\end{align*}
By using Lemma~\ref{TVD_convolution} and the construction of $\wt\theta_{\BDM}$, we first obtain that
\begin{align*}
T & ~:=~   \big \|\wt\Pi_{\btheta} - \calN(K\wh\theta_{\BDM}(m^*), K^2\tau^2_{n, N}(m^*)) \big\|_{\TV} \\
 & ~\leq~   \big\| \Pi_{\btheta}^{(1)} - \calN(\wh \theta_\BDM^{(1)}(m^*), \tau^2_{\nK, \NK}(m^*)) \big\|_{\TV} + \dots + \big\|\Pi_{\btheta}^{(K)} - \calN(\wh \theta_\BDM^{(K)}(m^*), \tau^2_{\nK, \NK}(m^*))  \big\|_{\TV} \\
& \ \  ~~~~ := \ T_1 + \dots + T_K,
\end{align*}
where $\nK = n/K,\NK = N/K$ and for $k = 1, \dots , K$, $T_k = \big\|\Pi_{\btheta}^{(k)} - \calN(\wh \theta_\BDM^{(k)}(m^*), \tau^2_{\nK, \NK}(m^*))  \big\|_{\TV}$,
\begin{align*}
\wh \theta_\BDM^{(k)}(m^*) \ = \  \frac{1}{\nK} \sum_{i \in \calI_k} \{Y_i - m^*(\boldX_i)\} + \frac{1}{\NK} \sum_{i \in \calJ_k} m^*(\boldX_i) \ \text{ and } \ \tau^2_{\nK,\NK}(m^*) \ = \ \frac{\sigma^2_{1}(m^*)}{\nK} + \frac{\sigma^2_{2}(m^*)}{\NK}.
\end{align*}
Under the assumptions of Theorem~\ref{main_thm}, we can apply Theorem~\ref{bvm_on_first_half_data} to each of the $\TV$ distances $T_1, \dots, T_K$ defined above. This gives that $T_k \cvP 0$ under $\bbP_{\calD}$ for $k = 1, \dots , K$. Since each $\TV$ distance converges to 0 in probability and $K$ is {\it fixed}, we finally obtain that $T \cvP 0$ under $\bbP_{\calD}$. Hence, by using the invariance property of the TV distance from Lemma~\ref{TV_invariance}, we first observe that
\begin{align*}
    \big\|\Pi_{\btheta} - \calN(\wh\theta_{\BDM}(m^*), \tau^2_{n, N}(m^*)) \big\|_{\TV} \ = \  \big\|\wt \Pi_{\btheta} - \calN(K \hspace{0.5mm}\wh\theta_{\BDM}(m^*), K^2 \hspace{0.5mm}\tau^2_{n, N}(m^*) ) \big\|_{\TV}.
\end{align*}
Since we already proved that the RHS of the equality above converges to $0$ in probability w.r.t. $\bbP_{\calD}$, we immediately conclude that $
\big\|\Pi_{\btheta} - \calN(\wh\theta_{\BDM}(m^*), \tau^2_{n, N}(m^*))
\big\|_{\TV}  \to  0$ in probability w.r.t. $\bbP_{\calD}$.
\qeds

\subsection{Proof of Corollary~\ref{corollory_asymp_equiv_postmean}}\label{proof_of_corr_asymp_equiv_pmean}

We first define a new random variable $Z_n \equiv \sqrt{n}(\hspace{0.5mm} \theta_{\BDM} - \theta_0 \hspace{0.5mm})$ with corresponding posterior distribution $\calP_n(\calD)$ where $\theta_{\BDM}$ is as defined in \eqref{eq:FINALposteriorBDMI}. Let $\calP(\calD)$ be the corresponding limiting Normal distribution with mean $\sqrt{n}\{\hspace{0.5mm}\wh \theta_{\BDM}(m^*) - \theta_0 \hspace{0.5mm} \}$ and variance $n \hspace{0.5mm} \tau^2_{n, N}(m^*)$ obtained from Theorem~\ref{main_thm} (after applying the appropriate scaling and location shifts, in particular, using Lemma~\ref{TV_invariance}) for the posterior $\calP_n(\calD)$ and let $Z \sim \calP(\calD)$. Note that the distributions $\calP_n \equiv \calP_n(\calD)$ and $\calP \equiv \calP(\calD)$ are random through the data $\calD$. Next, observe that
\begin{align*}
 & \sqrt{n}\{\hspace{0.5mm} \wh\theta_{\BDM}(\tm_{\CF}) - \theta_0 \hspace{0.5mm} \} - \sqrt{n}\{\hspace{0.5mm} \wh \theta_{\BDM}(m^*) - \theta_0 \hspace{0.5mm} \} \ = \ o_{\bbP_{\calD}}(1) \\
 & ~~ \Leftrightarrow \ |\sqrt{n} \hspace{0.5mm}  \{ \hspace{0.5mm} \wh\theta_{\BDM}(\tm_{\CF}) - \theta_0 \hspace{0.5mm} \} - \sqrt{n}\hspace{0.5mm} \{ \hspace{0.5mm} \wh \theta_{\BDM}(m^*) - \theta_0 \hspace{0.5mm} \} | \ \to \ 0 \ \text{ in probability under } \ \bbP_{\calD} \\
 & ~~ \Leftrightarrow \ | \bbE_{Z_n \sim \calP_n(\calD)}(Z_n \mid \calD) - \bbE_{Z \sim \calP(\calD)}(Z \mid \calD) |  \ \to \ 0 \  \mbox{ in probability under $\bbP_{\calD}$},
 \end{align*}
where the last line uses the constructions of the random variables $Z_n$ and $Z$. To use the exact formula of the TV distance between two Normal distributions with the same variance (refer to \eqref{eq:TV_two_gaussians}, specifically, see Lemma~\ref{TV_2normal}), we now consider the following Normal distribution: $ \wt\calP \equiv \wt\calP(\calD):= \calN(\sqrt{n} \hspace{0.5mm}\{\wh\theta_{\BDM}(\tm_{\CF}) - \theta_0\}, n \hspace{0.5mm} \tau^2_{n, N}(m^*))$. We note that the distributions $\wt\calP$ and $\calP_n$ have the same mean $\sqrt{n} \hspace{0.5mm} \{\hspace{0.5mm}  \wh\theta_{\BDM}(\tm_{\CF}) - \theta_0 \hspace{0.5mm} \}$, while the distributions $\wt\calP$ and $\calP$ have the same variance $n \hspace{0.5mm} \tau^2_{n,N}(m^*)$. Let $\wt Z \sim \wt\calP(\calD) $. Then, we observe that
\begin{align*}
    |\bbE_{Z_n \sim \calP_n}(Z_n | \calD) - \bbE_{Z \sim \calP}(Z | \calD)| & ~ = ~   |\bbE_{Z_n \sim \calP_n}(Z_n | \calD) - \bbE_{\wt Z \sim \wt\calP}(\wt Z | \calD) + \bbE_{\wt Z \sim \wt\calP}(\wt Z | \calD) - \bbE_{Z \sim \calP}(Z | \calD)| \\
    &  ~ = ~ |\bbE_{\wt Z \sim \wt\calP}(\wt Z | \calD) - \bbE_{Z \sim \calP}(Z | \calD)|,
\end{align*}
where the last step uses the fact $\bbE_{Z_n \sim \calP_n}(Z_n | \calD) = \bbE_{\wt Z \sim \wt\calP}(\wt Z | \calD)$. Since both distributions $\wt\calP$ and $\calP$ are Normal with the same variance $n\hspace{0.5mm} \tau^2_{n,N}(m^*)$, by the invariance property of the TV distance from Lemma~\ref{TV_invariance}, we have:
\begin{align}
  \|\calP - \wt\calP \|_{\TV} & ~\equiv ~  \|\calP(\calD) - \wt\calP(\calD) \|_{\TV} \nonumber \\
 & ~ = ~  \big \|\calN(\sqrt{n}\{\wh\theta_{\BDM}(\tm_{\CF}) - \theta_0 \}, n \tau^2_{n, N}(m^*)) - \calN(\sqrt{n}\{\wh\theta_{\BDM}(m^*) - \theta_0 \}, n \tau^2_{n, N}(m^*)) \big\|_{\TV} \nonumber \\
 &  ~ = ~   \big \| \calN(\alpha, 1) - \calN(0, 1) \big \|_{\TV}, \nonumber
\end{align}
where $\alpha = [\hspace{0.5mm} \sqrt{n}\{\wh\theta_{\BDM}(\tm_{\CF}) - \theta_0\} - \sqrt{n}\{\wh\theta_{\BDM}(m^*) - \theta_0\} \hspace{0.5mm}]/\sqrt{n \tau^2_{n, N}(m^*)}$. Then, by Lemma~\ref{TV_2normal},
\begin{align} \label{eq:TV_two_gaussians}
\|\calP(\calD) - \wt\calP(\calD) \|_{\TV} ~ = ~ 2\hspace{0.5mm} \Phi\left(\frac{|\sqrt{n}\{\wh\theta_{\BDM}(\tm_{\CF}) - \theta_0\} - \sqrt{n}\{\wh\theta_{\BDM}(m^*) - \theta_0\}|}{2 \hspace{0.5mm}  \sqrt{n \hspace{0.5mm}  \tau^2_{n, N}(m^*)}}\right) - 1,
\end{align}
where $\Phi(\cdot)$ is the CDF of the standard Normal distribution $\calN(0, 1)$. Since the term
 $n \hspace{0.5mm} \tau^2_{n, N}(m^*) = \sigma^2_{1}(m^*) + (n/N) \sigma^2_{2}(m^*)$ in the denominator (on the RHS in \eqref{eq:TV_two_gaussians}) is greater than and away from zero, \eqref{eq:TV_two_gaussians} implies that to complete the proof of Corollary~\ref{corollory_asymp_equiv_postmean}, it is enough to show $\|\calP(\calD) - \wt\calP(\calD) \|_{\TV} \cvP 0$ under $\bbP_{\calD}$. Towards that, we first use the same approach used in the proof of Theorem~\ref{main_thm} in Section~\ref{proof_main_thm}. More explicitly, we write both of the Normal distributions $\wt \calP(\calD)$ and $\calP(\calD)$ as convolutions of $K$ Normal distributions as follows:
\begin{align}
  \wt \calP(\calD) ~=~ & \calN(\sqrt{n}\{\hspace{0.5mm}\wh\theta_{\BDM}(\tm_{\CF}) - \theta_0 \hspace{0.5mm}\}, n \tau^2_{n, N}(m^*)) ~ = ~ \wt \calP^{(1)}(\wt \calD_1) * \dots * \wt \calP^{(K)}(\wt \calD_K), ~ \mbox{ and} \nonumber \\
  \calP(\calD) ~=~ & \calN(\sqrt{n}\{\hspace{0.5mm} \wh\theta_{\BDM}(m^*) - \theta_0 \hspace{0.5mm} \}, n \tau^2_{n, N}(m^*)) ~ = ~ \calP^{(1)}(\wt \calD_1) *\dots * \calP^{(K)}(\wt \calD_K), \nonumber
\end{align}
where for $k = 1, \dots, K$, $\wt \calP^{(k)}(\wt \calD_k) := \calN(\sqrt{n} \hspace{0.5mm} \{\wh\theta_{\BDM}^{(k)}(\tm_k)- \theta_0\}/K, n \hspace{0.5mm} \tau^2_{\nK, \NK}(m^*)/K^2)$ and $\calP^{(k)}(\wt \calD_k) := \calN(\sqrt{n} \hspace{0.5mm} \{\wh\theta_{\BDM}^{(k)}(m^*)- \theta_0\}/K, n \hspace{0.5mm} \tau^2_{\nK, \NK}(m^*)/K^2)$ with the parameters are as given in \eqref{supp:notations_half_data} and in Theorem~\ref{bvm_on_first_half_data}.

Then, by applying Lemma~\ref{TVD_convolution}, we obtain that
\begin{align*}
\big\| \hspace{0.5mm} \calP(\calD) - \wt\calP(\calD) \hspace{0.5mm} \big \|_{\TV} ~ \leq ~ \big \|\hspace{0.5mm} \calP^{(1)}(\wt\calD_1) - \wt\calP^{(1)}(\wt\calD_1) \hspace{0.5mm} \big \|_{\TV} ~ + ~ \cdots ~ + ~ \big\| \hspace{0.5mm} \calP^{(K)}(\wt\calD_K) - \wt\calP^{(K)}(\wt\calD_K) \hspace{0.5mm} \big\|_{\TV}.
\end{align*}
Furthermore, in the proof Theorem~\ref{bvm_on_first_half_data} (refer to Section~\ref{proof_bvm_half_data}), we have already established a result showing that $T_{22} = \big \| \calN(\wh\theta_{\BDM}^{(1)}(\tm_1), \tau^2_{\nK, \NK}(m^*)) - \calN(\wh\theta_{\BDM}^{(1)}(m^*), \tau^2_{\nK, \NK}(m^*)) \big\|_{\TV} \cvP 0$ under $\bbP_{\wt\calD_1}$.

Therefore, by using the invariance property of the TV distance from Lemma~\ref{TV_invariance}, for each $k = 1, \dots, K$, we obtain that $\big \| \calP^{(k)}(\wt\calD_k) - \wt\calP^{(k)}(\wt\calD_k) \big\|_{\TV} \cvP 0$ in under $\bbP_{\wt \calD_k}$. This immediately leads to the conclusion that $\big \|\calP(\calD) - \wt\calP(\calD) \big\|_{\TV} \cvP 0$ under $\bbP_{\calD}$. Hence, by using the equality in \eqref{eq:TV_two_gaussians} (and that the denominator on the RHS in \eqref{eq:TV_two_gaussians} is bounded away from zero),
we conclude that $|\sqrt{n}\hspace{0.5mm}\{\hspace{0.5mm}\wh\theta_{\BDM}(\tm_{\CF}) - \theta_0 \hspace{0.5mm} \} - \sqrt{n}\hspace{0.5mm}\{\hspace{0.5mm} \wh \theta_{\BDM}(m^*) - \theta_0 \hspace{0.5mm}\} | \cvP 0$ under $\bbP_{\calD}$, which completes the proof of the result. \qeds

\subsection{Proof of Theorem~\ref{bvm_standard}}\label{proof_bvm_standard}

For notational simplicity, we set $k = 1$ w.l.o.g. and present the proof below for $k = 1$.

Let $Q = \calN \big(\wh \theta_{\BDM}^{(1)}(m^*), \tau^2_{\nK, \NK}(m^*)\big)$. Then let $\pi_{\btheta}^{(1)}(\cdot)$ and $q(\cdot)$ be the pdfs of the distributions $\wt\Pi_{\btheta}^{(1)}$ and $Q$, respectively. Then, by the integral representation of TV distance, we have
\begin{align*}
\big \|\hspace{0.5mm} \wt \Pi_{\btheta}^{(1)} - Q \hspace{0.5mm}   \big \|_{\TV} & ~=~ \frac{1}{2} \int \big |\pi_{\btheta}^{(1)}(\theta) - q(\theta) \big| \hspace{0.7mm} \dd \theta ~=~ \frac{1}{2} \int \left| \int \pi(\theta \mid \tm_1, \calD_1) \hspace{0.5mm} \pi_{\mbm}^{(1)}(\tm_1)\hspace{0.5mm} \dd \tm_1  - q(\theta) \right| \hspace{0.3mm} \dd \theta \\
&  ~\leq~  \frac{1}{2} \int \int \big| \pi(\theta \mid \tm_1, \calD_1) \hspace{0.5mm} \pi_{\mbm}^{(1)}(\tm_1) - q(\theta) \big|  \hspace{0.7mm} \dd \tm_1 \hspace{0.7mm} \dd \theta.
\end{align*}
Then, by using Fubini's theorem and the integral representation of the TV distance, we obtain that
\begin{align*}
\big \|\hspace{0.5mm} \wt\Pi_{\btheta}^{(1)} - Q  \hspace{0.5mm} \big\|_{\TV} & ~\leq~  \frac{1}{2}  \int \int | \pi(\theta \mid \tm_1, \calD_1)  - q(\theta)|  \hspace{0.7mm} \dd \theta \hspace{0.7mm} \pi_\mbm^{(1)}(\tm_1) \hspace{0.7mm} \dd \tm_1 \\
& ~=~ \int \big\|\hspace{0.5mm} \Pi_{(\btheta \mid \tm_1, \calD_1)} - Q \hspace{0.5mm}\big\|_{\TV} \hspace{0.5mm} \pi_\mbm^{(1)}(\tm_1) \hspace{0.7mm} \dd \tm_1 \\
& ~ = ~  \bbE_{\tm_1 \sim \Pi_{\mbm}^{(1)}} \{\hspace{0.5mm} \| \hspace{0.5mm} \Pi_{(\btheta \mid \tm_1, \calD_1)}  - Q \hspace{0.5mm} \|_{\TV} \mid \calS_1 \hspace{0.5mm} \} ~:=~ \bbE_{\tm_1 \sim \Pi_{\mbm}^{(1)}}\{\hspace{0.5mm} T \mid \calS_1 \hspace{0.5mm} \}.
\end{align*}
Then, let $P$ be a Normal distribution with mean $\wh \theta_{\BDM}^{(1)}(\tm_1)$ and variance $\tau_{\nK,\NK}^2(\tm_1)$, denoted as $P \equiv \calN(\wh \theta_{\BDM}^{(1)}(\tm_1), \tau_{\nK,\NK}^2(\tm_1))$. Then, by applying the triangle inequality, we observe that
\begin{align*}
 \bbE_{\tm_1 \sim \Pi_{\mbm}^{(1)}}(T  \mid \calS_1) & ~\leq~  \bbE_{\tm_1 \sim \Pi_{\mbm}^{(1)}}(\| \Pi_{(\btheta \mid \tm_1, \calD_1)}  - P \|_{\TV} \mid  \calS_1) +  \bbE_{\tm_1 \sim \Pi_{\mbm}^{(1)}}(\| P - Q \|_{\TV} \mid \calS_1) ~:=~ T_1 + T_2.
\end{align*}
Thus, it is enough to show
that both $T_1$ and $T_2$ converge to 0 in probability w.r.t. $\bbP_{\wt\calD_1}$ to complete the proof.

We first consider
$T_1$. By Proposition~\ref{prop_half_data}, we have $\Pi_{(\btheta \mid \tm_1, \calD_1)}$ is a convolution of two $t$-distributions $t_{\nu_{\nK}}(\mu_{\nK}(\tm_1), \wh \sigma^2_{1, \nK}(\tm_1)/\nK)$ and $t_{\nu_{\NK}}(\mu_{\NK}(\tm_1), \wh \sigma^2_{2, \NK}(\tm_1)/\NK)$, where the parameters are as defined in \eqref{supp:notations_half_data} (by setting $k = 1$ therein). By using the invariance property of the TV distance (see Lemma~\ref{TV_invariance}), we observe the following:
\begin{align*}
T_1 & ~=~ \bbE_{\tm_1 \sim \Pi_{\mbm}^{(1)}} \big\{ \hspace{0.5mm} \|\hspace{0.5mm} \Pi_{(\btheta \mid \tm_1, \calD_1)}  - P \hspace{0.5mm} \|_{\TV} \mid \calS_1 \hspace{0.5mm} \big\} \\
& ~\leq~  \bbE_{\tm_1 \sim \Pi_{\mbm}^{(1)}} \big \{\hspace{0.5mm} \big \|\hspace{0.5mm} t_{\nu_{\nK}}(\mu_{\nK}(\tm_1), \wh \sigma^2_{1, \nK}(\tm_1)/\nK) - \calN(\mu_{\nK}(\tm_1), \sigma^2_{1}(\tm_1)/\nK)\hspace{0.5mm} \big \|_{\TV}\mid \calS_1 \hspace{0.5mm} \big \}  \nonumber \\
& ~~~~+~  \bbE_{\tm_1 \sim \Pi_{\mbm}^{(1)}} \big \{ \hspace{0.5mm} \big \|\hspace{0.5mm} t_{\nu_{\NK}}(\mu_{\NK}(\tm_1), \wh \sigma^2_{2, \NK}(\tm_1)/\NK) - \calN(\mu_{\NK}(\tm_1), \sigma^2_{2}(\tm_1)/\NK)\hspace{0.5mm} \big \|_{\TV} \mid \calS_1 \hspace{0.5mm} \big\} \nonumber\\
& ~=~  \bbE_{\tm_1 \sim \Pi_{\mbm}^{(1)}} \big \{ \hspace{0.5mm} \big \|\hspace{0.5mm} t_{\nu_{\nK}}(0,\wh\sigma^2_{1, \nK}(\tm_1)) - \calN(0, \sigma^2_{1}(\tm_1)) \hspace{0.5mm} \big\|_{\TV} \mid \calS_1 \hspace{0.5mm} \big\} \\
& ~~~~+~ \bbE_{\tm_1 \sim \Pi_{\mbm}^{(1)}} \big\{\hspace{0.5mm} \big \| \hspace{0.5mm} t_{\nu_{\NK}}(0, \wh\sigma^2_{2, \NK}(\tm_1)) - \calN(0, \sigma^2_{2}(\tm_1)) \hspace{0.5mm} \big\|_{\TV} \mid \calS_1 \hspace{0.5mm} \big \} \\
&  ~:=~ \bbE_{\tm_1 \sim \Pi_{\mbm}^{(1)}} \{ \hspace{0.5mm} T_{11} \mid \calS_1 \hspace{0.5mm} \} + \bbE_{\tm_1 \sim \Pi_{\mbm}^{(1)}}\{ \hspace{0.5mm} T_{12} \mid \calS_1 \hspace{0.5mm} \}, ~~\mbox{where}
\end{align*}
$T_{11} := \big\|t_{\nu_{\nK}}(0,\wh\sigma^2_{1, \nK}(\tm_1)) - \calN(0, \sigma^2_{1}(\tm_1)) \big\|_{\TV}$ and $T_{12} := \big \|t_{\nu_{\NK}}(0, \wh\sigma^2_{2, \NK}(\tm_1)) - \calN(0, \sigma^2_{2}(\tm_1)) \big \|_{\TV}$. By the definition of the $\TV$ distance, we have $0 \leq T_{11} \leq 1$ and  $0 \leq T_{12} \leq 1$. Thus if we show both $T_{11}$ and $T_{12}$ converge to 0 in probability w.r.t. $\bbP_{\calS_1}$, then by applying the DCT (or Lemma~\ref{chernozhukov_lemma6_1} (b)) we conclude that $T_1$ converges to 0 in probability under $\bbP_{\wt\calD_1}$. We recall that in the proof of Theorem~\ref{bvm_on_first_half_data} (see Section~\ref{proof_bvm_half_data}) we have already established that both $T_{11}$ and $T_{12}$ converge to 0 in probability $\bbP_{\wt\calD_1}$. Therefore, by following the same steps in the proof of Theorem~\ref{bvm_on_first_half_data} for the analysis of $T_{11}$ and $T_{12}$, we can conclude that both $T_{11}$ and $T_{12}$ converge to 0 in probability $\bbP_{\wt\calD_1}$ which implies that $T_{1}$ converges to 0 in probability w.r.t. $\bbP_{\wt\calD_1}$.  \qeds

\par\medskip
Next, we consider $T_2$. By using the triangle inequality, we first observe that
\begin{align*}
T_2 & ~=~ \bbE_{\tm_1 \sim \Pi_{\mbm}^{(1)}} \big\{ \big \|\calN(\wh \theta_{\BDM}^{(1)}(\tm_1), \tau_{\nK, \NK}^2(\tm_1)) - \calN(\wh \theta_{\BDM}^{(1)}(m^*), \tau_{\nK, \NK}^2(m^*)) \big\|_{\TV} \mid \calS_1 \big\}\\
& ~\leq ~ \bbE_{\tm_1 \sim \Pi_{\mbm}^{(1)}} \big\{ \big\|\calN(\wh \theta_{\BDM}^{(1)}(\tm_1), \tau_{\nK, \NK}^2(\tm_1)) - \calN(\wh \theta_{\BDM}^{(1)}(\tm_1), \tau_{\nK, \NK}^2(m^*)) \big\|_{\TV} \mid \calS_1 \big\} \\
& ~~~~+ ~ \bbE_{\tm_1 \sim \Pi_{\mbm}^{(1)}} \big\{ \big\|\calN(\wh \theta_{\BDM}^{(1)}(\tm_1), \tau_{\nK, \NK}^2(m^*)) - \calN(\wh \theta_{\BDM}^{(1)}(m^*), \tau_{\nK, \NK}^2(m^*)) \big\|_{\TV} \mid \calS_1 \big\}
\end{align*}
Further, by using the invariance property of the TV distance in Lemma~\ref{TV_invariance}, we obtain that
\begin{align*}
T_2 &  ~\leq ~ \bbE_{\tm_1 \sim \Pi_{\mbm}^{(1)}}\big\{ \big\|\calN(0, \tau_{\nK, \NK}^2(\tm_1)) - \calN(0, \tau_{\nK, \NK}^2(m^*)) \big \|_{\TV} \mid \calS_1 \big\} \\
& ~~~~~+ ~ \bbE_{\tm_1 \sim \Pi_{\mbm}^{(1)}} \big\{ \big\|\calN(\alpha, 1) - \calN(0, 1) \big\|_{\TV} \mid \calS_1 \big\} \\
& ~ := ~ T_{21} + T_{22}, ~~\mbox{where $\alpha = \sqrt{\nK} \hspace{0.5mm} \{\wh \theta_{\BDM}^{(1)}(\tm_1) - \wh \theta_{\BDM}^{(1)}(m^*)\}/\tau_{\nK, \NK}(m^*)$.}
\end{align*}
We first consider $T_{21} := \bbE_{\tm_1 \sim \Pi_{\mbm}^{(1)}} \big \{ \big\| \calN(0, \tau_{\nK, \NK}^2(\tm_1)) - \calN(0, \tau_{\nK, \NK}^2(m^*)) \big\|_{\TV} \hspace{0.5mm} | \hspace{0.5mm} \calS_1 \big\} $. Then by following the same algebraic steps in the proof of Theorem~\ref{bvm_on_first_half_data} (see Section~\ref{proof_bvm_half_data}), we have that
\begin{align*}
\big\| \calN(0, \tau_{\nK, \NK}^2(\tm_1)) - \calN(0, \tau_{\nK, \NK}^2(m^*)) \|_{\TV} ~\leq ~ C\hspace{0.5mm} \big\| \tm_1(\boldX) - m^*(\boldX)\|_{\bbL_2(\bbP_{\boldX})},
\end{align*}
for some fixed constant $C < \infty$. This implies that
$T_{21} \leq \bbE_{\tm_1 \sim \Pi_{\mbm}^{(1)}}(C\hspace{0.5mm} \| \tm_1(\boldX) - m^*(\boldX)\|_{\bbL_2(\bbP_{\boldX})}  | \hspace{0.5mm} \calS_1).
$
Hence, using the nuisance Bayes risk condition in Theorem~\ref{bvm_standard}, we conclude $T_{21} \cvP 0$ w.r.t. $\bbP_{\wt\calD_1}$. \qeds

\par\medskip
We now consider $T_{22} = \bbE_{\tm_1 \sim \Pi_{\mbm}^{(1)}}(\|\calN(\alpha, 1) - \calN(0, 1)\|_{\TV} \mid \calS_1)$, where $\alpha = \sqrt{\nK}\{\wh \theta_{\BDM}^{(1)}(\tm_1) - \wh \theta_{\BDM}^{(1)}(m^*)\}/\tau_{\nK, \NK}(m^*)$. We first observe that by using Lemma~\ref{TV_2normal}, we obtain that
\begin{align*}
    T_{22} ~\leq~ \bbE_{\tm_1 \sim \Pi_{\mbm}^{(1)}}  \left(\frac{|\sqrt{\nK} \hspace{0.5mm} \{\wh \theta_{\BDM}^{(1)}(\tm_1) - \wh \theta_{\BDM}^{(1)}(m^*)\}|}{\sqrt{2\hspace{0.5mm} \pi \hspace{0.5mm}  \nK \hspace{0.5mm} \tau_{\nK, \NK}^2(m^*)}} \hspace{0.7mm}\big| \hspace{0.5mm} \calS_1\right).
\end{align*}
Since the denominator $\sqrt{2\pi \hspace{0.3mm} \nK \hspace{0.3mm} \tau_{\nK, \NK}^2(m^*)}$ (on the RHS of the inequality above) is a non--random quantity which is greater than and away from 0, for some constant $0 < C < \infty$, we obtain that
$$
T_{22} ~\leq~ C \hspace{0.7mm} \bbE_{\tm_1 \sim \Pi_{\mbm}^{(1)}} [ \hspace{0.5mm} |\sqrt{\nK} \hspace{0.7mm} \{\wh \theta_{\BDM}^{(1)}(\tm_1) - \wh \theta_{\BDM}^{(1)}(m^*)\} \mid \calS_1 \hspace{0.5mm} ].
$$
By writing the terms $\wh \theta_{\BDM}^{(1)}(\tm_1)$ and $\wh \theta_{\BDM}^{(1)}(m^*)$ explicitly and using the triangle inequality, we obtain the following bound for $T_{22}$:
\begin{equation*}
T_{22} ~\leq~ C \left( \bbE_{\tm_1 \sim \Pi_{\mbm}^{(1)}} [ \hspace{0.5mm} |\bbG_{\nK}^{(1)} \{ \tm_1(\boldX) - m^*(\boldX) \} |\hspace{0.5mm}  | \hspace{0.5mm} \calS_1 \hspace{0.5mm} ] + \sqrt{\frac{n}{N}} \hspace{0.5mm} \bbE_{\tm_1 \sim \Pi_{\mbm}^{(1)}} [ \hspace{0.5mm} |\bbG_{\NK}^{(1)} \{ \tm_1(\boldX) - m^*(\boldX) \} |\hspace{0.5mm}  | \hspace{0.5mm} \calS_1 \hspace{0.5mm} ] \right),
\end{equation*}
where we recall the notations $\bbG_{\nK}^{(k)}(\cdot)$ and $\bbG_{\NK}^{(k)}(\cdot)$ as defined in Section~\ref{appendix} (and here take $k=1$). Further, it is clear that the analyses of $\bbE_{\tm_1 \sim \Pi_{\mbm}^{(1)}} [ \hspace{0.5mm} |\bbG_{\nK}^{(1)} \{ \tm_1(\boldX) - m^*(\boldX) \} | \mid \calS_1 \hspace{0.5mm} ]$ and $\bbE_{\tm_1 \sim \Pi_{\mbm}^{(1)}} [ \hspace{0.5mm} |\bbG_{\NK}^{(1)} \{ \tm_1(\boldX) - m^*(\boldX) \} | \mid \calS_1 \hspace{0.5mm} ]$ will follow the same steps by their definitions. Therefore, it suffices to show that $\bbE_{\tm_1 \sim \Pi_{\mbm}^{(1)}} [ \hspace{0.5mm} |\bbG_{\nK}^{(1)} \{ \tm_1(\boldX) - m^*(\boldX) \} |\hspace{0.5mm}  | \hspace{0.5mm} \calS_1 \hspace{0.5mm} ] \to 0$ in probability w.r.t. $\bbP_{\wt\calD_1}$ to conclude that $T_{22} \to 0$ in probability w.r.t. $\bbP_{\wt\calD_1}$.

By using the definition of convergence in probability, we want to show that: for any constant $t > 0$,
\begin{align*}
    \bbP_{\wt\calD_1} \left( \bbE_{\tm_1 \sim \Pi_{\mbm}^{(1)}} \big [ \hspace{0.5mm} |\bbG_{\nK}^{(1)} \{ \tm_1(\boldX) - m^*(\boldX) \} | \mid \calS_1 \hspace{0.5mm} \big ] > t \right) \ \to \ 0.
\end{align*}
Towards this, for notational convenience, let us define $\calZ_{\nK}(\tm_1, \calS_1):= [\bbG_{\nK}^{(1)} \{ \tm_1(\boldX) - m^*(\boldX)\} \mid \calS_1]$. Since $\calS_1 \ind \calD_1$ by the construction of the h-BDMI procedure (see Section~\ref{standard_bayesian_approach}), we observe that
\begin{align}
\bbP_{\wt\calD_1}[\hspace{0.5mm} \bbE_{\tm_1 \sim \Pi_{\mbm}^{(1)}} \{\hspace{0.5mm} |\calZ_{\nK}(\tm_1, \calS_1)| \hspace{0.5mm} \hspace{0.5mm} \} > t \hspace{0.5mm} ] & ~=~ \bbE_{\wt\calD_1} [\hspace{0.5mm} \mathbf{1} ( \bbE_{\tm_1 \sim \Pi_{\mbm}^{(1)}} \{ \hspace{0.5mm} |\calZ_{\nK}(\tm_1, \calS_1)|\hspace{0.5mm} \} > t ) \hspace{0.5mm} ] \nonumber \\
& ~=~ \bbE_{\calS_1} ( \hspace{0.5mm}\bbE_{\calD_1} [ \hspace{0.5mm} \mathbf{1} ( \bbE_{\tm_1 \sim \Pi_{\mbm}^{(1)}} \hspace{0.5mm} \{ \hspace{0.5mm}|\calZ_{\nK}(\tm_1, \calS_1)| \hspace{0.5mm} \hspace{0.5mm} \} > t) \mid \calS_1 \hspace{0.5mm} ] \hspace{0.5mm} ) \nonumber \\
& ~=~ \bbE_{\calS_1} \left( \hspace{0.5mm} \bbP_{\calD_1} [ \hspace{0.5mm} \bbE_{\tm_1 \sim \Pi_{\mbm}^{(1)}} \{ \hspace{0.5mm} |\calZ_{\nK}(\tm_1, \calS_1)| \hspace{0.5mm} \} > t \mid \calS_1 \hspace{0.5mm} ] \hspace{0.5mm} \right),
\label{eq:equality_to_prove_T22}
\end{align}
where $\mathbf{1}(\cdot)$ denotes the indicator function. Since $0 ~\leq ~ \bbP_{\calD_1} [ \hspace{0.5mm} \bbE_{\tm_1 \sim \Pi_{\mbm}^{(1)}} \{ \hspace{0.5mm} |\calZ_{\nK}(\tm_1, \calS_1)| \hspace{0.5mm}\} > t \mid \calS_1 ] ~\leq ~ 1$, if we show that $\bbP_{\calD_1} [ \hspace{0.5mm} \bbE_{\tm_1 \sim \Pi_{\mbm}^{(1)}} \{ \hspace{0.5mm} |\calZ_{\nK}(\tm_1, \calS_1)| \hspace{0.5mm} \} > t \hspace{0.5mm} | \hspace{0.5mm} \calS_1 \hspace{0.5mm} ] \cvP 0$ w.r.t. $\bbP_{\calS_1}$, then by applying the DCT (or Lemma~\ref{chernozhukov_lemma6_1}), we obtain that $\bbE_{\calS_1}\left(\hspace{0.5mm} \bbP_{\calD_1}[ \bbE_{\tm_1 \sim \Pi_{\mbm}^{(1)}} \{ \hspace{0.5mm} |\calZ_{\nK}(\tm_1, \calS_1)| \hspace{0.5mm} \} > t \mid \calS_1 ] \hspace{0.5mm} \right) \cvP 0$ under $\bbP_{\wt \calD_1}$.

Further, we also observe that
\begin{align*}
\bbP_{\calD_1}\left[ \bbE_{\tm_1 \sim \Pi_{\mbm}^{(1)}} \{ \hspace{0.5mm} |\calZ_{\nK}(\hspace{0.5mm} \tm_1, \calS_1 \hspace{0.5mm})| \hspace{0.5mm} \} > t \mid \calS_1 \right] & ~=~ \bbP_{\calD_1}\left( \big [ \bbE_{\tm_1 \sim \Pi_{\mbm}^{(1)}} \{ \hspace{0.5mm} |\calZ_{\nK}(\tm_1, \calS_1)| \hspace{0.5mm} \} \big ]^2 > t^2 \mid \calS_1 \right) \\
& ~\leq~ \bbP_{\calD_1}\left[ \bbE_{\tm_1 \sim \Pi_{\mbm}^{(1)}} \{\hspace{0.5mm} \calZ_{\nK}^2(\tm_1, \calS_1) \hspace{0.5mm} \} > t^2 \mid \calS_1 \right]  \\
& ~\leq~ t^{-2}~ \bbE_{\calD_1}\left[ \bbE_{\tm_1 \sim \Pi_{\mbm}^{(1)}} \{ \hspace{0.5mm} \calZ_{\nK}^2(\tm_1, \calS_1) \hspace{0.5mm} \} \mid \calS_1 \right] \\
& ~=~ t^{-2}~ \bbE_{\tm_1 \sim \Pi_{\mbm}^{(1)}} \left[\bbE_{\calD_1}\{\hspace{0.5mm} \calZ_{\nK}^2(\hspace{0.5mm} \tm_1, \calS_1) \hspace{0.5mm} \} \mid \calS_1 \right],
\end{align*}
where the last three steps come from Cauchy--Schwarz inequality, Markov's inequality, and Fubini's theorem, respectively. Next, by the construction of $\calZ_{\nK}(\tm_1, \calS_1) \equiv [\bbG_{\nK}^{(1)}\{ \tm_1(\boldX) - m^*(\boldX) \} \mid \calS_1]$, we note that {\it given} $\tm_1$, it is a sum of centered $\sqrt{\nK}$-scaled {\it independent} random variables. This implies that $\bbE_{\calD_1}\{ \calZ_{\nK}(\tm_1, \calS_1) \hspace{0.5mm}|\hspace{0.5mm} \tm_1\} = 0$, and further, $\Var_{\calD_1}\{\calZ_{\nK}(\tm_1, \calS_1) \hspace{0.5mm}|\hspace{0.5mm} \tm_1\} =  \bbE_{\calD_1}\{\calZ^2_{\nK}(\tm_1, \calS_1) \hspace{0.5mm}|\hspace{0.5mm} \tm_1\}$. Next, by utilizing the definition of $\calZ_{\nK}(\tm_1, \calS_1)$, we calculate that $\Var_{\calD_1}\{\calZ_{\nK}(\tm_1, \calS_1) \hspace{0.5mm}|\hspace{0.5mm} \tm_1\} = \Var_{\boldX}[\{ \tm_1(\boldX) - m^*(\boldX)\} \hspace{0.5mm}|\hspace{0.5mm} \tm_1]$.

Therefore, by using the observations above, we finally obtain that
\begin{align*}
\bbP_{\calD_1} [ \hspace{0.5mm} \bbE_{\tm_1 \sim \Pi_{\mbm}^{(1)}} \{ \hspace{0.5mm} |\calZ_{\nK}(\tm_1, \calS_1) | \hspace{0.5mm} \} > t \mid \calS_1 \hspace{0.5mm} ] & ~\leq ~ t^{-2}\hspace{0.5mm} \bbE_{\tm_1 \sim \Pi_{\mbm}^{(1)}} \left( \hspace{0.5mm} \Var_{\boldX} [ \{ \tm_1(\boldX) - m^*(\boldX) \} \mid \tm_1] \mid \calS_1 \hspace{0.5mm} \right) \\
& ~\leq~ t^{-2}\hspace{0.5mm} \bbE_{\tm_1 \sim \Pi_{\mbm}^{(1)}} \{ \hspace{0.5mm} \| \tm_1(\boldX) - m^*(\boldX)\|^2_{\bbL_2(\bbP_{\boldX})} \mid \calS_1 \hspace{0.5mm} \}.
\end{align*}
Then, by using the nuisance Bayes risk condition given in Theorem~\ref{bvm_standard}, we directly have that the RHS of the inequality above converges to zero in probability w.r.t. $\bbP_{\calS_1}$ which implies that $\bbP_{\calD_1} [ \hspace{0.5mm} \bbE_{\tm_1 \sim \Pi_{\mbm}^{(1)}}\{|\calZ_{\nK}(\tm_1, \calS_1)|\} > t \mid \calS_1 \hspace{0.5mm} ] \cvP 0$ under $\bbP_{\calS_1}$. Next, by using the DCT (or Lemma~\ref{chernozhukov_lemma6_1} (b)), we obtain that
$$
\bbE_{\calS_1} \left( \hspace{0.5mm} \bbP_{\calD_1}[ \hspace{0.5mm} \bbE_{\tm_1 \sim \Pi_{\mbm}^{(1)}} \{ |\calZ_{\nK}( \tm_1, \calS_1 )| \} \hspace{0.5mm}  > t \mid \calS_1 \hspace{0.5mm}] \hspace{0.5mm} \right) ~\cvP~ 0 ~\text{ under }\, \bbP_{\wt \calD_1}.
$$
By using the equality in \eqref{eq:equality_to_prove_T22} and recalling the definition of $\calZ_{\nK}( \tm_1, \calS_1 )$ therein, this equivalently implies that $\bbE_{\tm_1 \sim \Pi_{\mbm}^{(1)}} [ \hspace{0.5mm} |\bbG_{\nK}^{(1)}\{ \tm_1(\boldX) - m^*(\boldX) \} | \mid \calS_1 \hspace{0.5mm} ] \to 0$ in probability under $\bbP_{\wt\calD_1}$, as $n \to \infty$.

\par\medskip
Similarly, by following the same steps and calculations above, but this time for the second term $\bbE_{\tm_1 \sim \Pi_{\mbm}^{(1)}} [ \hspace{0.5mm} |\bbG_{\NK}^{(1)} \{ \tm_1(\boldX) - m^*(\boldX) \} | \mid \calS_1 \hspace{0.5mm} ] $, we obtain that $\bbE_{\tm_1 \sim \Pi_{\mbm}^{(1)}} [ \hspace{0.5mm} |\bbG_{\NK}^{(1)} \{ \tm_1(\boldX) - m^*(\boldX) \} | \mid \calS_1 \hspace{0.5mm} ]  \cvP 0$ under $\bbP_{\wt\calD_1}$.
Recalling the bound
obtained for $T_{22}$ above, and since as $n, N \to \infty$, $n/N \to c \in [0, 1)$, we then conclude that $T_{22} \cvP 0$ under $\bbP_{\wt \calD_1}$ as $n, N \to \infty$. Hence, this completes the entire
proof.
\qeds

\section[Proofs of the remaining results]{Proofs of the remaining results: Preliminary and intermediate lemmas}\label{supp:prelim_proof}

In this section, we provide proofs of the preliminary results (Lemmas~\ref{TV_invariance}--\ref{chernozhukov_lemma6_1}) employed in the proofs of the main results in Section~\ref{appendix}, as well as proof of Lemma~\ref{lemma_half_of_data} introduced in the course of proving Theorem~\ref{bvm_on_first_half_data}. Note also that two of the preliminary Lemmas~\ref{klartag} and~\ref{chernozhukov_lemma6_1} are directly adopted from existing papers.

\subsection{Proof of Lemma~\ref{TV_invariance}} We give a proof for the sake of completeness. We have
\begin{align*}
\|P^{\mu, \sigma} - Q^{\mu, \sigma} \|_\TV ~ = ~ & \frac{1}{2} \int |p^{\mu, \sigma}(x) - q^{\mu, \sigma}(x)| \hspace{0.5mm} \dd x
~ = ~ \frac{1}{2} \int \left|\frac{1}{\sigma} \hspace{0.5mm} p\left(\frac{x - \mu}{\sigma} \right) - \frac{1}{\sigma} \hspace{0.5mm} q\left(\frac{x - \mu}{\sigma} \right) \right|\hspace{0.5mm} \dd x  \\
~ = ~ & \frac{1}{2} \int |p(t) - q(t)| \hspace{0.5mm}\dd t ~ = ~ \| P - Q \|_{\TV},
\end{align*}
where going from the first to the second line, we make a change of variable $t = (x - \mu)/\sigma$. \qeds
\subsection{Proof of Lemma~\ref{TV_2normal}} By using the invariance property of the TV distance from Lemma~\ref{TV_invariance}, $\|P - Q \|_{\TV} = \|\calN(\alpha, 1) - \calN(0, 1) \|_{\TV}$, where $\alpha = (\mu_1 - \mu_2)/\sigma$. The result now follows from Lemma 4 of \citet{bontemps2011bernstein}.
\qeds
\subsection{Proof of Lemma~\ref{TV_tnormal}} Since the TV distance is invariant under scaling and location-shift (see Lemma~\ref{TV_invariance}) and both the $t$-distribution $t_\nu(\mu, \sigma^2)$  and the Normal distribution $\calN(\mu, \sigma^2)$ belong to a location-scale family, we have that $\|t_\nu(\mu, \sigma^2) - \calN(\mu, \sigma^2) \|_{\TV} = \|t_\nu - \calN(0, 1) \|_{\TV}$, where $t_\nu \equiv t_\nu(0, 1)$. This implies that the TV distance is free of the parameters $\mu$ and $\sigma^2$. Further, we note that the $t$-distribution $t_{\nu}(\mu, \sigma^2)$ can be expressed as a precision mixture of a Gaussian distribution \citep{west1987scale}. In particular, suppose $X \mid W \sim \calN(\mu, W^{-1} \sigma^2)$, and $W \sim \text{Gamma}(\nu/2, \nu/2)$ with the pdf $f_{W}(\cdot)$, then, $X \sim t_{\nu}(\mu, \sigma^2)$. Since $f_{W}(\cdot)$ is a pdf, we can write the TV distance above as follows:
\begin{align*}
\|t_\nu - \calN(0, 1) \|_{\TV}  & ~=~  \frac{1}{2} \int |t_\nu(x; 0, 1) - \calN(x; 0, 1)|\hspace{0.5mm} \dd x \\
& ~=~ \frac{1}{2} \int \left| \int \{f_{X \mid W}(x) - \calN(x; 0, 1)\}\hspace{0.5mm} f_{W}(w) \hspace{0.5mm} \dd w \right| \dd x,
\end{align*}
where $f_{X \mid W}(\cdot)$ is the pdf of the conditional random variable $X \hspace{0.5mm}|\hspace{0.5mm} W$ having a Normal $\calN\big(0, W^{-1}\big)$ distribution.

Further, we obtain that
\begin{align*}
    \|t_\nu - \calN(0, 1) \|_{\TV}
    & ~\leq~ \frac{1}{2}  \int \int | f_{X \mid W}(x) -\calN(x; 0, 1)| f_{W}(w)  \hspace{0.5mm} \dd w \hspace{0.5mm} \dd x \\
    & ~=~ \frac{1}{2} \int \int | f_{X \mid W}(x) -\calN(x; 0, 1) | \hspace{0.5mm}\dd x \hspace{0.5mm}f_{W}(w) \hspace{0.5mm}\dd w ,
\end{align*}
where the last step uses Fubini's theorem to change the order of the integrals.

Next, by following the integral representation of the TV distance, we further obtain that
\begin{align*}
\|t_\nu - \calN(0, 1) \|_{\TV} ~\leq~ \int \|\calN(0, w^{-1}) - \calN(0,1) \|_{\TV} \hspace{0.8mm} f_{W}(w) \hspace{0.5mm} \dd w.
\end{align*}
Then, by using Lemma~\ref{klartag}, there is a constant $C > 0$ such that $\|\calN(0, w^{-1}) - \calN(0,1) \|_{\TV} \leq C |w -1|$. Since $W \sim \text{Gamma}(\nu/2, \nu/2)$, by using the Cauchy--Schwarz inequality, for some $C_0 > 0$, we have
$
\|t_\nu - \calN(0, 1) \|_{\TV}  ~\leq~ C \int |w - 1| \hspace{0.5mm} f_{W}(w) \hspace{0.7mm} \dd w ~\leq~ C \hspace{0.5mm} [\bbE_{W}(W - 1)^2]^{1/2} ~=~ C_0/\sqrt{\nu}.
$ \qeds

\subsection{Proof of Lemma~\ref{TVD_convolution}} The proof follows from the definition of the convolution operator and the triangle inequality. By the triangle inequality and Fubini's theorem, we observe that
\begin{align*}
&\| P - Q \|_{\TV} ~=~ \frac{1}{2} \int |p(z) - q(z)| \hspace{0.7mm}\dd z ~=~ \frac{1}{2} \int \left| \int p_1(x) p_2(z - x) \hspace{0.7mm}\dd x - \int q_1(x)q_2(z- x) \hspace{0.7mm} \dd x \right| \dd z \\
& ~\leq~ \frac{1}{2} \int \int  \left\{ | p_1(x) p_2(z - x) - q_1(x)p_2(z - x)| ~+~ | q_1(x)p_2(z - x) - q_1(x)q_2(z- x)|  \right\} \hspace{0.7mm} \dd x \hspace{0.7mm} \dd z \\
& ~=~  \frac{1}{2} \int \int |p_1(x) - q_1(x)|p_2(z - x) \hspace{0.7mm} \dd z \hspace{0.7mm}  \dd x ~+~ \frac{1}{2} \int \int q_1(x)|p_2(z - x) - q_2(z- x)|  \hspace{0.7mm} \dd z  \hspace{0.7mm} \dd x \\
 & ~=~  \frac{1}{2} \int|p_1(x) - q_1(x)| \hspace{0.7mm}\dd x + \frac{1}{2} \int |p_2(w) - q_2(w)| \hspace{0.7mm} \dd w ~=~\|P_1 - Q_1 \|_{\TV} ~+~ \|P_2 - Q
_2 \|_{\TV}.
\end{align*}
Hence, we have obtained the desired inequality. This completes the proof. \qeds

\subsection{Proof of Lemma~\ref{lemma_half_of_data}}\label{proof_of_lemma_half_of_data}

We start with writing the distribution $\bbP_{\tm_1}$ explicitly. Since the randomness of $\tm_1$ comes from both the data $\calS_1$ and the posterior distribution $\Pi_{\mbm}^{(1)}(\cdot) \equiv \Pi_{\mbm}^{(1)}(\cdot;\calS_1)$ itself, we have that $\bbP_{\tm_1} = \bbP_{\calS_1} \otimes \Pi_{\mbm}^{(1)}$. Then, to establish the result, by the definition of convergence in probability, we need to show that for any $t > 0$, $\bbP_{\tm_1}\{\| m^*(\boldX) - \tm_1(\boldX) \|_{\bbL_2(\bbP_{\boldX})} > t\} \to 0$ as $n \to \infty$. Towards this goal, we first observe that
\begin{align}
& \bbP_{\tm_1}\{\hspace{0.5mm} \| m^*(\boldX) - \tm_1(\boldX) \|_{\bbL_2(\bbP_{\boldX})} > t \hspace{0.5mm}\}  ~=~ \bbE_{\tm_1}\{\mathbf{1}(\hspace{0.5mm} \| m^*(\boldX) - \tm_1(\boldX) \|_{\bbL_2(\bbP_{\boldX})} > t \hspace{0.5mm})\}\nonumber \\
& ~=~ \bbE_{\calS_1}[\hspace{0.5mm} \bbE_{\mbm \mid \calS_1}\{\hspace{0.5mm} \mathbf{1}(\hspace{0.5mm} \| m^*(\boldX) - \tm_1(\boldX) \|_{\bbL_2(\bbP_{\boldX})} > t \hspace{0.5mm}) \mid \calS_1  \hspace{0.5mm}\} \hspace{0.5mm}]
\nonumber \\
& ~=~ \bbE_{\calS_1}[\hspace{0.5mm} \Pi_{\mbm}^{(1)}\{ \hspace{0.5mm}\| m^*(\boldX) - \tm_1(\boldX) \|_{\bbL_2(\bbP_{\boldX})} > t \mid \calS_1 \hspace{0.5mm}\} \hspace{0.5mm}] ~:=~ \bbE_{\calS_1}[\hspace{0.5mm}  T\{\hspace{0.5mm} \calS_1(n), t \hspace{0.5mm}\} \hspace{0.5mm} ], \label{equivalence_Pm} \\
& ~~~ \quad \mbox{ where} ~~  T\{\calS_1(n), t\} \ := \ \Pi_{\mbm}^{(1)} \{\hspace{0.5mm} \| m^*(\boldX) - \tm_1(\boldX) \|_{\bbL_2(\bbP_{\boldX})} > t \mid \calS_1 \hspace{0.5mm} \}, \label{def:TS1nt}
\end{align}
and $\mathbf{1}(\cdot)$ denotes the indicator function, and the second step uses the law of iterated expectation. We note that the notation $T\{\calS_1(n), t\}$ above indicates the dependence on the size of $\calS_1$ (note that its size $n - n/K$
is of the same order as $n$) and the given $t >0$. Further, since $T\{\calS_1(n), t\}$ itself is a probability as in \eqref{def:TS1nt}, $T\{\calS_1(n), t\} \in [0, 1]$, and moreover, \eqref{def:TS1nt} implies that: for any $0 < t_1 < t_2$,
\begin{align}
& T\{\calS_1(n), t_1\} ~\geq~ T\{\calS_1(n), t_2\}, ~~\mbox{so that} ~~\bbP_{\calS_1}[\hspace{0.5mm} T\{\calS_1(n), t_1\} \geq T\{\calS_1(n), t_2\} \hspace{0.5mm}] \hspace{0.75mm} = \hspace{0.75mm} 1  ~~ \forall \ t_1 < t_2. \label{eq:T1Snt:monotonicity}
\end{align}
Then, since $T\{\calS_1(n), t\}$ is bounded, an application of the DCT (or Lemma~\ref{chernozhukov_lemma6_1}) ensures that it suffices to show $T\{\calS_1(n), t\} \cvP 0$ under $\bbP_{\calS_1}$. Now, recall the NPCC in Assumption~\ref{assumption_for_half_fold} (ii): for some $a_n \to 0$, $\Pi_{\mbm}^{(k)} \{ m: \| m^*(\boldX) - m(\boldX) \|_{\bbL_2(\bbP_{\boldX})} > a_n \mid \calS_1 \} \cvP 0$ under $\bbP_{\calS_1}$, as $n \to \infty$ (we here take $k = 1$).

Then, using \eqref{def:TS1nt}, we can {\it rewrite} the NPCC as follows: as $n \to \infty$, for some $a_n \to 0$,
\begin{align}
    T\{\calS_1(n), a_n\} ~=~ o_{\bbP_{\calS_1}}(1), ~~\mbox{or equivalently,}~~
    \bbP_{\calS_1}[\hspace{0.7mm} T\{\calS_1(n), a_n\} > \gamma \hspace{0.7mm}] \to 0 ~ \mbox{ for any } ~ \gamma > 0. \label{eq:NPCC:altform}
\end{align}
The RHS above equivalently says that for any $\delta > 0$ there exists a $n_{\gamma, \delta}$ such that for any $n \geq n_{\gamma, \delta}$, $\bbP_{\calS_1}[\hspace{0.7mm} T\{\calS_1(n), a_n\} > \gamma \hspace{0.7mm}] < \delta$. Note that the double index in $n_{\gamma, \delta}$ indicates the dependence on both $\gamma$ and $\delta$.

Also, we observe that for the given $t > 0$, since $a_n \to 0$, there exists $n_{t}$ such that for all $n \geq n_t$, $a_n < t$ (and recall that $a_n \geq 0$ by definition), almost surely (a.s.) w.r.t. $\bbP_{\calS_1}$ (i.e., $\bbP_{\calS_1}[ \hspace{0.5mm} T\{\calS_1(n), a_n\} \geq
T\{\calS_1(n), t\} \hspace{0.5mm}] = 1$) for all $n \geq n_t$.

Now, by using the definition of convergence in probability, we ultimately need to show that for any $\varepsilon > 0$, $\bbP_{\calS_1}[\hspace{0.5mm} T\{S_1(n), t\} > \varepsilon \hspace{0.5mm} ] \to 0$ as $n \to \infty$, or equivalently, for any $\delta > 0$, there exists a $n_{\varepsilon, \delta}^*$ such that for any $n \geq n_{\varepsilon, \delta}^*$, $\bbP_{\calS_1}[ \hspace{0.5mm} T\{S_1(n), t\} > \varepsilon \hspace{0.5mm}] < \delta$.

Toward showing this, we let $n_{\varepsilon, \delta}^* := \max\{ n_{\gamma, \delta}, n_{t}\}$ (by recalling the terms from the observations above and also setting $\gamma = \varepsilon$). Then, for any $n \geq n_{\varepsilon, \delta}^*$, by the total law of probability, we have that
\begin{align*}
    \bbP_{\calS_1}[ T\{\calS_1(n), t\} > \varepsilon ]  ~ = ~ & \ \bbP_{\calS_1}[ \hspace{0.7mm} T\{\calS_1(n), t\} > \varepsilon , T\{\calS_1(n), a_n\} > \varepsilon] \\
    &~+  \bbP_{\calS_1}[ T\{\calS_1(n), t\} > \varepsilon , ~\hspace{0.5mm} T\{\calS_1(n), a_n\} \leq  \varepsilon \hspace{0.7mm}] \\
 \leq ~ & \ \bbP_{\calS_1}[ \hspace{0.7mm} T\{\calS_1(n), a_n\} > \varepsilon \hspace{0.7mm} ] + \bbP_{\calS_1}[ \hspace{0.7mm} T\{\calS_1(n), a_n\} \leq T\{\calS_1(n), t\} \hspace{0.7mm} ] ~ < ~ \delta,
\end{align*}
where the last step uses the following observations: $\bbP_{\calS_1}[ \hspace{0.4mm} T\{\calS_1(n), a_n\} > \varepsilon \hspace{0.4mm} ] < \delta$, since $n \geq n_{\varepsilon, \delta}^* \geq n_{\gamma, \delta}$ (with $\gamma \equiv \epsilon$) using \eqref{eq:NPCC:altform}, and $\bbP_{\calS_1}[ \hspace{0.7mm} T\{\calS_1(n), a_n\} \leq T\{\calS_1(n), t\} \hspace{0.7mm} ] = 0$ using \eqref{eq:T1Snt:monotonicity}, since $n \geq n_{\varepsilon, \delta}^* \geq n_t$ (referring to the discussions and implications above). Hence, for the given $\varepsilon > 0$ and $t,\delta > 0$, we have: $\bbP_{\calS_1}[ \hspace{0.7mm} T\{\calS_1(n), t\} > \varepsilon \hspace{0.7mm} ] < \delta$ for any $n \geq n_{\varepsilon, \delta}^*$, which equivalently gives $T\{\calS_1(n), t\} = o_{\bbP_{\calS_1}}(1)$.

Finally, using the equality in \eqref{equivalence_Pm}, we can now apply the DCT (or Lemma~\ref{chernozhukov_lemma6_1}) to conclude that $\bbP_{\tm_1}\{\| m^*(\boldX) - \tm_1(\boldX) \|_{\bbL_2(\bbP_{\boldX})} > t\} \to 0$ for all $t > 0$, i.e., $\| m^*(\boldX) - \tm_1(\boldX) \| _{\bbL_2(\bbP_{\boldX})} = o_{\bbP_{\tm_1}}(1)$, as claimed.
\qeds
\end{document}